\documentclass[10pt, conference, letterpaper]{IEEEtran}

\usepackage{subcaption}
\usepackage[table,xcdraw]{xcolor}
\usepackage{multirow}
\usepackage{url}
\usepackage{makecell}
\usepackage{verbatim}

\IEEEoverridecommandlockouts
\usepackage{cite}
\usepackage{amsmath,amssymb,amsfonts}
\usepackage{algorithmic}
\usepackage{graphicx}
\usepackage{textcomp}

\usepackage{fancyhdr}
\def\BibTeX{{\rm B\kern-.05em{\sc i\kern-.025em b}\kern-.08em
T\kern-.1667em\lower.7ex\hbox{E}\kern-.125emX}}

\definecolor{light-gray}{gray}{0.9}
\definecolor{dark-gray}{gray}{0.5}

\begin{document}

\title{Internet Traffic Volumes Are Not Gaussian - They Are Log-Normal: An 18-Year Longitudinal Study With Implications for Modelling and Prediction (Complete Version)}

\author{\IEEEauthorblockN{Mohammed Alasmar}
\IEEEauthorblockA{\textit{Department of Informatics} \\
\textit{University of Sussex}\\
Brighton, UK \\
m.alasmar@sussex.ac.uk}
\and
\IEEEauthorblockN{Richard Clegg}
\IEEEauthorblockA{\textit{School of Computer Science } \\
\textit{Queen Mary University of London}\\
London, UK \\
r.clegg@qmul.ac.uk }
\and
\IEEEauthorblockN{Nickolay Zakhleniuk}
\IEEEauthorblockA{\textit{School of Computer Science} \\
\textit{University of Essex}\\
Colchester, UK\\
naz@essex.ac.uk}
\and
\IEEEauthorblockN{George Parisis}
\IEEEauthorblockA{\textit{Department of Informatics} \\
	\textit{University of Sussex}\\
	Brighton, UK  \\
	g.parisis@sussex.ac.uk}
}

\newenvironment{newtext}{\par\color{red}}{\par}

\maketitle
\thispagestyle{plain}
\pagestyle{plain}
\begin{abstract}
	\footnote{This paper has been accepted for publication at 
		IEEE/ACM Transactions on Networking 02/2021}Getting good statistical models of traffic on network links is a well-known, often-studied problem.  A lot of attention has been given to correlation patterns and flow duration.  The distribution of the amount of traffic per unit time is an equally important but less studied problem.  We study a large number of traffic traces from many different networks including academic, commercial and residential networks using state-of-the-art statistical techniques.  We show that traffic obeys the log-normal distribution which is a better fit than the Gaussian distribution commonly claimed in the literature. We also investigate an alternative heavy-tailed distribution (the Weibull) and show that its performance is better than Gaussian but worse than log-normal.  We examine anomalous traces which exhibit a poor fit for all distributions tried and show that this is often due to traffic outages or links that hit maximum capacity. We demonstrate that the data we look at is stationary if we consider samples of 15-minute long or even  1-hour long. This gives confidence that we can use the distributions for estimation and modelling purposes.
	We demonstrate the utility of our findings in two contexts: predicting that the proportion of time traffic will exceed a given level (for service level agreement or link capacity estimation) and predicting 95th percentile pricing.  We also show that the log-normal distribution is a better predictor than Gaussian or Weibull distributions in both contexts.
\end{abstract}


\section{Introduction}
\label{introduction}
Internet traffic characterisation is an important problem for network researchers and vendors. The subject has a long history. Early works~\cite{selfSimilarity95, self-sim97} discovered that the correlation structure of traffic exhibits self-similarity and that the durations of individual flows of packets show heavy-tails~\cite{heavy-tailed-2010-trans}. These works were later challenged and refined (see Section~\ref{sec:related} for a summary). By comparison, the distribution of the amount of traffic present on a link in a given time period has seen comparatively less research interest. This is surprising as correct traffic statistics can be extremely useful in network planning. In this paper we use a rigorous statistical approach to fitting a statistical distribution to the amount of traffic within a given time period. Formally, we choose some timescale $T$ and let $X_i$ be the amount of traffic seen in the time period $[iT, (i+1)T)$. We investigate the distribution of the random variable $X$ over a wide range of values of $T$. We show that the distribution of the variable has considerable implications for network planning; for assessing how often a link is over capacity and in particular for service level agreements (SLAs), and for traffic pricing, particularly using the 95th percentile scheme~\cite{95percentileIMC}.

Previous authors have claimed that $X$ has a normal (or Gaussian) distribution~\cite{Gaussian-everywhere,proveGaussianIFIP,Gaussian-revisited}. Others claim $X$ is Gaussian plus a tail associated with bursts~\cite{2014-ifip-conf,12-GLOBECOM2002}.  All these studies are based on straightforward goodness-of-fit tests (e.g. Quantile-Quantile (Q-Q) plots) and relevant correlation tests that are used to assess how well captured traffic traces are fitted to Gaussian or heavy-tailed distributions. As discussed in \cite{clauset}, these statistical approaches can produce a substantially inaccurate assessment about whether samples follow a Gaussian/heavy-tailed or not. This is because the difference in these distributions lies in the behaviour of the tail where there can be relatively few samples, therefore large amounts of data and careful statistical handling are both important to determine the correct distribution~\cite{clauset}. 

In this paper, we use a well-established statistical methodology~\cite{clauset} to show that a log-normal\footnote{A variable $X$ has a log-normal distribution if its logarithm is normally distributed $\ln(X) \sim N(\mu,\sigma^2)$ where $\mu \in \mathbb{R}$ is the mean and $\sigma > 0$ is the standard deviation of the distribution (see Table 1 in~\cite{clauset}).} distribution is a better fit than Gaussian or Weibull\footnote{A variable $X$ has a Weibull distribution with parameters $k > 0$ (known as shape) and $\lambda > 0$ (known as scale) if its probability density function follows $f(x) = \frac{k}{\lambda}\left(\frac{x}{\lambda}\right)^{k-1} \exp(-(x/\lambda)^k)$ when $x \geq 0$ and is $0$ otherwise.} for the vast majority of traces.  This holds over a wide range of timescales $T$ (from $5$ ms to $5$ sec)~\cite{our-infocom-paper}. The question becomes whether the log-normal model remains applicable for all traffic volumes that have the same time periods and aggregation timescales used for testing this model. Stationarity tests can answer this question as stationarity from a probabilistic point of view means that the distribution remains unchanged for whatever shift of time window~\cite{Lauks2011, stationaryModelingLS, NonstationarityIP}. The vast majority of modelling techniques developed for volume-based traffic profiling imply the assumption of statistical stationarity which is often taken for granted without being explicitly validated~\cite{2014-ifip-conf,Gaussian-everywhere,proveGaussianIFIP}. In contrast, we extensively test all studied time series for stationarity using state of the art techniques and examine their trend and seasonality components.  The majority of the 15-minute and 1-hour long traces in the dataset are stationary when aggregated at timescales of $500$ ms to $5$ sec.
 
This paper is the most comprehensive investigation of this phenomenon the authors know about. We study a large number of publicly available traces from a diverse set of locations (including commercial, academic and residential networks) with different link speeds and spanning the last 18 years. There is a small number of anomalous traces in our datasets where the distribution deviates from log-normal and we find that this occurs when a link spends considerable time either having an outage or completely at maximum capacity. These anomalous traces can be presented using a bimodal distribution.

We show how often a link following a given distribution will be over a given capacity and show that our approach improves greatly on results which assume that traffic follows a Gaussian distribution. We further show that if an ISP wishes to estimate future transit bills that use the 95th percentile billing scheme, then the log-normal is a better model than the Gaussian distribution. 

The structure of the paper is as follows. In Section~\ref{sec:dataset} we describe the  datasets used. In Section~\ref{sec:fitting} we describe our best-practice procedure for fitting traffic and demonstrate that log-normal is the best fit distribution among all studied distributions for our traces under a variety of circumstances. We examine those few traces that do not follow this distribution and find it occurs when a link spends considerable time either having an outage or being completely at maximum capacity. In Section~\ref{sec:stationarity} we test all studied time series for stationarity at different timescales. In Section~\ref{sec:provision} we demonstrate that the log-normal distribution is the most suitable for estimating how often a link is over capacity. In Section~\ref{sec:pricing} we show that the log-normal distribution provides good estimates when looking at 95th percentile pricing. In Section~\ref{sec:related} we give related work. Finally, Section~\ref{sec:conclusion} gives our conclusions.

\begin{figure*}[t]
	\setlength{\belowcaptionskip}{-1pt}
	\centering
	\subcaptionbox{CAIDA traces}[.30\linewidth][c]{%
		\includegraphics[scale=0.27]{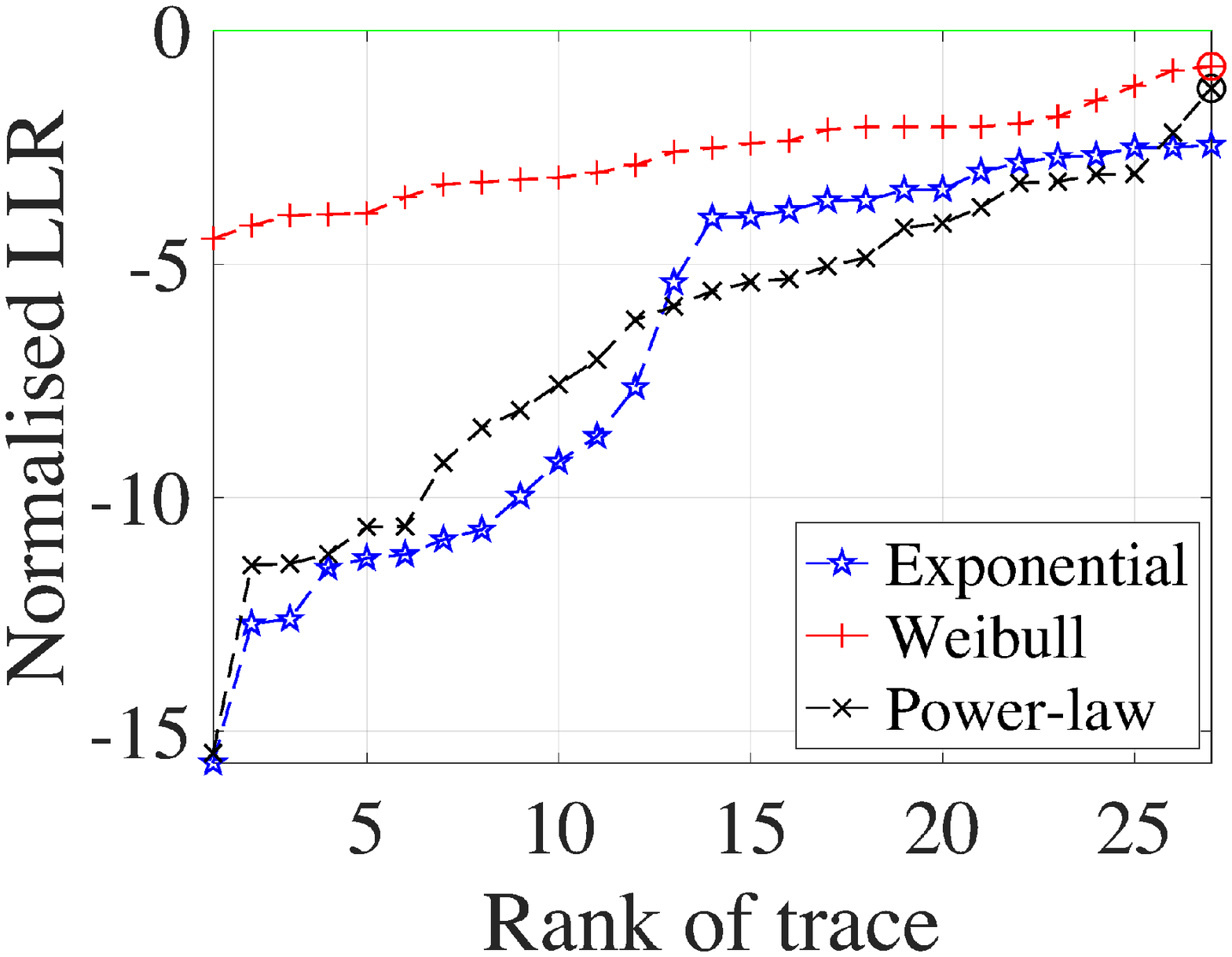}}\quad
	\subcaptionbox{Waikato traces}[.30\linewidth][c]{%
		\includegraphics[scale=0.27]{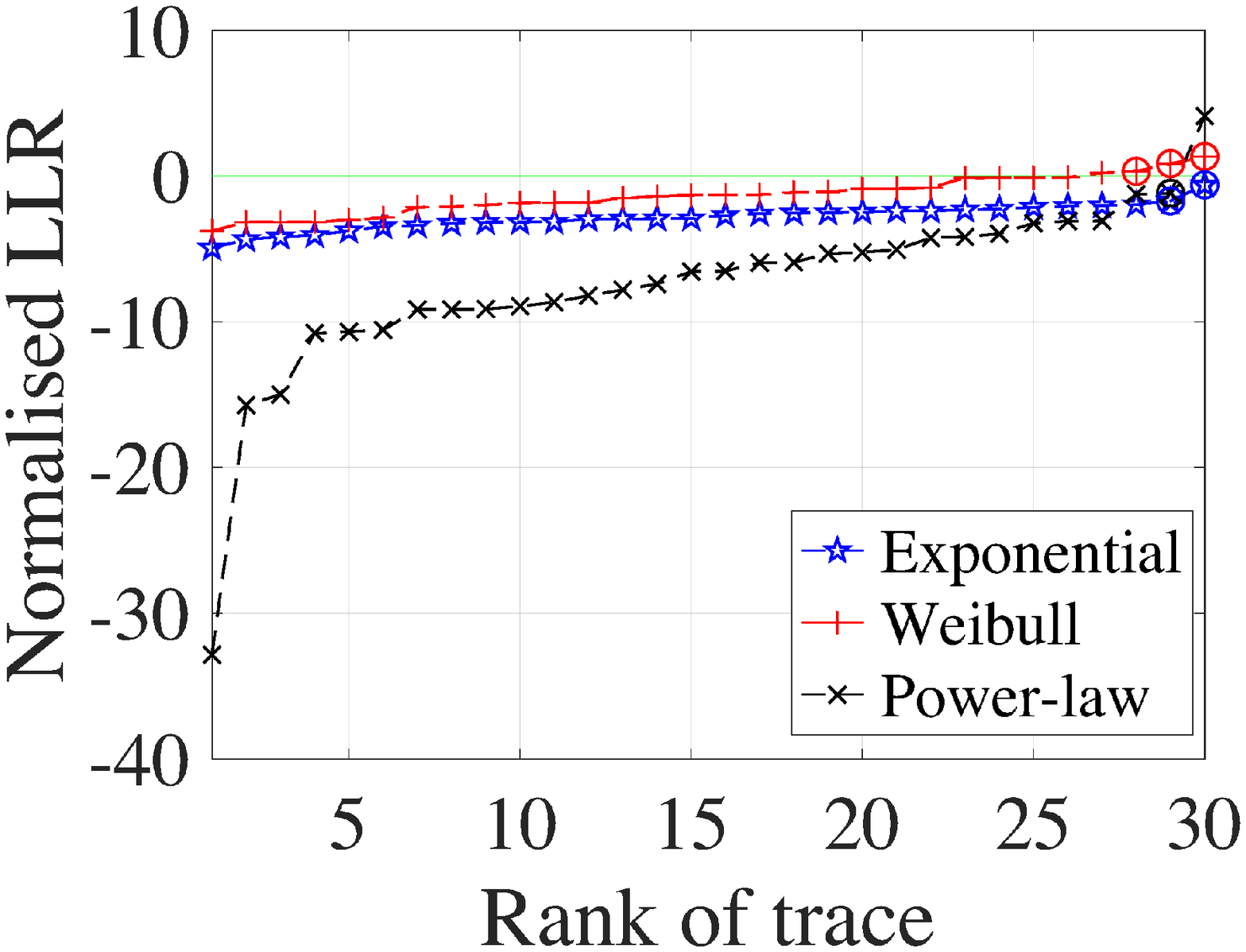}}\quad
	\subcaptionbox{Auckland traces}[.30\linewidth][c]{%
		\includegraphics[scale=0.27]{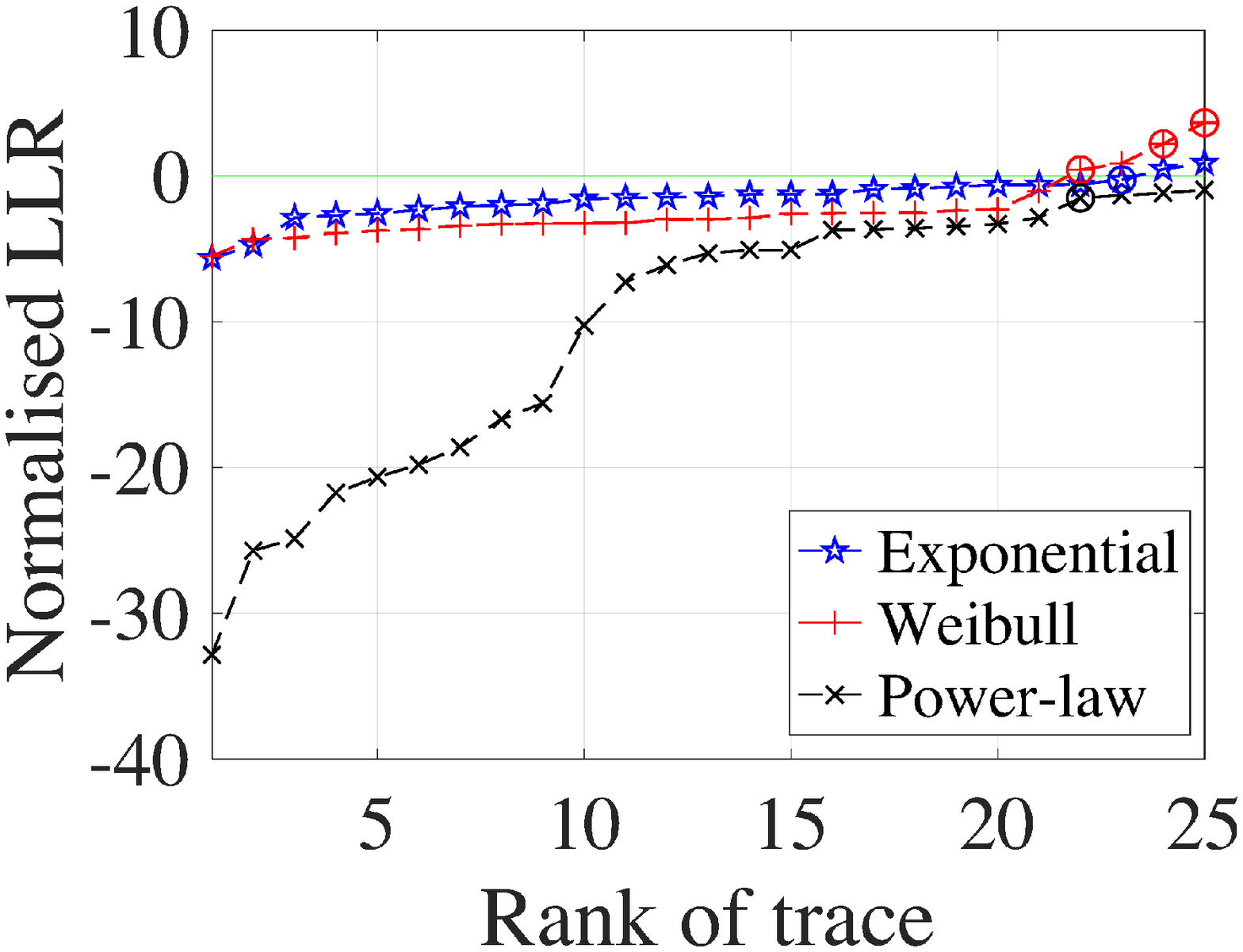}}
	\bigskip
	\setlength{\belowcaptionskip}{-8pt}
	\subcaptionbox{Twente traces}[.30\linewidth][c]{%
		\includegraphics[scale=0.27]{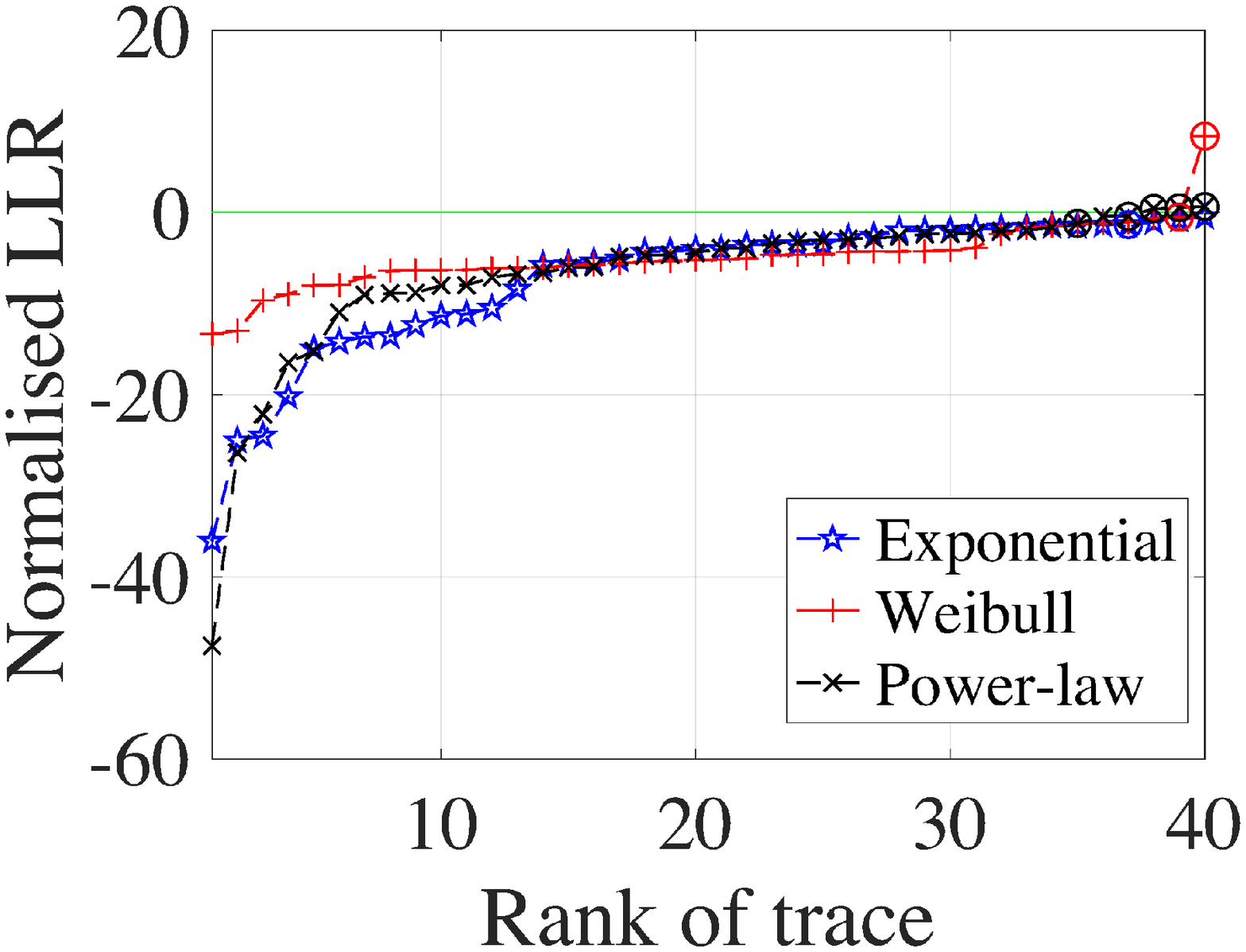}}\quad
	\subcaptionbox{MAWI traces}[.30\linewidth][c]{%
		\includegraphics[scale=0.27]{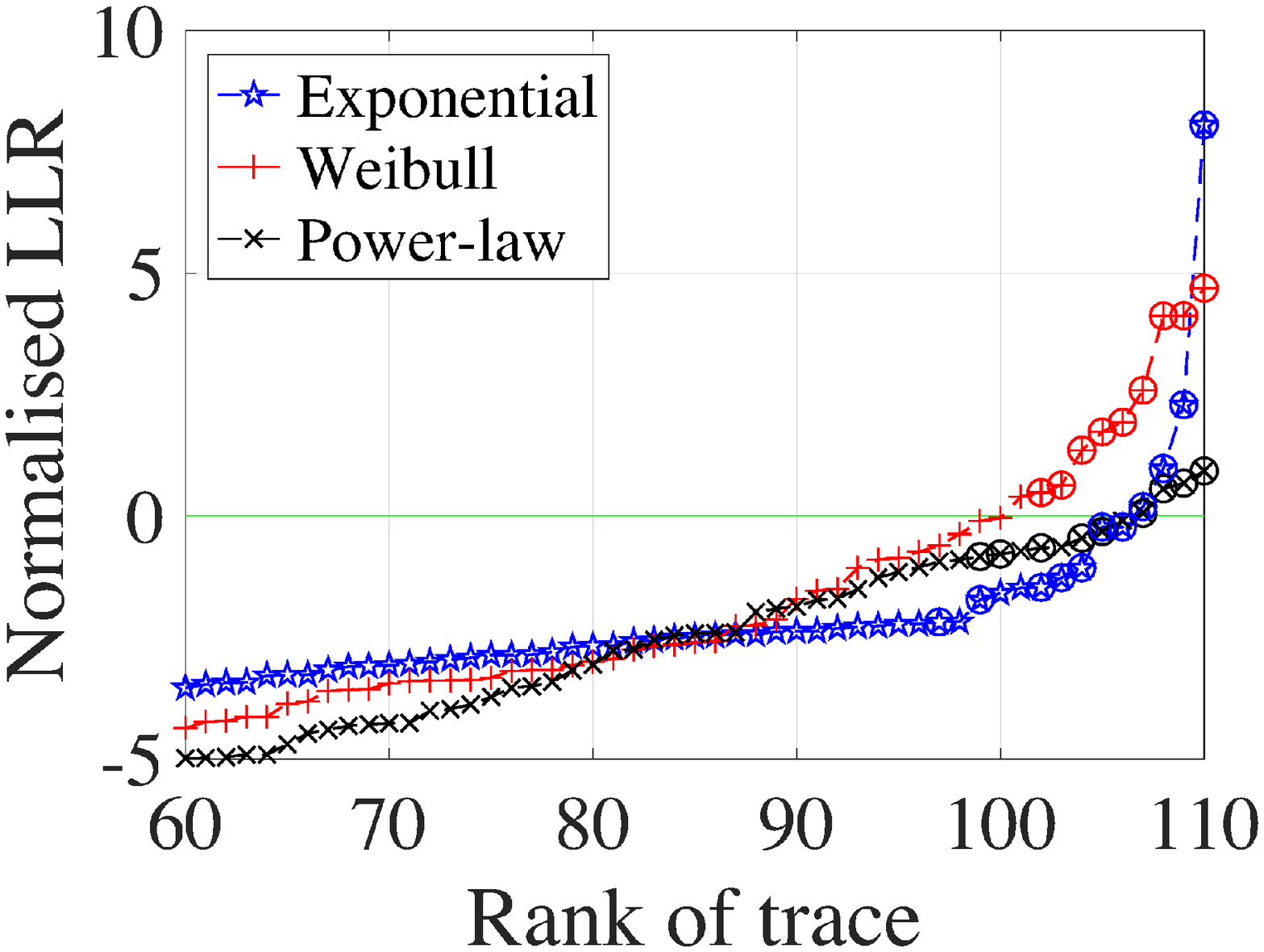}}\quad
	\caption{Normalised Log-Likelihood Ratio ($\Re$) test results for all studied traces and candidate distributions when running $fit.distCompare$(alternative, log-normal). Aggregation timescale $T$ is 100 msec.  Circled points in the plots are the ones with $p$-value greater than $0.1$, i.e. test is inconclusive with respect to assessing which of the reference and candidate distributions is a better fit to the traffic data.}
	\label{LRtestResults} 	
\end{figure*}

\section{Network Traffic Traces}
\label{sec:dataset}

A key contribution of our work stems from the spatial and temporal diversity of the studied traces. The studied dataset spans a period of 18 years and comprises $232$ traces\footnote{All Matlab and Python scripts used for the data analysis presented in this paper can be found on \url{https://github.com/mohammedalasmar/InternetTraces}}.

\noindent\textbf{CAIDA traces.} We have used $27$ CAIDA traces captured at an Internet data collection monitor which is located at an Equinix data centre in Chicago~\cite{caidaRef}. The data centre is connected to a backbone link of a Tier 1 ISP. The  monitor records an hour-long trace four times a year, usually from $13{:}00$ to $14{:}00$ UTC. The selection of a 15-minute trace from the respective 1-hour trace is done by first splitting the trace into four 15-minute subtraces and, subsequently, selecting one of these at random. Each trace contains billions of IPv4 packets, the headers of which are anonymised. The average captured data rate is 2.5~Gbps. At the time of capturing, the monitored link had a capacity of 10~Gbps. Traces were captured between $2013$ and $2016$. 
\break 
\noindent\textbf{MAWI traces.} The MAWI archive~\cite{mawiRef} consists of a collection of Internet traffic traces, captured within the WIDE backbone network that connects Japanese universities and research institutions to the Internet. Each trace consists of IP level traffic observed daily from $14{:}00$ to $14{:}15$ at a vantage point within WIDE. Traces include anonymised IP and MAC headers, along with an \textit{ntpd} timestamp~\cite{mawiRef}. We have looked at $110$ traces (each one being $15$-minute long). Traces were captured between $2014$ and $2020$. On average, each trace consists of 70 million packets; the average captured data rate is 422~Mbps. The monitored link had a capacity of 1~Gbps. For the stationarity tests presented in Section~\ref{sec:stationarity} we used a 24-hour long MAWI trace\footnote{\url{https://mawi.wide.ad.jp/mawi/ditl/ditl2018-G/}}. This trace was captured on 09/05/2018 at samplepoint-G which monitors a 10~Gbps link to DIX-IE\footnote{\url{http://two.wide.ad.jp/}}. 
\break 
\noindent\textbf{Twente University traces.} We used $40$ traffic traces captured at five different locations ($8$ traces from each location). Traces are diverse in terms of the link rates, types of users and capture time~\cite{Meent2010Traces}. Each trace is $15$-minute long. The first location is a residential network with a 300~Mbps link, which connects 2000 students (each one having a 100~Mbps access link); traces were captured in July 2002. The second location is a research institute network with a 1~Gbps link which connects 200 researchers (each one having a 100~Mbps access link); traces were captured between May and August 2003. The third location is at a large college with a 1~Gbps link which connects 1000 employees (each one having a 100~Mbps access link); traces were captured between February and July 2004. The fourth location is an ADSL access network with a 1~Gbps ADSL link used by hundreds of users (each one having a 256~Kbps to 8~Mbps access link); traces were captured between February and July 2004. The fifth location is an educational organisation with a 100~Mbps link connecting 135 students and employees (each one having a 100~Mbps access link);  traces were captured between May and June 2007.
\break 
\noindent\textbf{Waikato University VIII traces.} The Waikato dataset consists of traffic traces captured by the WAND group at the University of Waikato, New Zealand~\cite{waikatoRef}. The capture point is at the link interconnecting the University with the Internet. All of the traces were captured using software that was specifically developed for the Waikato capture point and a DAG 3 series hardware capture card. All IP addresses within the traces are anonymised. In our study, we have used $30$ traces captured between April 2011 and November 2011.
\break 
\noindent\textbf{Auckland University IX  traces.} The Auckland dataset consists of traffic traces captured by the WAND group at the University of Waikato~\cite{aucklandRef}. The traces were collected at the University of Auckland, New Zealand. The capture point is at the link interconnecting the University with the Internet. All IP addresses within the traces are anonymised. In our study, we have used $25$ traces captured in 2009.
\break

\begin{figure*}[t]
	\centering
	\subcaptionbox{CAIDA traces}[.18\linewidth][]{%
		\includegraphics[scale=0.18]{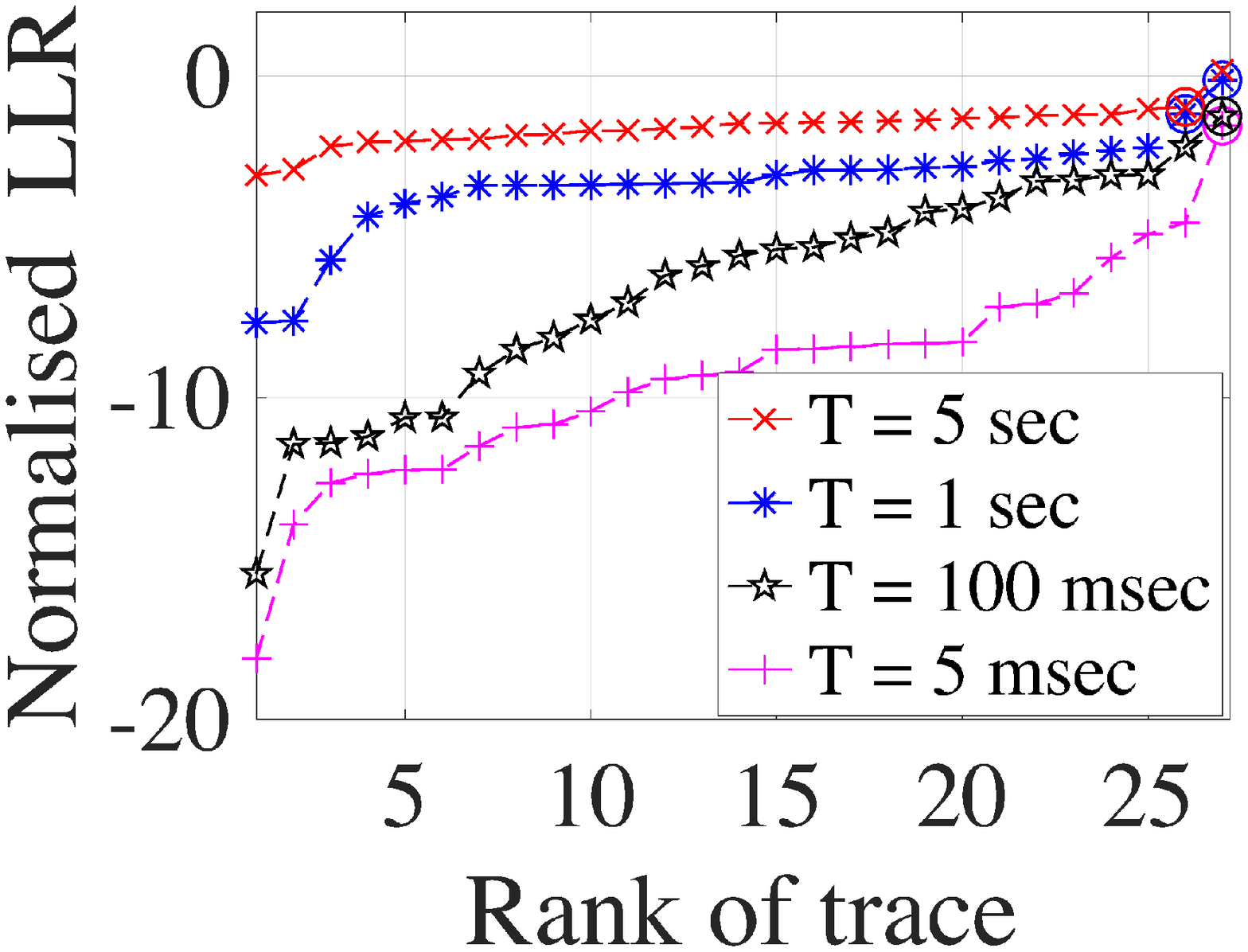}}\quad
	\subcaptionbox{Waikato traces}[.18\linewidth][]{%
		\includegraphics[scale=0.18]{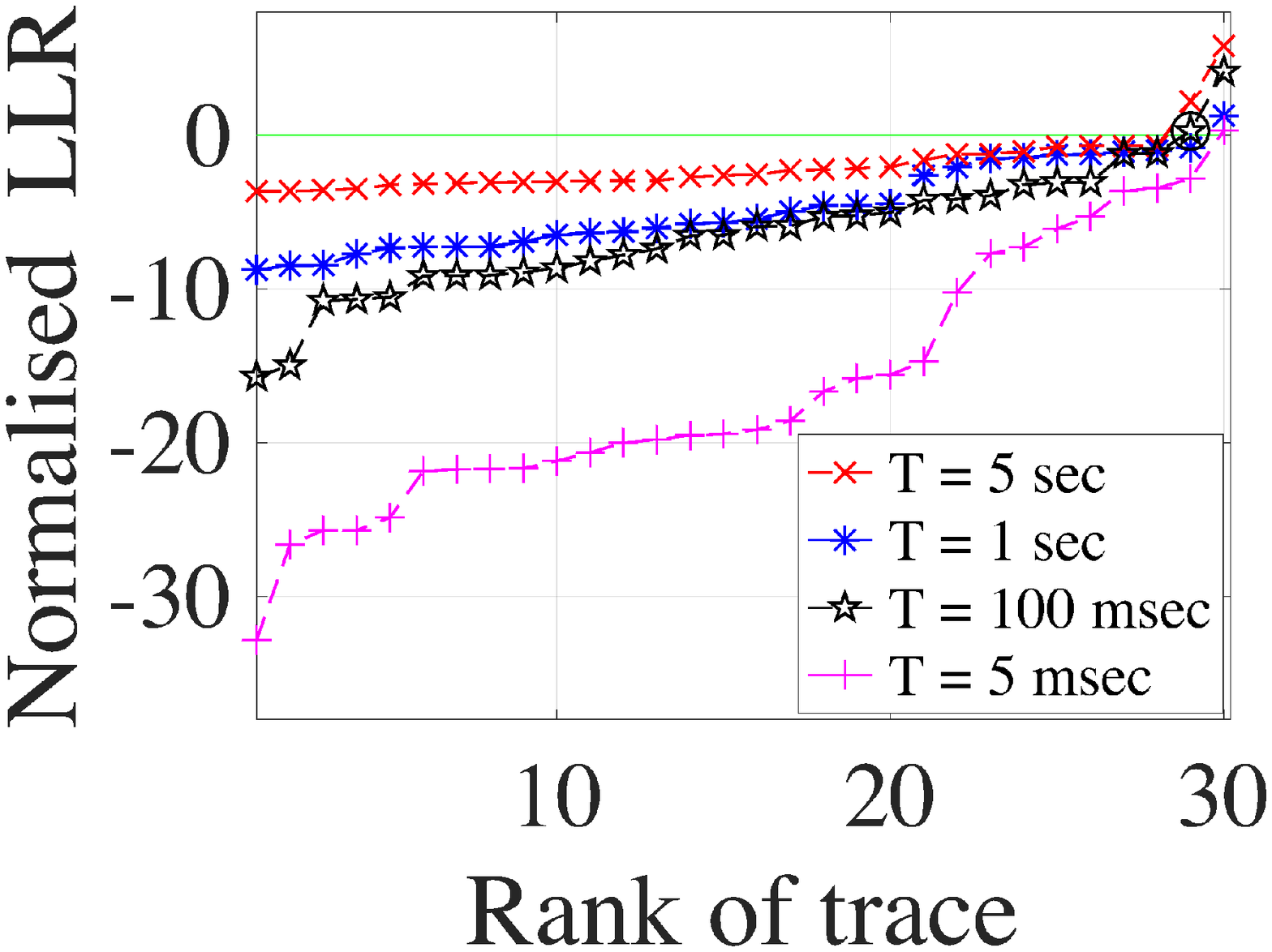}}\quad
	\subcaptionbox{Auckland traces}[.18\linewidth][]{%
		\includegraphics[scale=0.18]{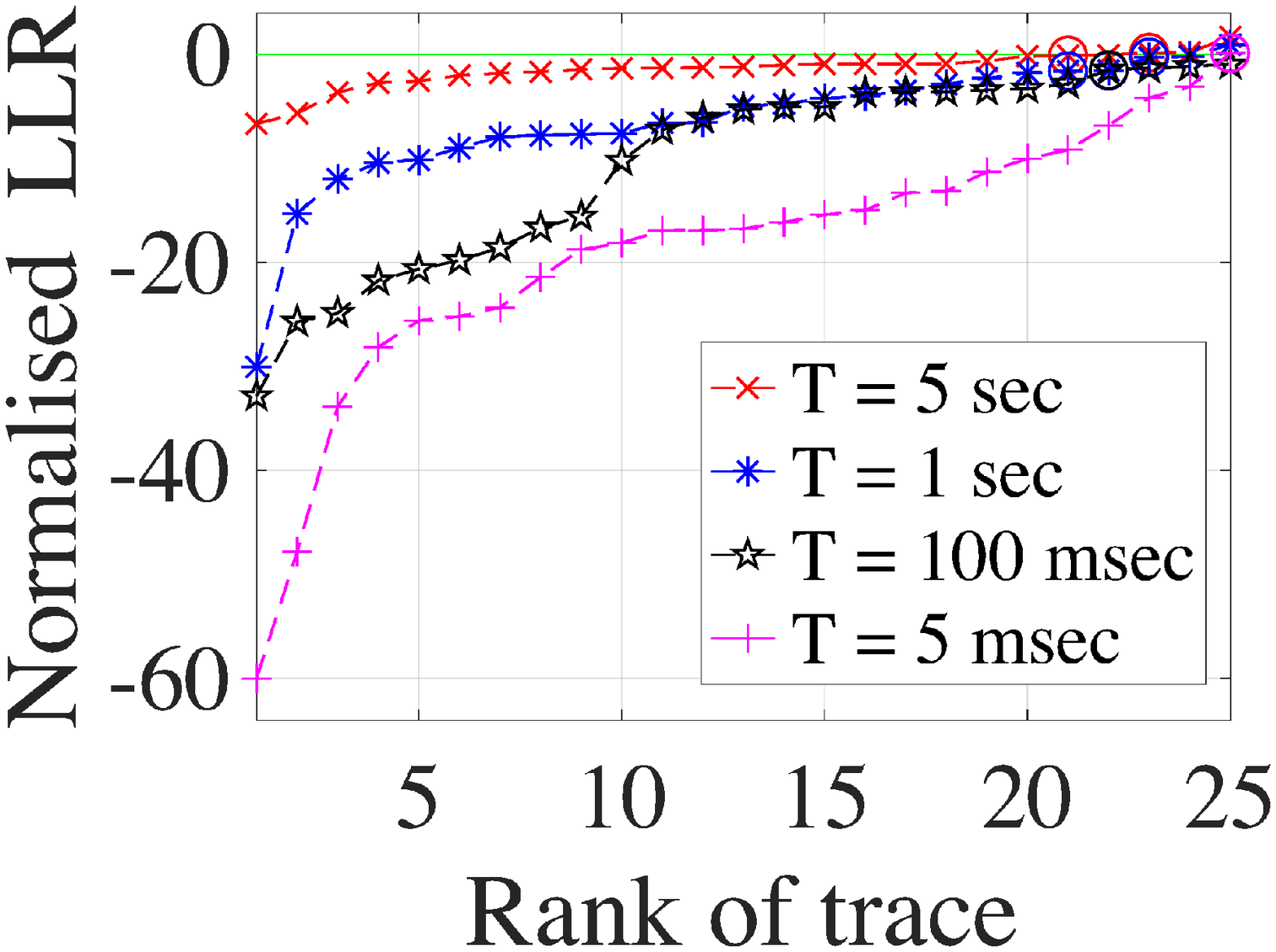}}\quad
	\subcaptionbox{Twente traces}[.18\linewidth][]{%
		\includegraphics[scale=0.18]{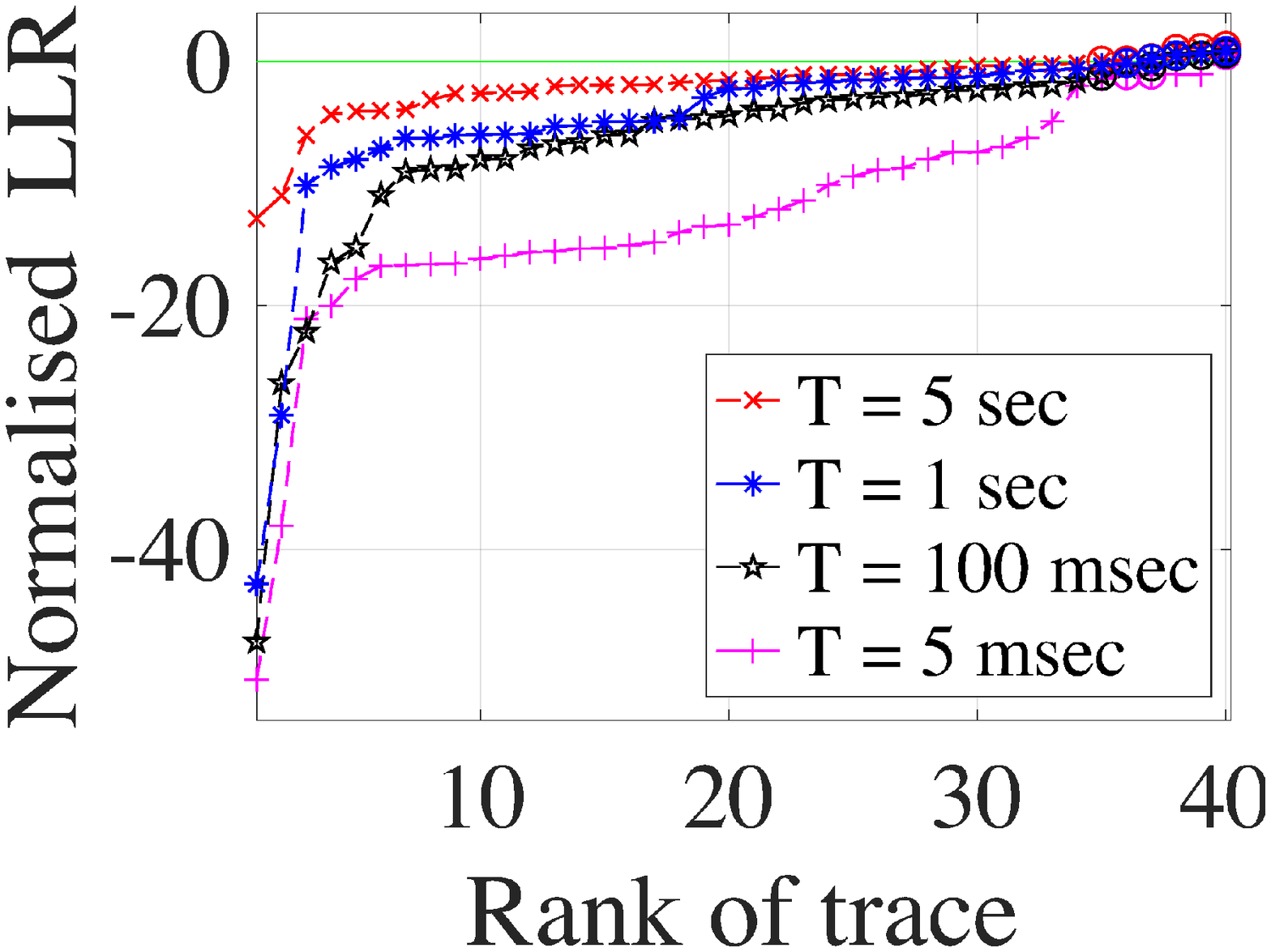}}\quad
	\subcaptionbox{MAWI traces}[.18\linewidth][]{%
		\includegraphics[scale=0.18]{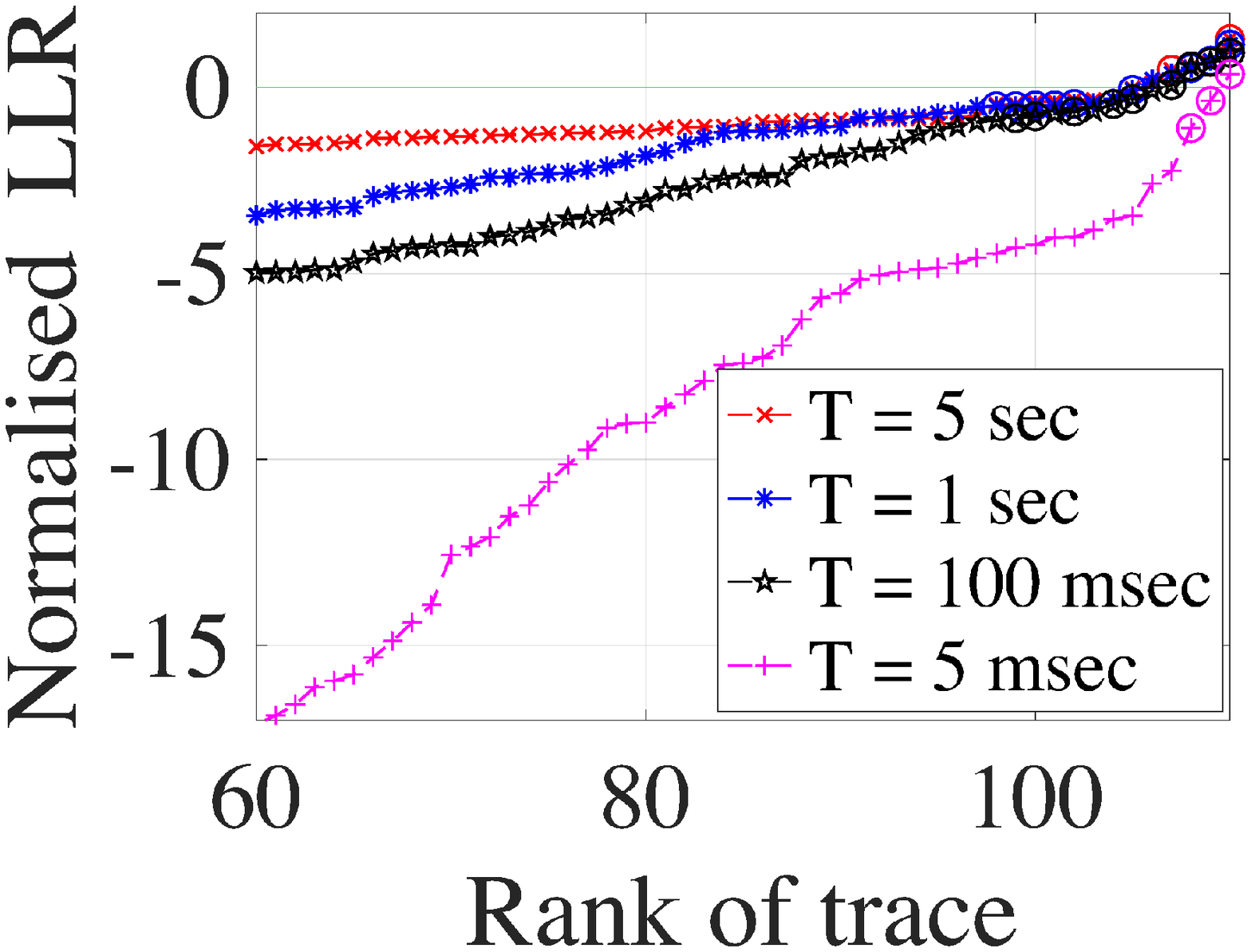}}
	\caption{Normalised Log-Likelihood Ratio ($\Re$) test results for all studied traces when running $fit.distCompare$(power-law, log-normal). Aggregation timescales are 5 sec, 1 sec, 100 msec and 5 msec. Circled points in the plots are the ones with $p$-value greater than $0.1$, i.e.  test is inconclusive with respect to assessing which of the reference and candidate distributions is a better fit to the traffic data.}
	\label{LRResultsDiffTValues}
\end{figure*} 

\begin{figure*}[t]
	\setlength{\belowcaptionskip}{-4pt}
	\centering
	\subcaptionbox{CAIDA traces}[.18\linewidth][]{%
		\includegraphics[scale=0.18]{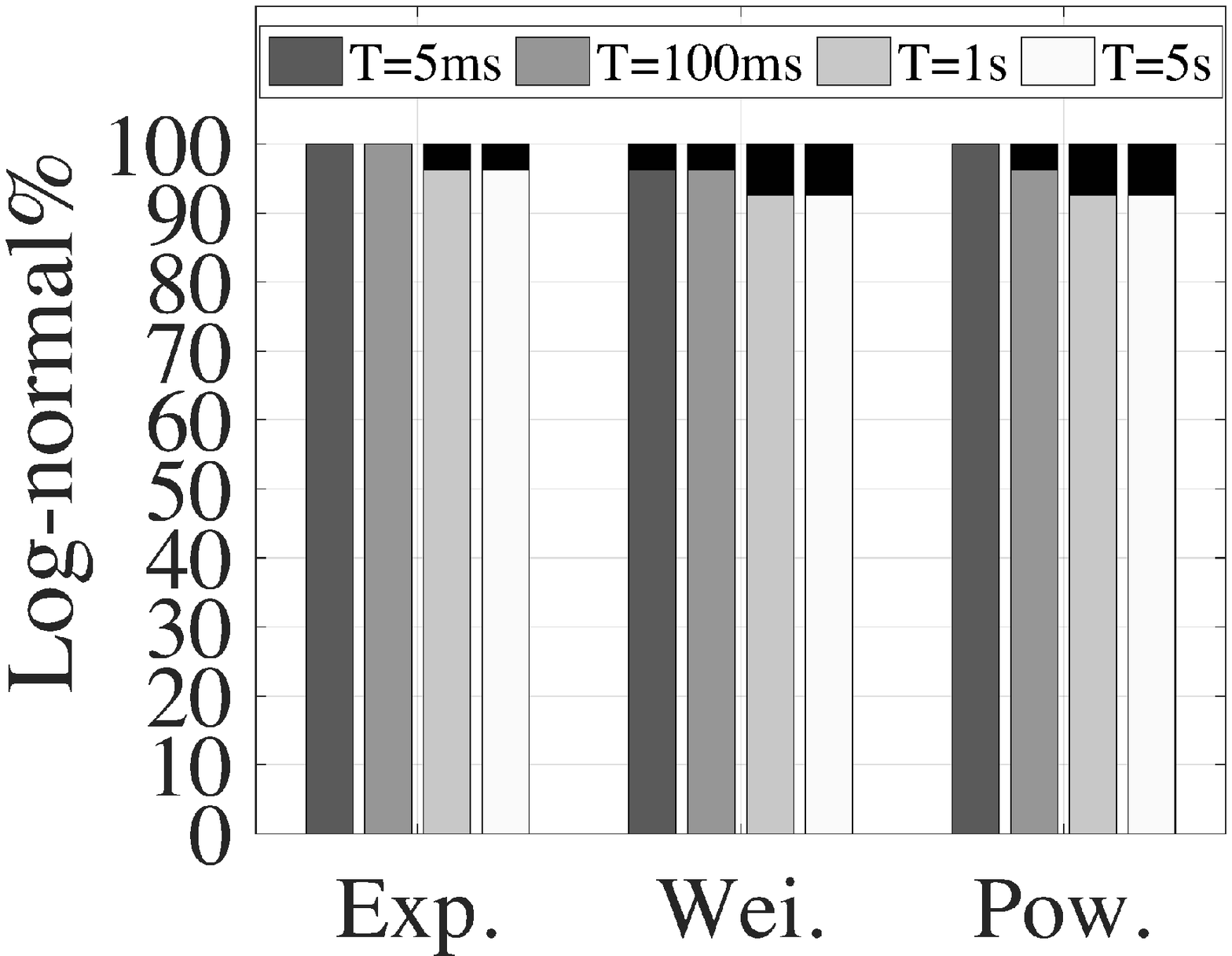}}\quad
	\subcaptionbox{Waikato traces}[.18\linewidth][]{%
		\includegraphics[scale=0.18]{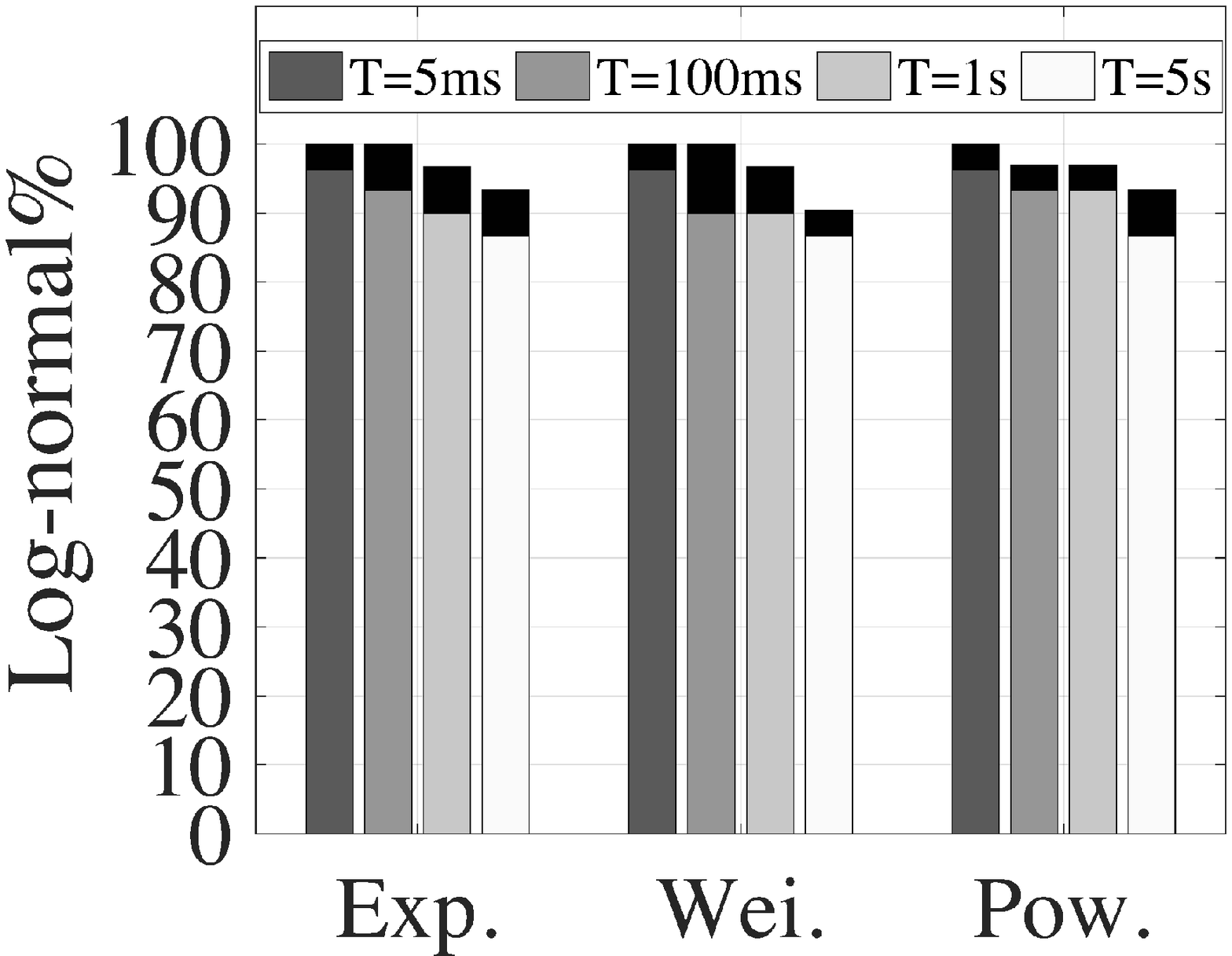}}\quad
	\subcaptionbox{Auckland traces}[.18\linewidth][]{%
		\includegraphics[scale=0.18]{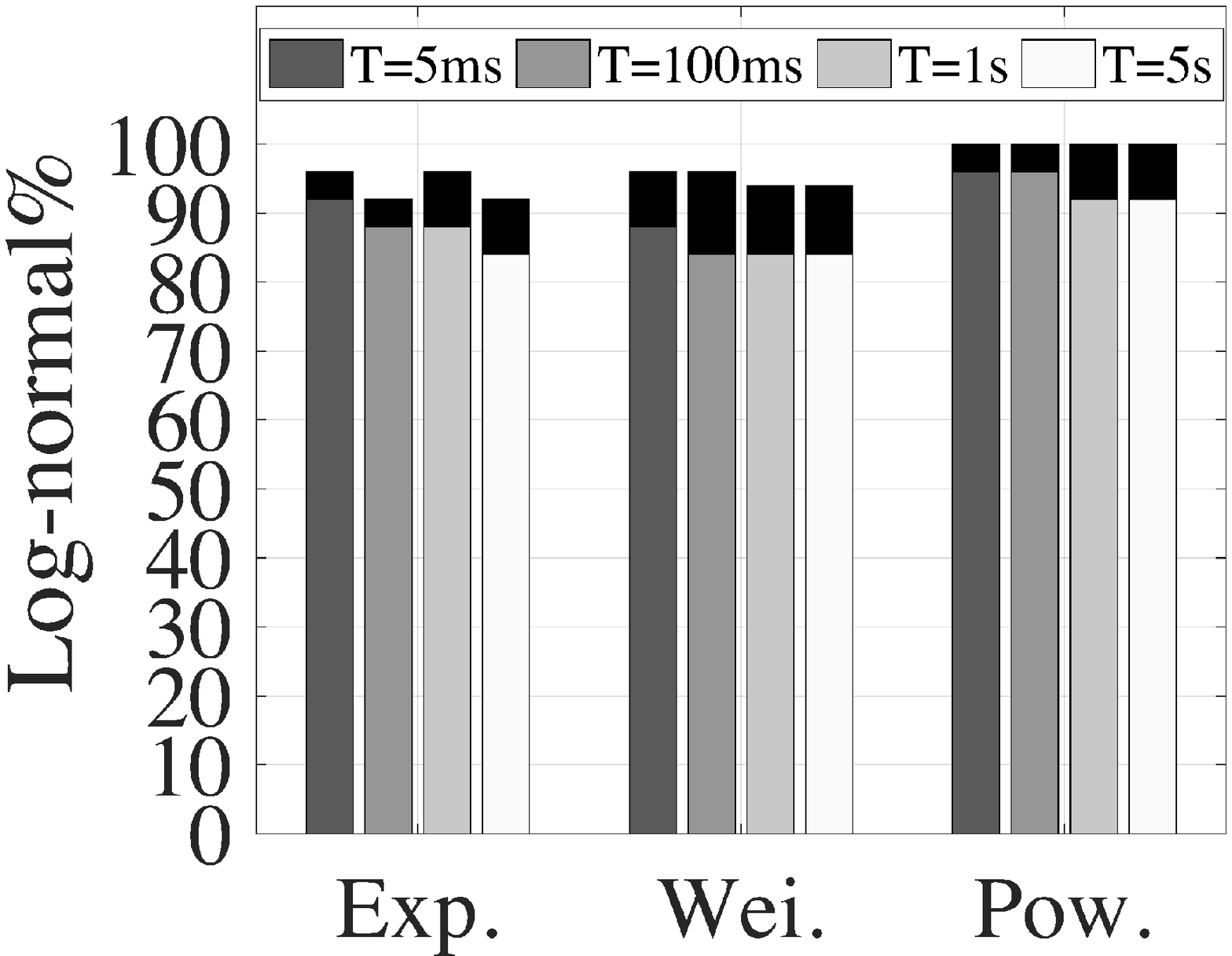}}\quad
	\subcaptionbox{Twente traces}[.18\linewidth][]{%
		\includegraphics[scale=0.18]{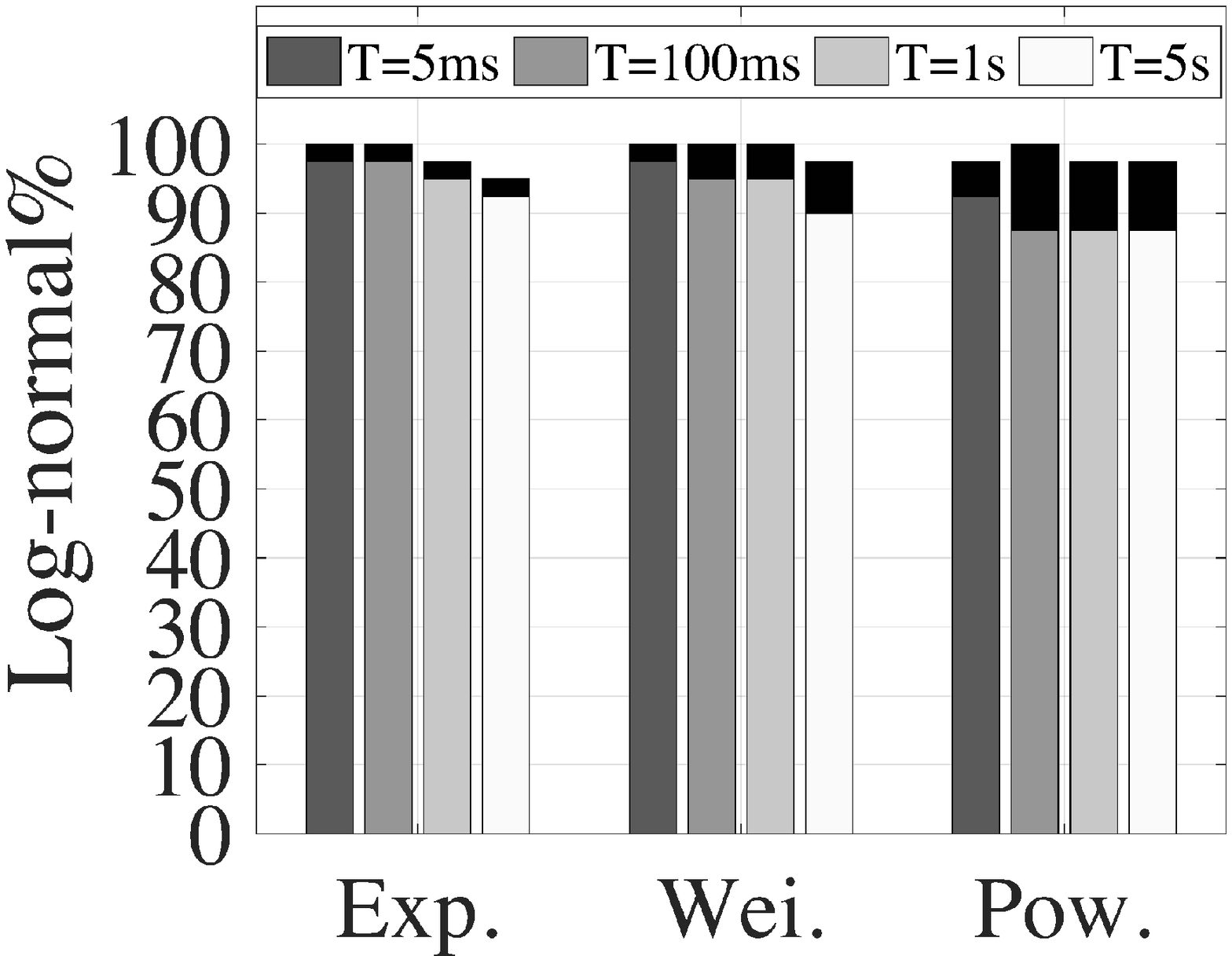}}\quad
	\subcaptionbox{MAWI traces}[.18\linewidth][]{%
		\includegraphics[scale=0.18]{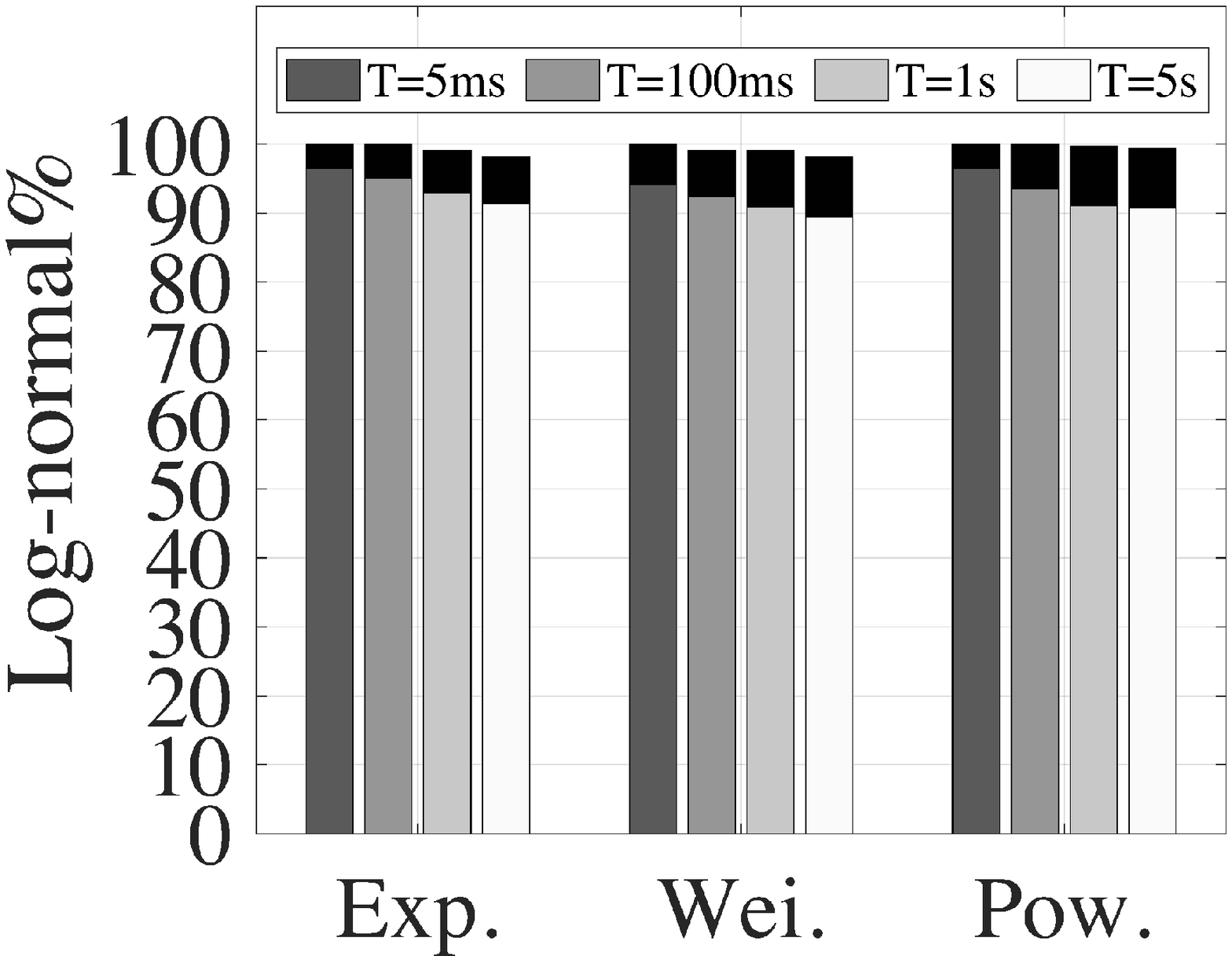}}
	
	\caption{The percentage of traces for which the log-normal distribution is a better fit compared to the alternative distribution exponential, Weibull or power-law  at aggregation timescales: $5$ msec, $100$ msec, $1$ sec and $5$ sec. Black areas represent the inconclusive results (i.e. $p$-value $>0.1$).}
	\label{LRResultsPercentages}
\end{figure*}

\section{Fitting a statistical distribution to Internet traffic data}
\label{sec:fitting}

In this section we present an extensive statistical analysis applied to the datasets described in the previous section. The aim is to find a good model for all studied traces for different aggregation time values. In contrast to the existing research (see Section~\ref{sec:related}), we are basing our analysis on the framework proposed by Clauset et al.~\cite{clauset}, a comprehensive statistical framework developed specifically for testing power-law  (or another reference distribution) behaviour in empirical data\footnote{We have used the source code discussed in~\cite{Alstott}. More details in \S 2.9~\cite{moThesis}.}. The framework combines maximum-likelihood fitting methods with goodness-of-fit tests based on the Kolmogorov--Smirnov statistic and likelihood ratios. The method reliably tests whether the reference distribution is the best model for a specific dataset, or, if not, whether an alternative statistical distribution is. The framework performs the tests described above as follows: (1) the parameters of the reference model are estimated for a given dataset; (2) the goodness-of-fit between the data and the the reference distribution is calculated, under the hypothesis that  the reference distribution   is  a good fit to the provided traffic samples. 

If the resulting $p$-value is greater than $0.1$ the hypothesis is accepted (i.e. the reference distribution is a plausible fit to the given data), otherwise the hypothesis is rejected; (3) alternative distributions are tested against the reference distribution as a better fit to the data by employing a likelihood ratio test.

In this paper, we use the log-normal as the reference distribution and compare it to the exponential, Weibull and power-law distributions (see Equation~\ref{llrlognormal}). This is in contrast to our initial results presented in \cite{our-infocom-paper}, where following Clauset method we used the power-law as the reference distribution. This is because (1) the power-law distribution failed to fit the vast majority of the studied datasets and (2), as shown in \cite{our-infocom-paper}, the log-normal distribution was a good fit for most of the traces. Indeed, Step 2 of the Clauset method showed that for the vast majority of the examined traces, the respective hypothesis was accepted; i.e. the log-normal distribution was a good fit. This is an encouraging result; here, we build on it by comparing the log-normal to three alternative distributions, exponential, Weibull and power-law, using Step 3 of the Clauset method, i.e. by performing the log-likelihood ratio (LLR) test, as follows:
\begin{equation}
\Re, p = \mathrm{fit.distCompare(alternative, lognormal)}  \label{llrlognormal}
\end{equation}
where $\Re$ is the normalised LLR\footnote{$\Re$ is calculated as $\mathcal{R} /(\sigma\sqrt{n}) $, where $\mathcal{R}$ is the log-likelihood ratio~\cite{clauset}. Note that if we run fit.distCompare(Y,X) and fit.distCompare(Z,X), then we would not be able to tell which of the Y and Z distributions is a better fit to the used data, even if both were better fits compared to X. This is because the normalised LLR value is measured by normalising LLR by $\sigma$ (the estimated standard deviation on LLR), which is a nonlinear operator.} between the alternative and log-normal  distributions, $p$ is the significance value for this test. $\Re$ is positive if the alternative distribution is a better fit for the data, and negative if the log-normal distribution is a better fit for the data. The further $\Re$ value is from zero, the better the fit is for one distribution over the other. A $p$-value less than $0.1$ means that the value of $\Re$ is a reliable indicator of which model (log-normal or alternative, depending on the sign of $\Re$) is the better fit to the data. In contrast, a $p$-value greater than $0.1$ means that there is nothing to be concluded from the likelihood ratio test.

\subsection{Fitting the log-normal distribution to Internet traffic data}
\label{fit-log-normal}

Figure~\ref{LRtestResults} shows the results of the LLR test for all $232$ traces when comparing the log-normal to the exponential, Weibull and power-law distributions. For this test we have aggregated traffic at a timescale $T=100$ msec. The points marked with a circle are the ones with $p>0.1$. It is clear that the log-normal distribution is the best fit for the studied traces; i.e. $\Re<0$ and $p<0.1$\footnote{For clarity, in Figures~\ref{LRtestResults}(e) and~\ref{LRResultsDiffTValues}(e) we only plot traces 60 -- 110. For traces 1 -- 59, $\Re$ is less than $0$ and the respective $p$-value is less than $0.1$; i.e. the log-normal distribution is the best fit for the respective trace.}. The log-normal distribution is not the best fit for $1$ out of $27$ CAIDA  traces, $3$ out $30$ Waikato traces, $3$ out of $25$ Auckland traces, $5$ out of $40$ Twente traces and $9$ out of $110$ MAWI traces i.e. $21$ out of $232$ are anomalous traces. Most of these traces have inconclusive results ($17$ out of $21$) while for few of them an alternative distribution is a better fit  ($4$ out of $21$). We examined these traces in more detail and discuss them in Section~\ref{anomalous}.

Identifying the log-normal distribution as the best fit for the vast majority of traffic traces at $T=100$ msec is very encouraging. This specific traffic aggregation timescale has been commonly studied in the literature~\cite{ResDimension,transaction2015}.  

\begin{figure*}[t!]
	\setlength{\belowcaptionskip}{-6pt}
	\minipage[t]{1\textwidth}
	\subcaptionbox{Anomalous trace}{%
		\includegraphics[scale=0.178]{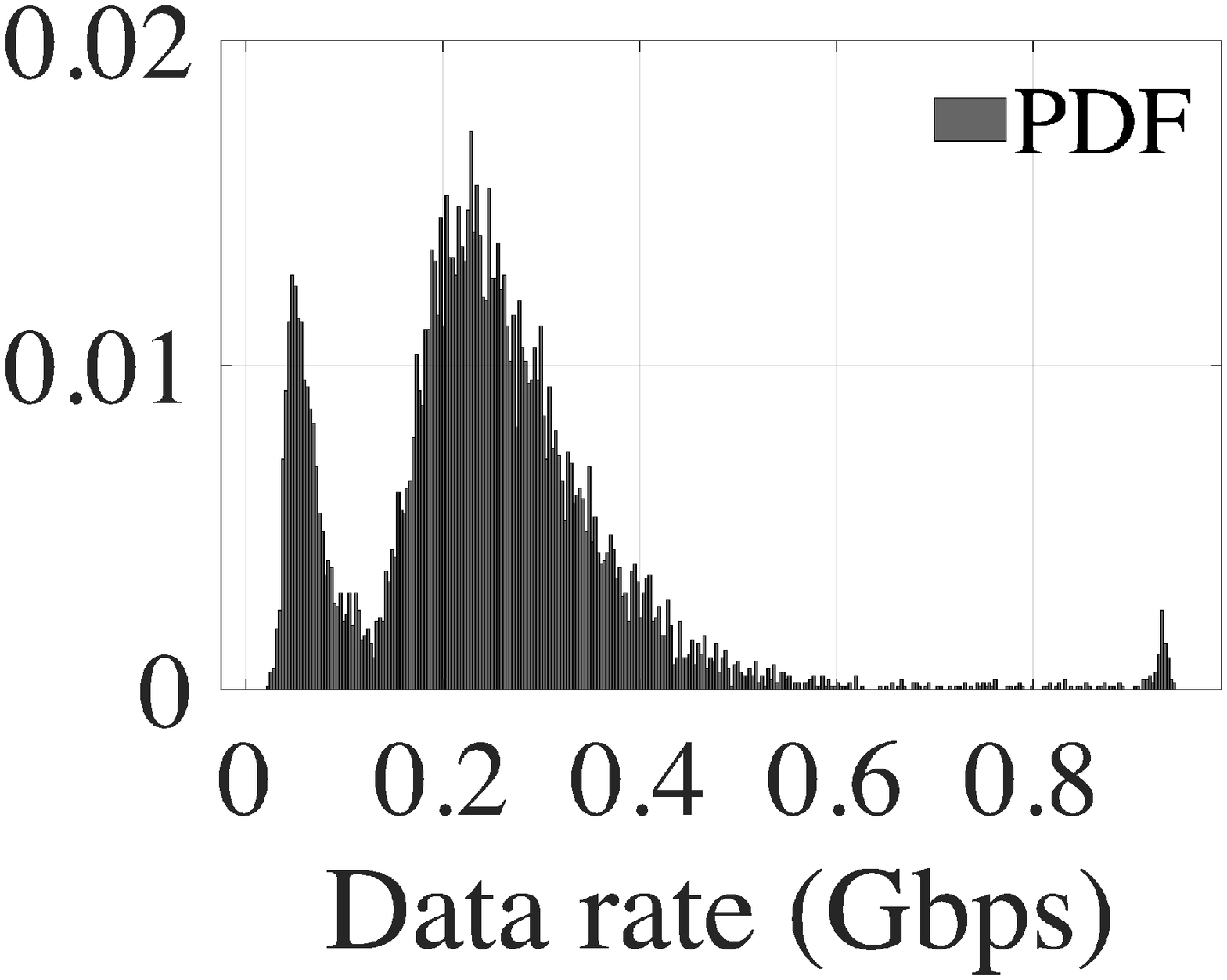}}\quad
	\subcaptionbox{Log-normal trace}{%
		\includegraphics[scale=0.178]{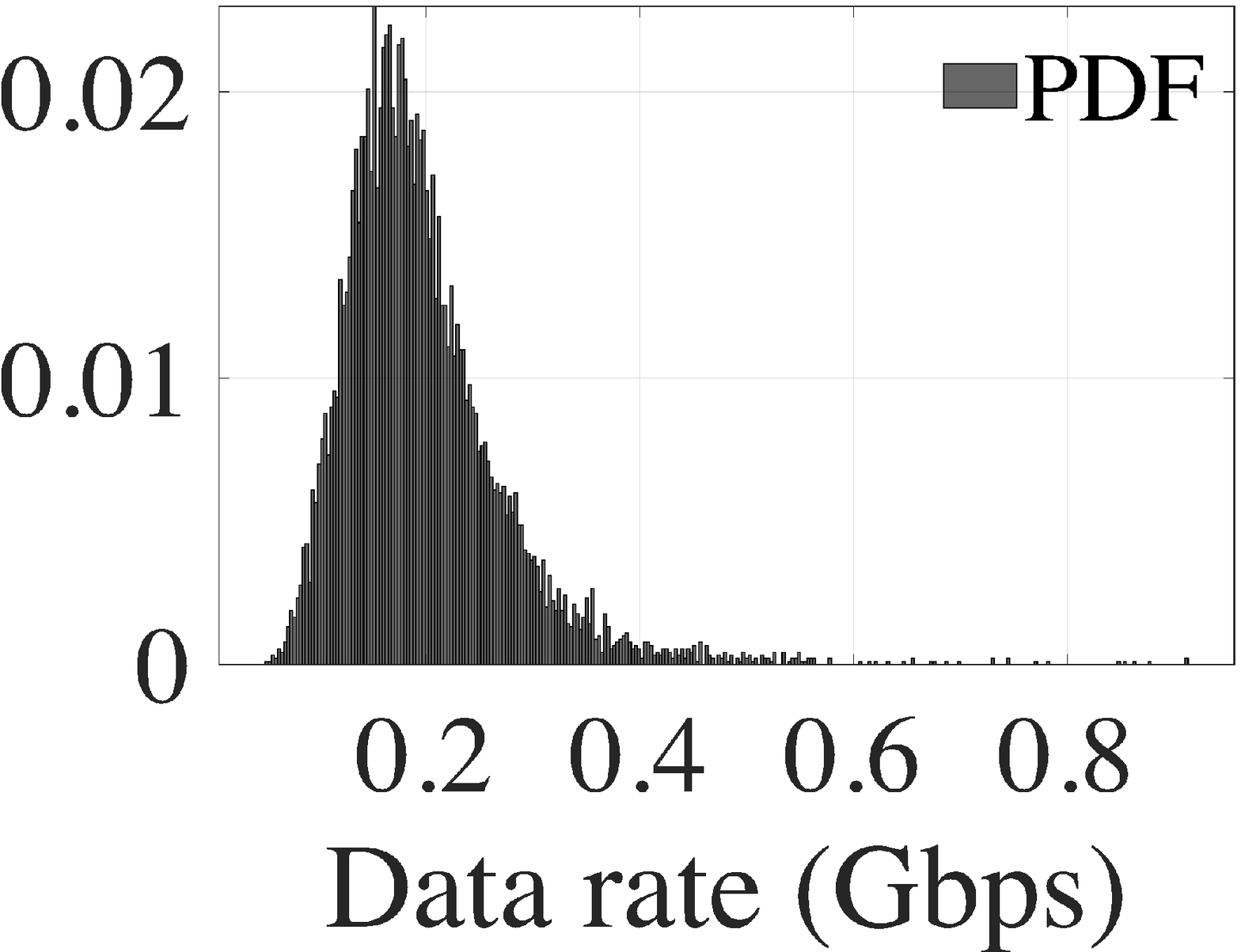}}\quad
	\subcaptionbox{PDF: 17:00-18:00}{%
		\includegraphics[scale=0.178]{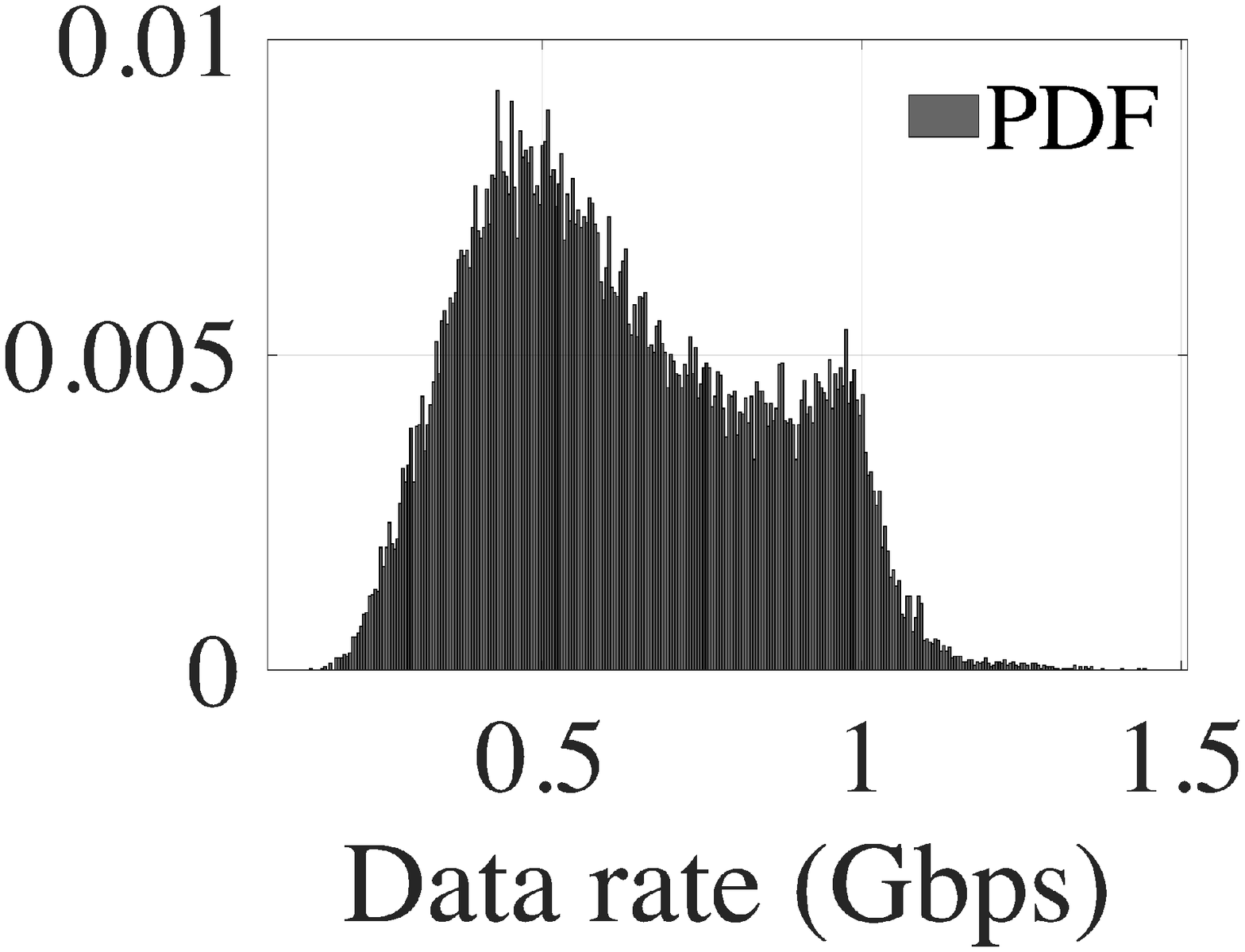}}\quad
	\subcaptionbox{PDF: 17:00-17:30}{%
		\includegraphics[scale=0.178]{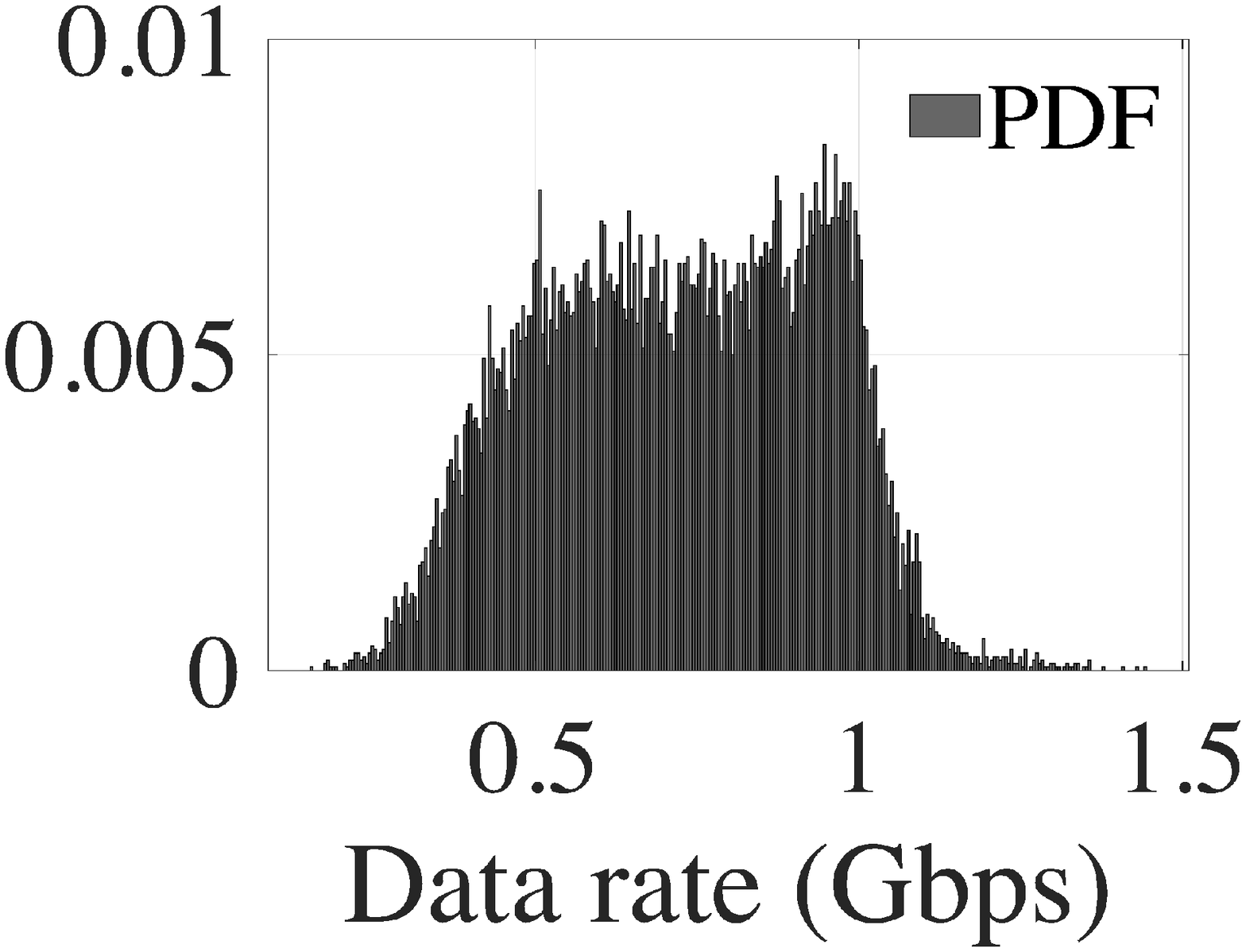}}\quad
	\subcaptionbox{PDF: 17:30-18:00}{%
		\includegraphics[scale=0.178]{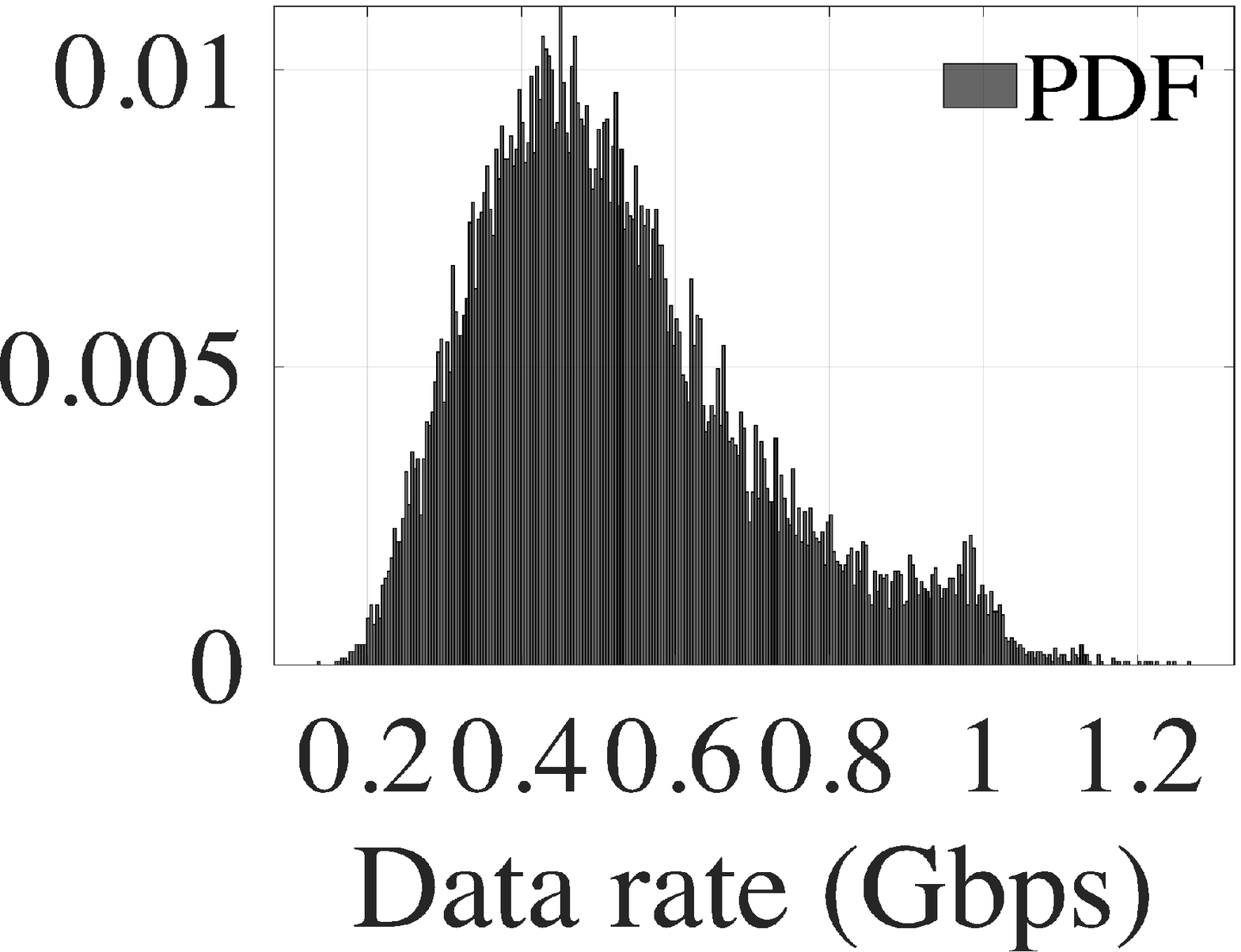}}\quad
	\caption{ PDF plots of some MAWI traces.}
	\label{anomalousTraces}
	\endminipage\hfill
\end{figure*}

Next we investigate which one of the studied distributions is the best model for a range of aggregation timescales. We run the fitting test again for the following pairs: (power-law, log-normal), (exponential, log-normal) and (Weibull, log-normal) for timescales between $5$ msec to $5$ sec. Figure~\ref{LRResultsDiffTValues} shows the results from Equation~\ref{llrlognormal} of the test between (power-law, log-normal). As reflected by the $\Re$ and $p$-values, the log-normal distribution is the best fit for the vast majority of captured traces at all examined timescales\footnote{Note that it is possible that the network traffic may not follow a log-normal distribution at very fine or coarse aggregation granularities.}. Due to lack of space, we do not show detailed results for the other two pairs; i.e. (exponential, log-normal) and (Weibull, log-normal). Instead, we present overall results, including the percentage of inconclusive tests, for all pairs of distributions, traces and timescales, in Figure~\ref{LRResultsPercentages}. The results show that the log-normal is the best fit for the vast majority of traces (overall: $95.19\%$ when  $T$=$5$ msec and $89.13\%$ when $T$=$5$ sec), while few tests are inconclusive (black areas in Figure~\ref{LRResultsPercentages}, overall: $4.11\%$ when  $T$=$5$ msec and $7.44\%$ when $T$=$5$ sec) and very few are in favour of the alternative distribution (the rest of the percentages, e.g. overall: $0.7\%$ when  $T$=$5$ msec and $3.43\%$ when $T$=$5$ sec). This is a strong result suggesting the generality of our observations and the potential for wide applicability of the log-normal model in practical applications.

We also examined Q-Q plots for a large number of traces\footnote{Due to lack of space, Q-Q plots are not included as we would have to present plots for each trace, separately.}. The log-normal distribution appeared to be a better fit than other tested distributions and no deviations from the expected pattern were observed in the body or tail of the distribution.

\subsection{Anomalous traces} 
\label{anomalous}
As mentioned in Section~\ref{fit-log-normal}, there is a small number of traces for which the log-normal distribution is not a good fit (none of the other examined distributions is, either). Figure~\ref{anomalousTraces}a shows the probability density function (PDF) plot for one of the $9$ anomalous MAWI traces. For comparison, Figure~\ref{anomalousTraces}b shows the PDF for another MAWI trace for which the log-normal distribution is a good fit. It is obvious from Figure~\ref{anomalousTraces}a that the link was either severely underutilised (see the large spike on the left part of the plot area) or fully utilised (see the smaller spike at the right part of the plot area) for higher data rates. All traces for which the log-normal distribution was not a good fit exhibited similar behaviour and (aggregated) traffic patterns. On the contrary, we did not observe any such behaviour for the majority of traces for which the log-normal distribution was the best fit. A likely explanation for the anomalous traces is that those traces contain either periods of over-capacity (traffic is at 100\% of link capacity) or periods where the link is broken (no traffic).

\subsection{Fitting the log-normal distribution to subtraces in the 24-hour long trace} 
We need to establish whether we can reliably say that the log-normal distribution is a good fit for any sample length of data, not only the 15-minute  long traces (this is discussed in details in Section~\ref{sec:stationarity}). We apply  Clauset test~\cite{clauset}  (discussed in Section~\ref{sec:fitting}) on longer and shorter traces as follows.  

Firstly, we apply this test on each subtrace (1-hour long) of the 24-hour long MAWI trace  {at timescale $T=100$ msec. Figure~\ref{mawi-24-power-law-results}a shows the results of the LLR test on 24 subtraces  when applying Equation~\ref{llrlognormal}. These results complement our results on the 15-minute long traces (see Figure~\ref{LRtestResults}) by showing that the log-normal distribution is the best fit for these subtraces.  Similar results are seen for other timescales between $T=5$ msec and $T=5$ sec. There is only one trace (trace id 17 in Figure~\ref{mawi-24-power-law-results}a) where neither the log-normal nor any of the other tested distributions provide a good fit; more specifically, Step 2 of the Clauset method yielded inconclusive results for all distributions. This trace was captured at time 17:00-18:00. The PDF of the 1-hour long trace has two peaks (Figure~\ref{anomalousTraces}c), and this could be fitted using a bimodal distribution (which we leave as future work). However, when dividing this trace into two 30-minute long subtraces  (Figures~\ref{anomalousTraces}d and e), the log-normal distribution was the best fit for each one of these, separately. Looking at the PDF plots, it is clear that for the first subtrace, the network was much busier compared to the second subtrace. We have no definitive explanation for why this might be, but one could speculate that this is the result of crossing the end of working day or because of some partial equipment failure. 
	
Secondly, we apply the Clauset method on small groups from a 1-hour long MAWI trace. We picked data points from this trace using different time windows: 2, 3, 4 and 5 minutes. This means that each group contains 30, 20, 15 and 12 subtraces, respectively. Figure~\ref{mawi-24-power-law-results}b shows the LLR test results on all these groups when using power-law as alternative to log-normal (same results are seen when using other alternative distributions i.e. exponential and Weibull). The results show that the log-normal distribution is the best fit among all tested distributions for all of these small subtrace groups at all tested windows.
	
In Section~\ref{sec:stationarity} we show that the majority of the 15-minute and 1-hour long traces in the dataset are stationary at all tested aggregation timescales. This gives confidence that the log-normal distribution can be used for estimation and modelling purposes as we show in Sections~\ref{sec:provision}\&\ref{sec:pricing}.

\subsection{Fitting the log-normal and Gaussian distributions using the correlation coefficient test}
\label{fit-correlation-coefficient-test}
	\begin{figure}
		\setlength{\belowcaptionskip}{-1pt}
		\centering
		\subcaptionbox{ 24 subtraces}{%
			\includegraphics[scale=0.215]{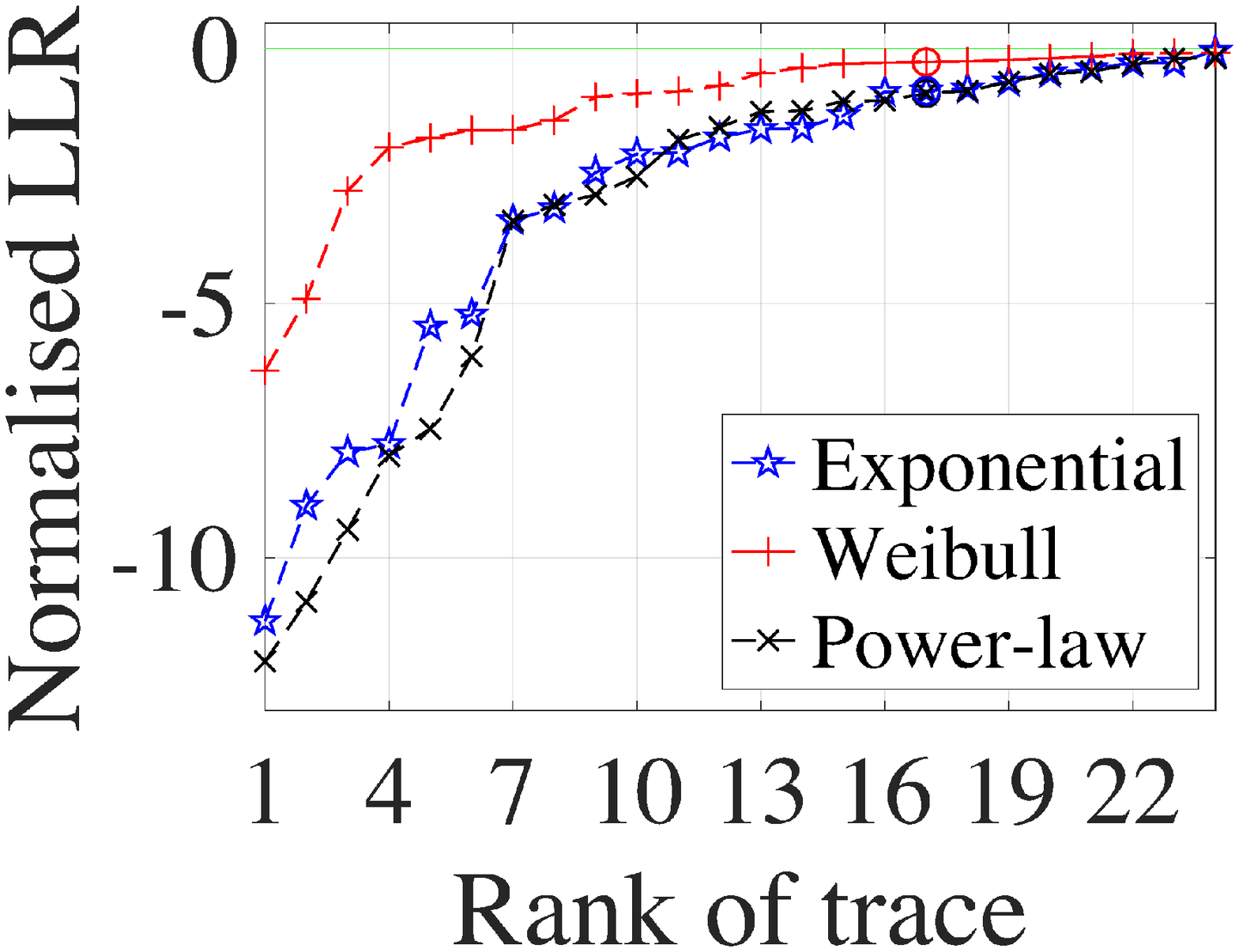}}\quad
		\subcaptionbox{windows in 1-hour long trace}{%
			\includegraphics[scale=0.215]{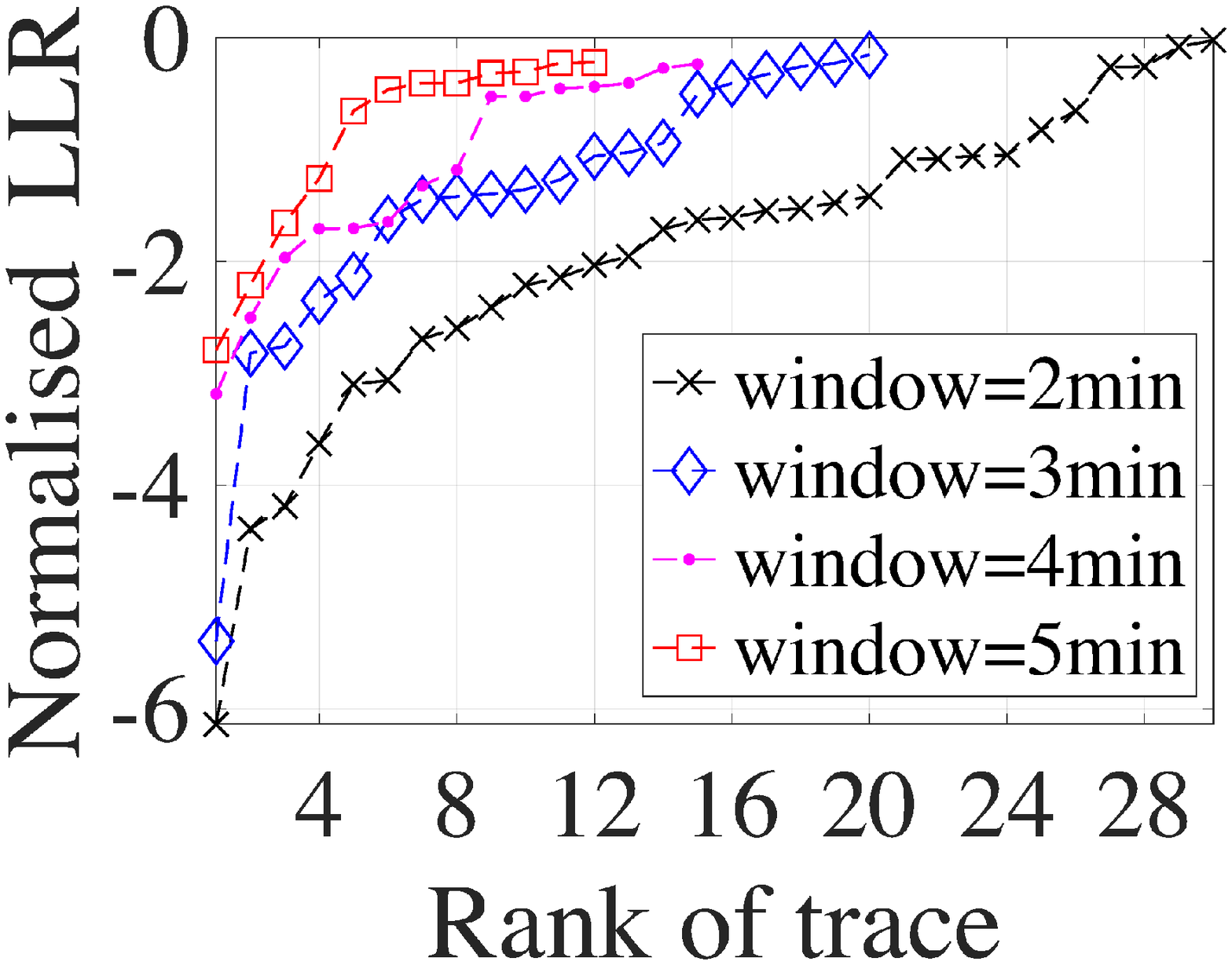}}\quad
		\caption{Normalised LLR test results ($T=100$ ms) for MAWI traces (a)  24 subtraces (b) windows in 1-hour trace.}
		\label{mawi-24-power-law-results} 	
	\end{figure}
	\begin{figure*}[t]
		\centering
		\subcaptionbox{CAIDA traces}[0.188\linewidth][ctth]{%
			\includegraphics[scale=0.184 ]{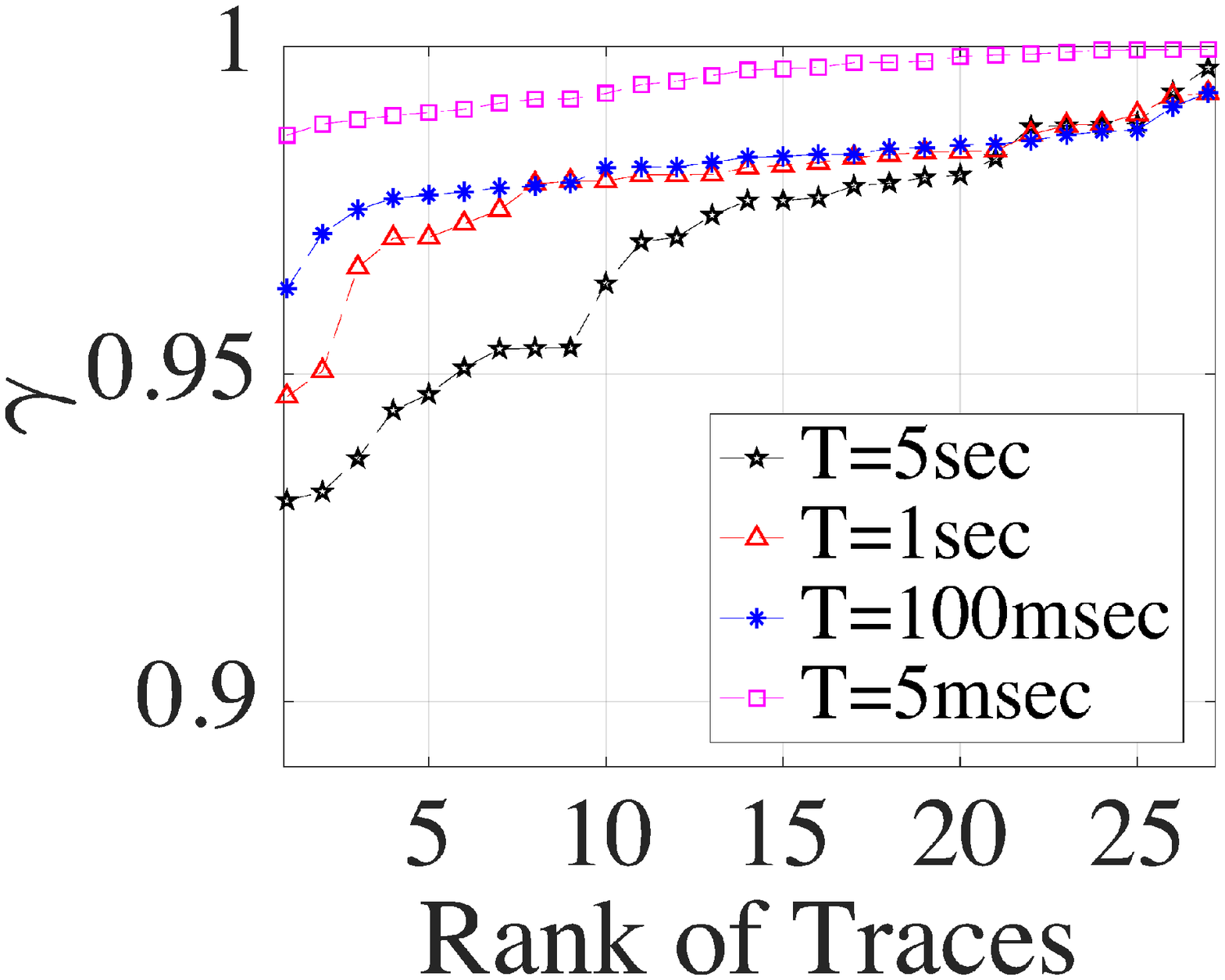}}\quad
		\subcaptionbox{Waikato traces}[0.188\linewidth][c]{%
			\includegraphics[scale=0.184]{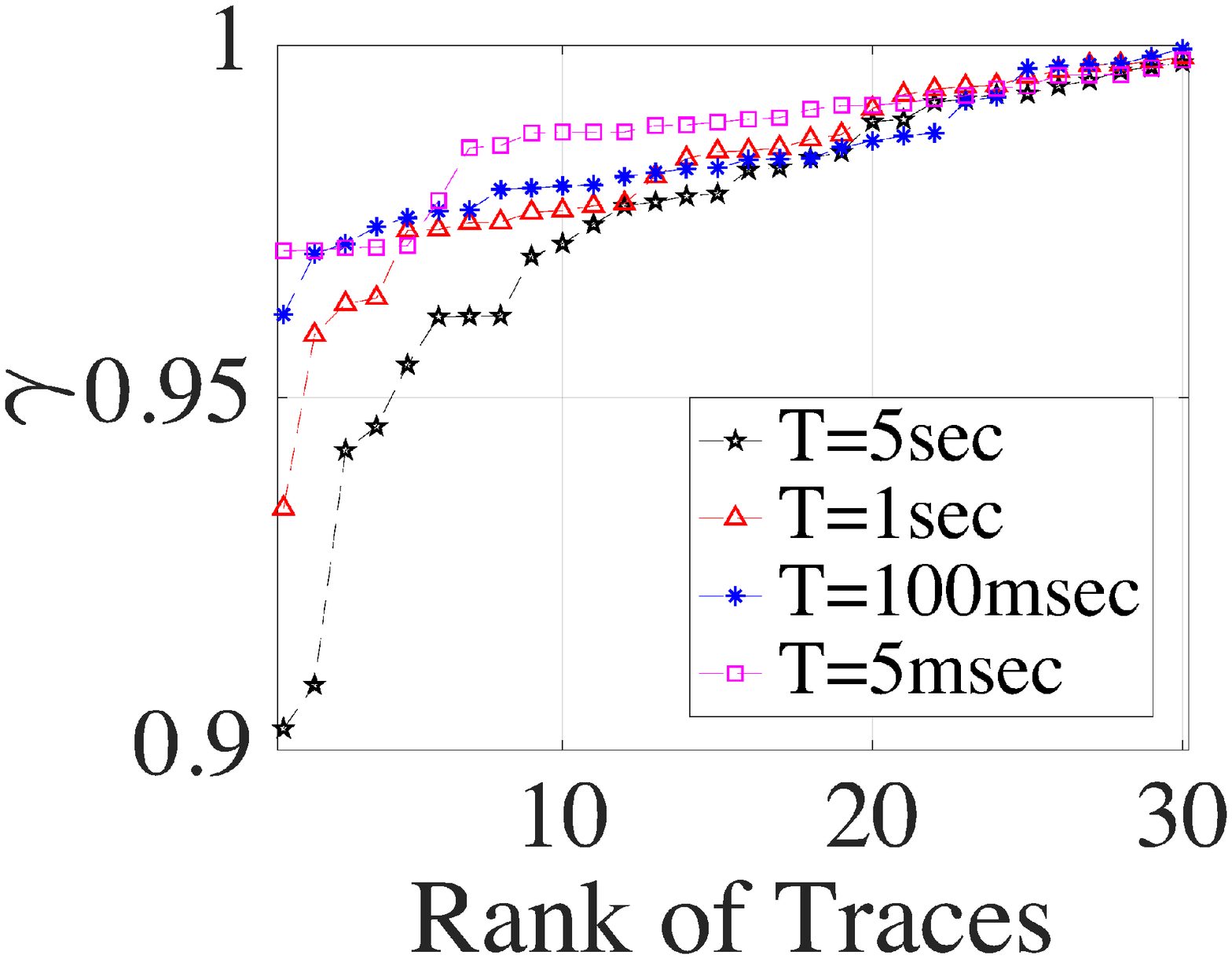}}\quad
		\subcaptionbox{Auckland traces}[0.188\linewidth][c]{%
			\includegraphics[scale=0.184]{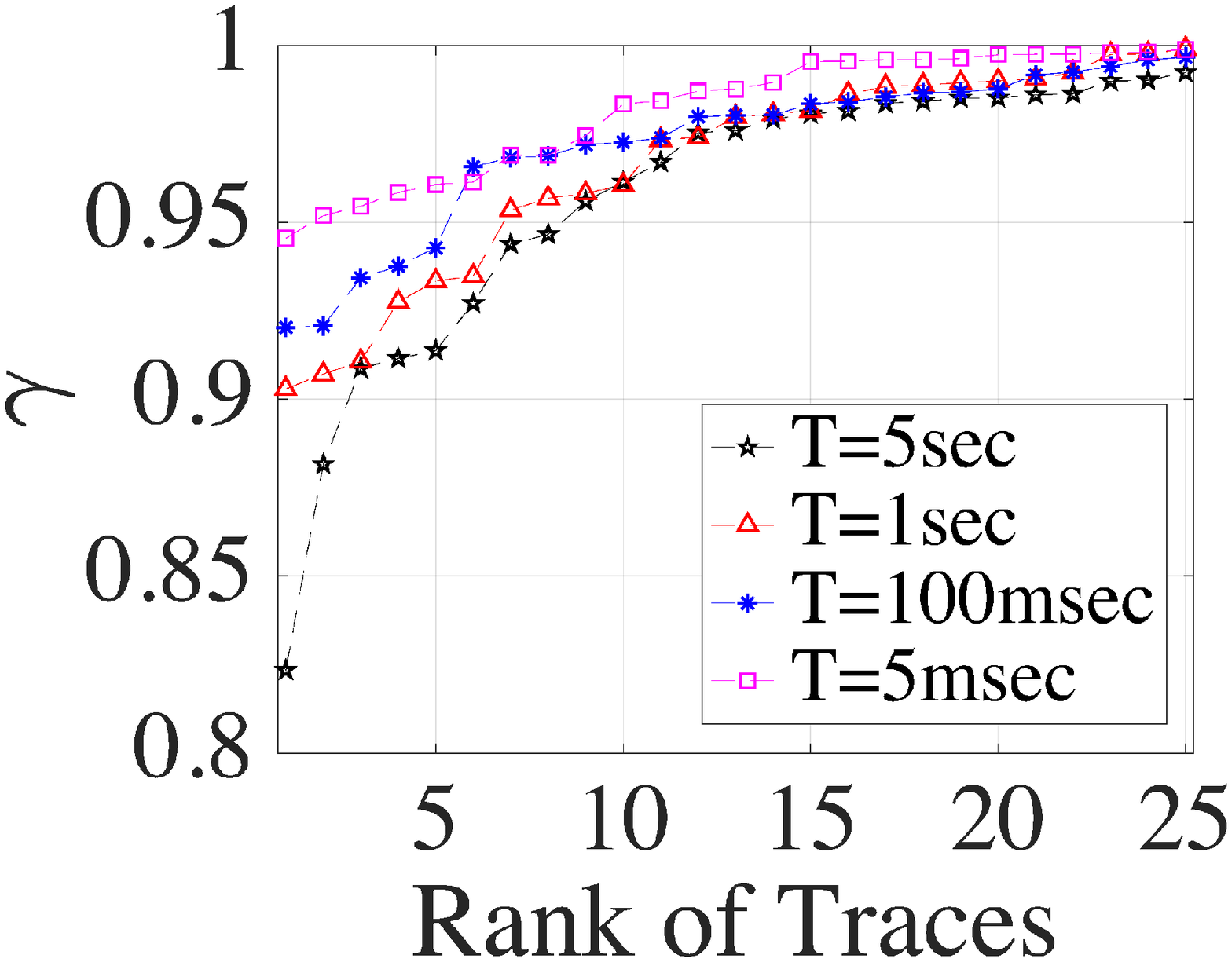}}
		\subcaptionbox{Twente traces}[0.188\linewidth][c]{%
			\includegraphics[scale=0.184 ]{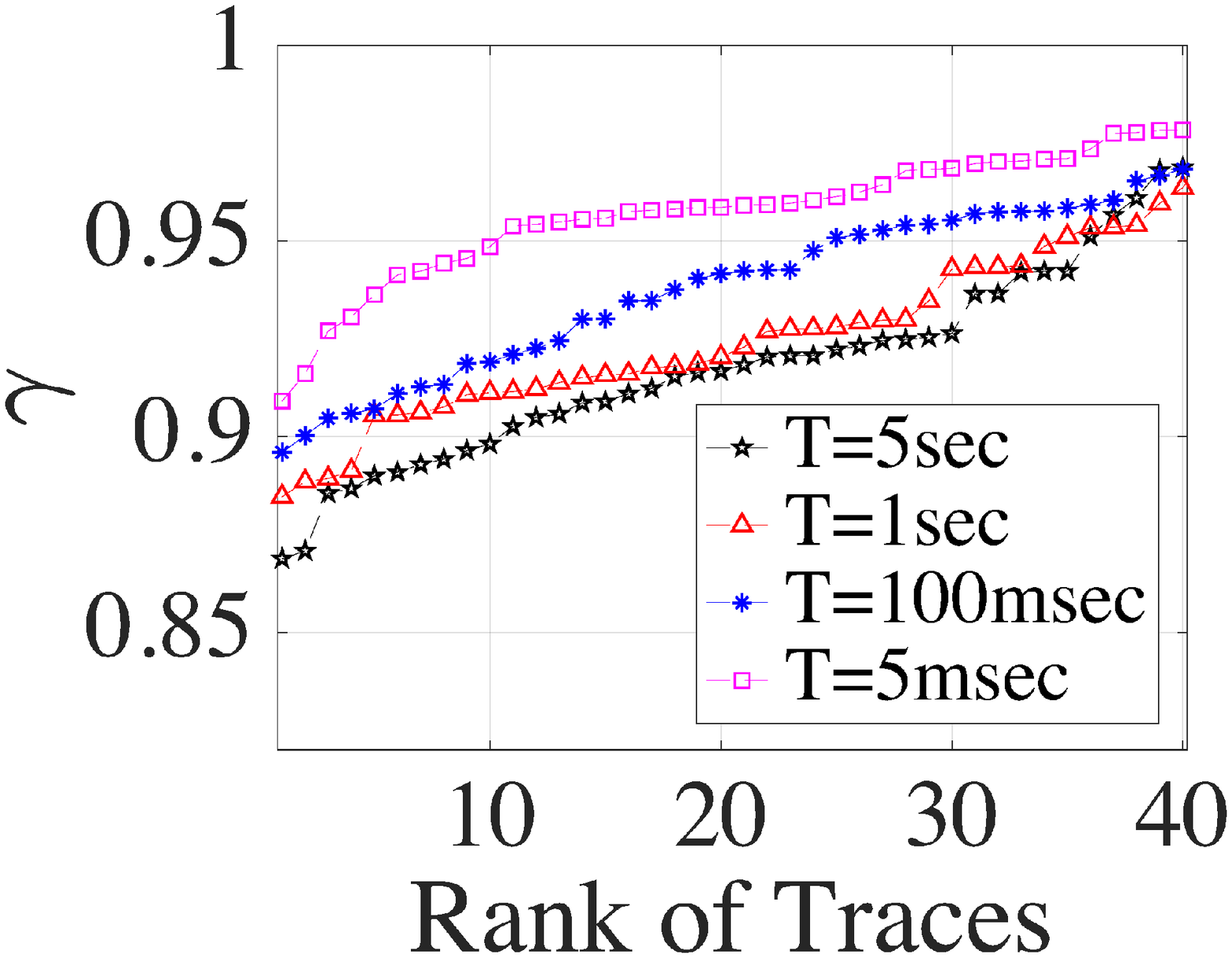}}
		\subcaptionbox{MAWI traces}[0.188\linewidth][c]{%
			\includegraphics[scale=0.184 ]{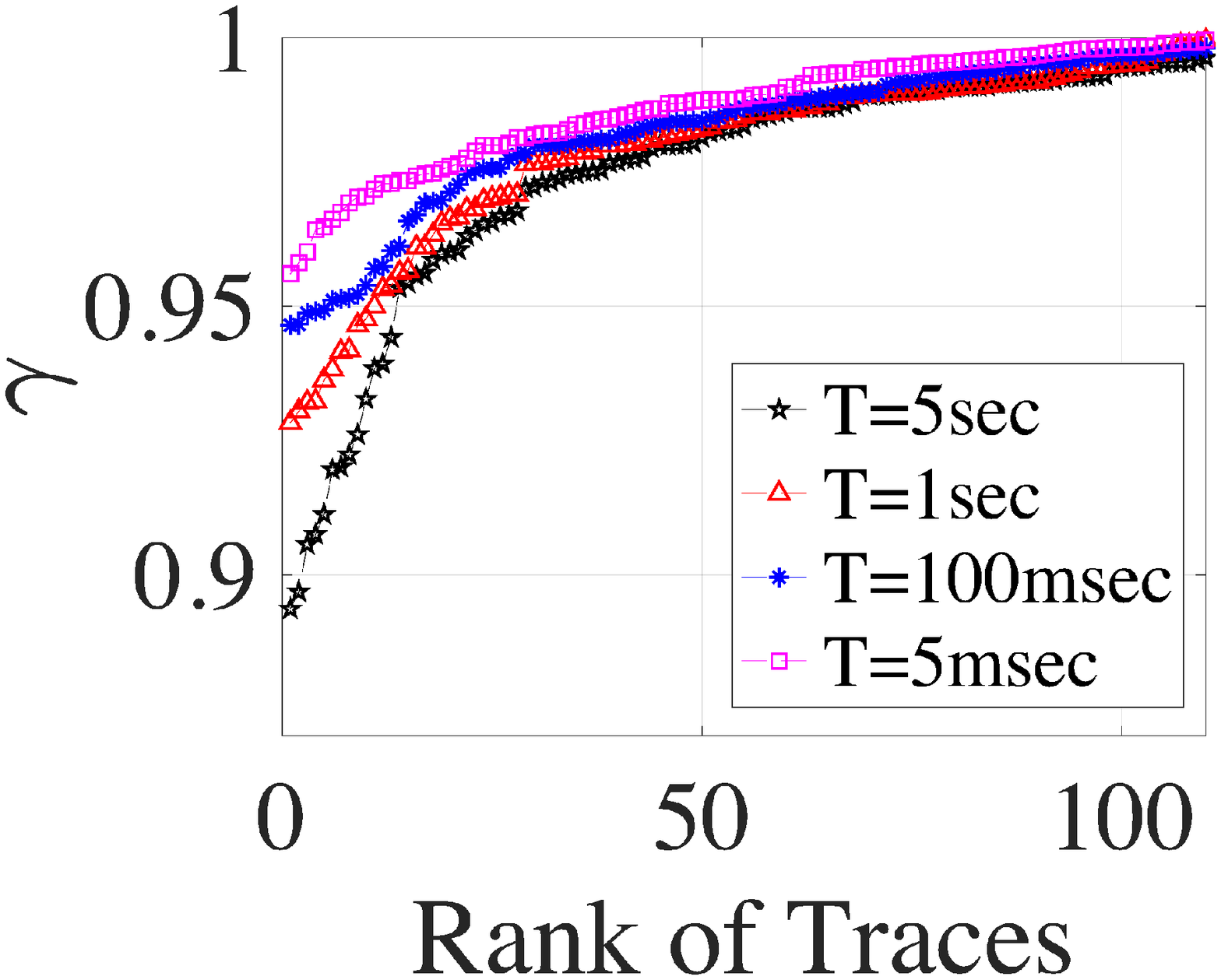}}	
		\subcaptionbox{CAIDA traces}[0.188\linewidth][ctth]{%
			\includegraphics[scale=0.184 ]{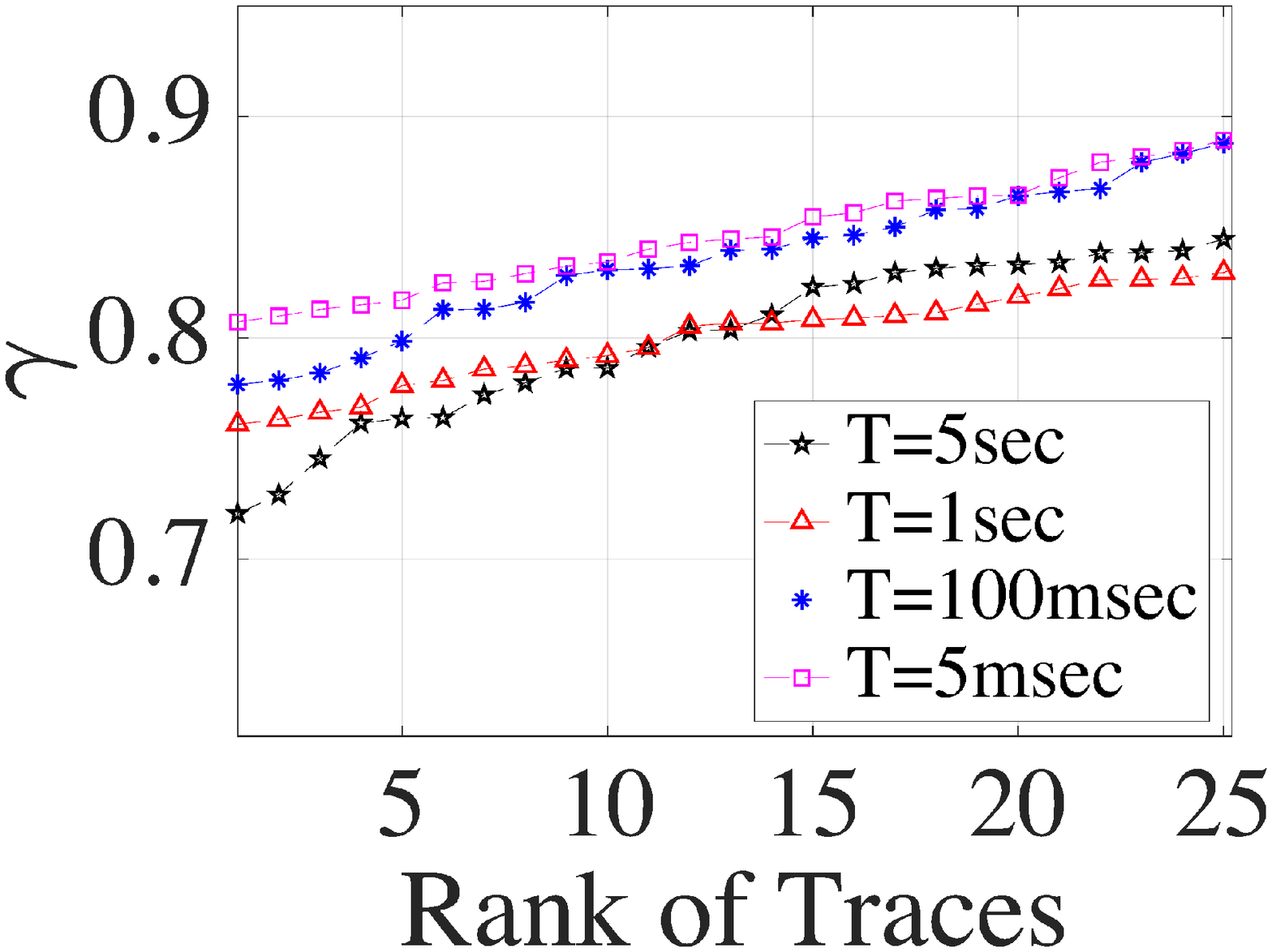}}\quad
		\subcaptionbox{Waikato traces}[0.188\linewidth][c]{%
			\includegraphics[scale=0.184 ]{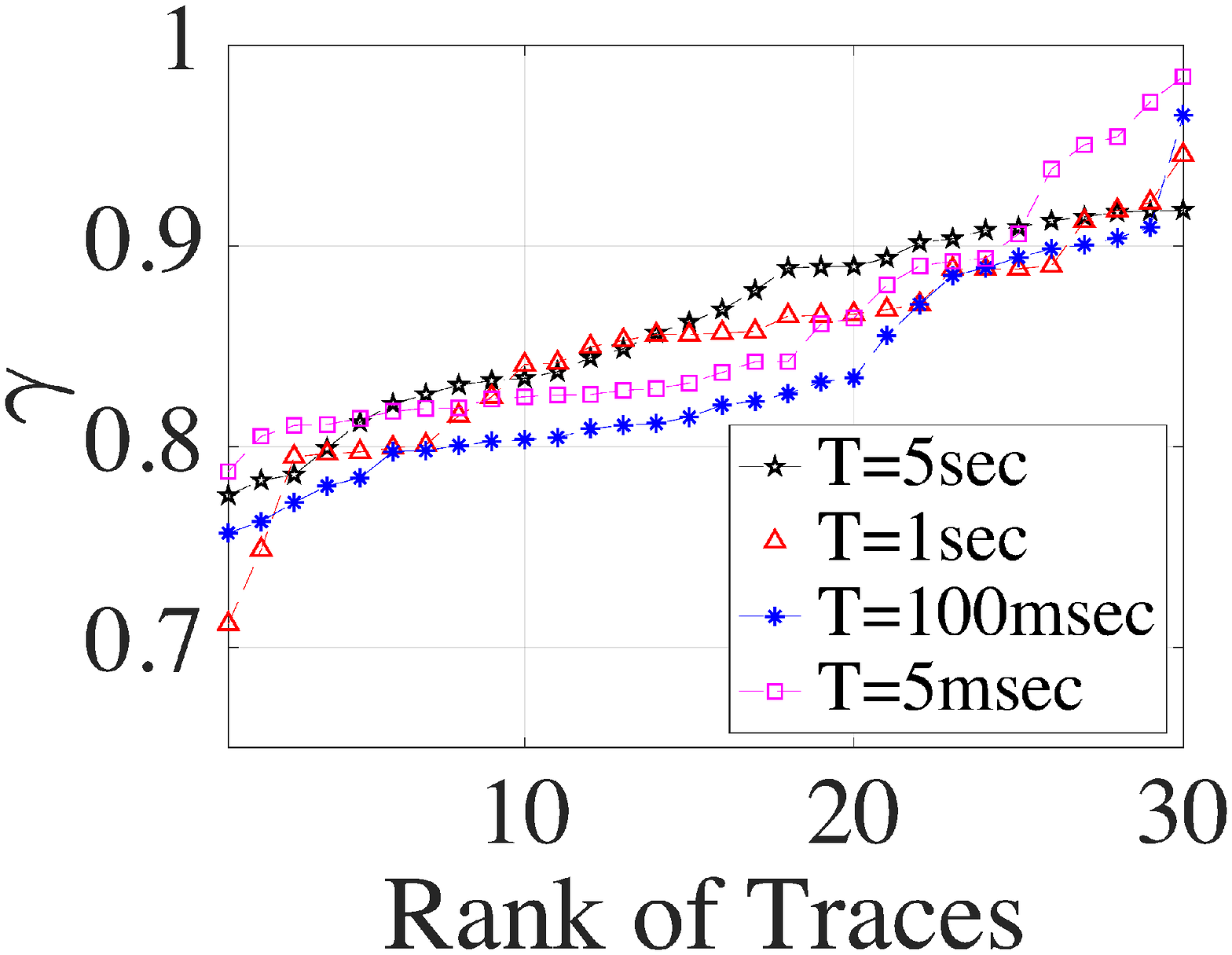}}\quad
		\subcaptionbox{Auckland traces}[0.188\linewidth][c]{%
			\includegraphics[scale=0.184 ]{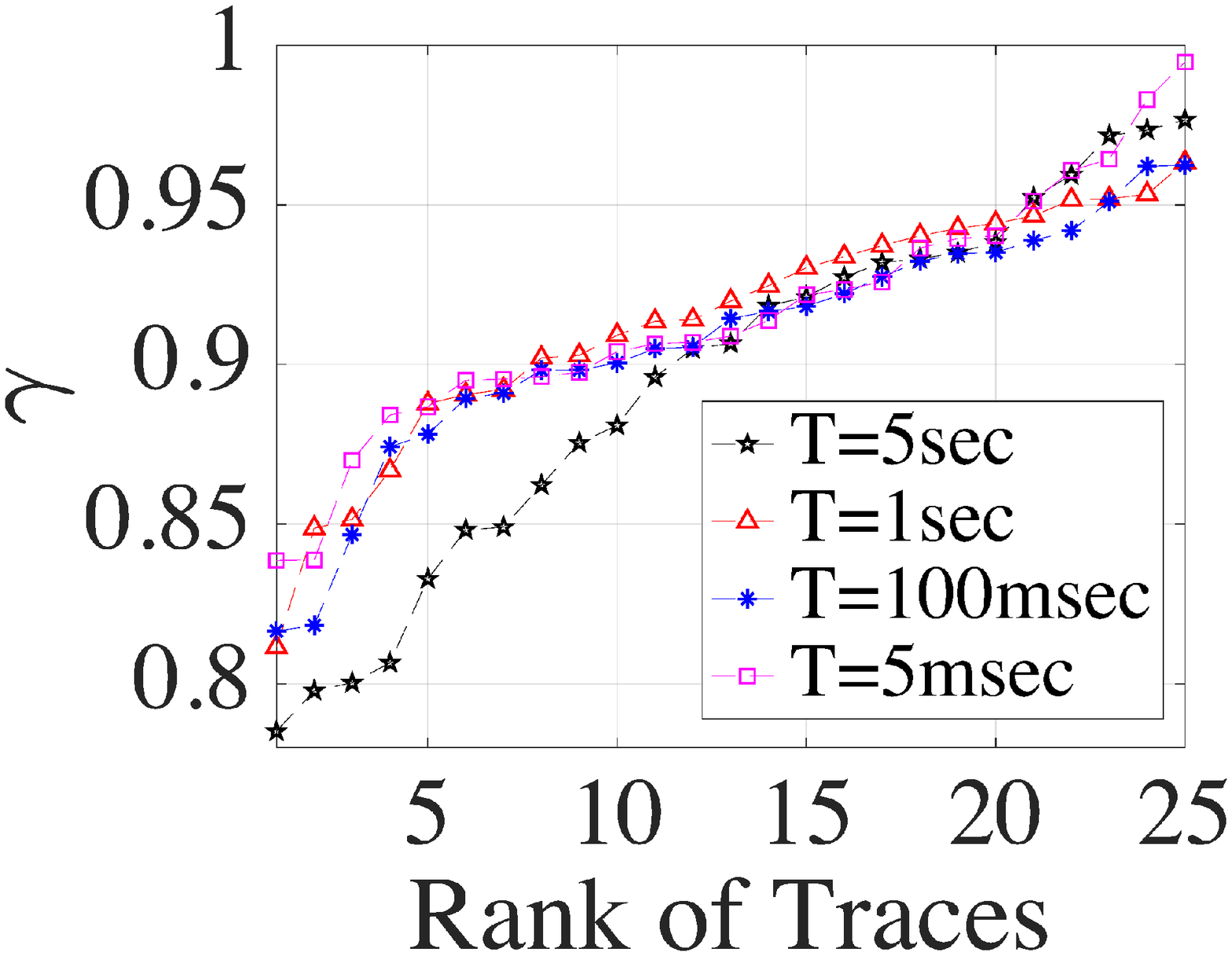}}
		\subcaptionbox{Twente traces}[0.188\linewidth][c]{%
			\includegraphics[scale=0.1834 ]{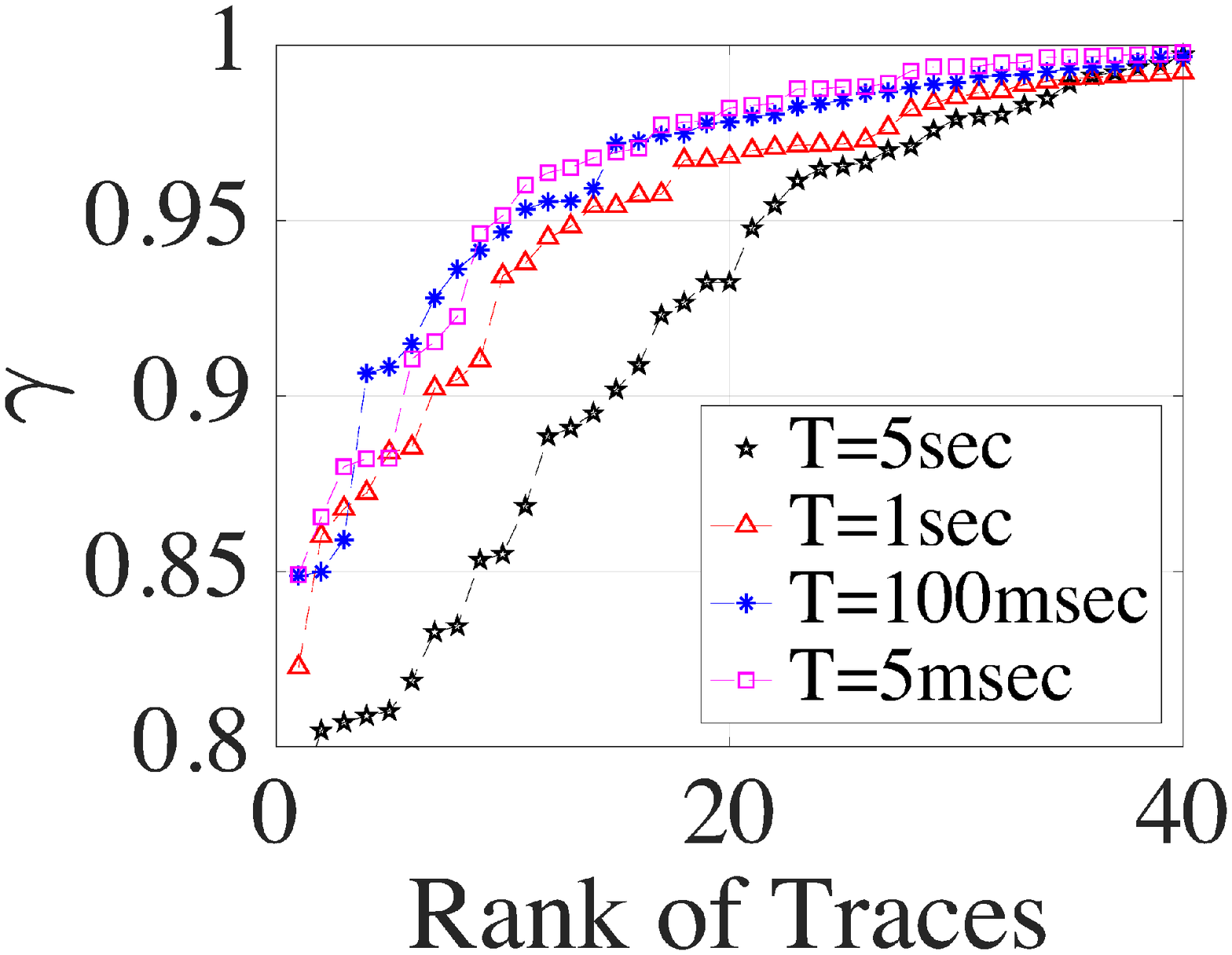}}
		\subcaptionbox{MAWI traces}[0.188\linewidth][c]{%
			\includegraphics[scale=0.184 ]{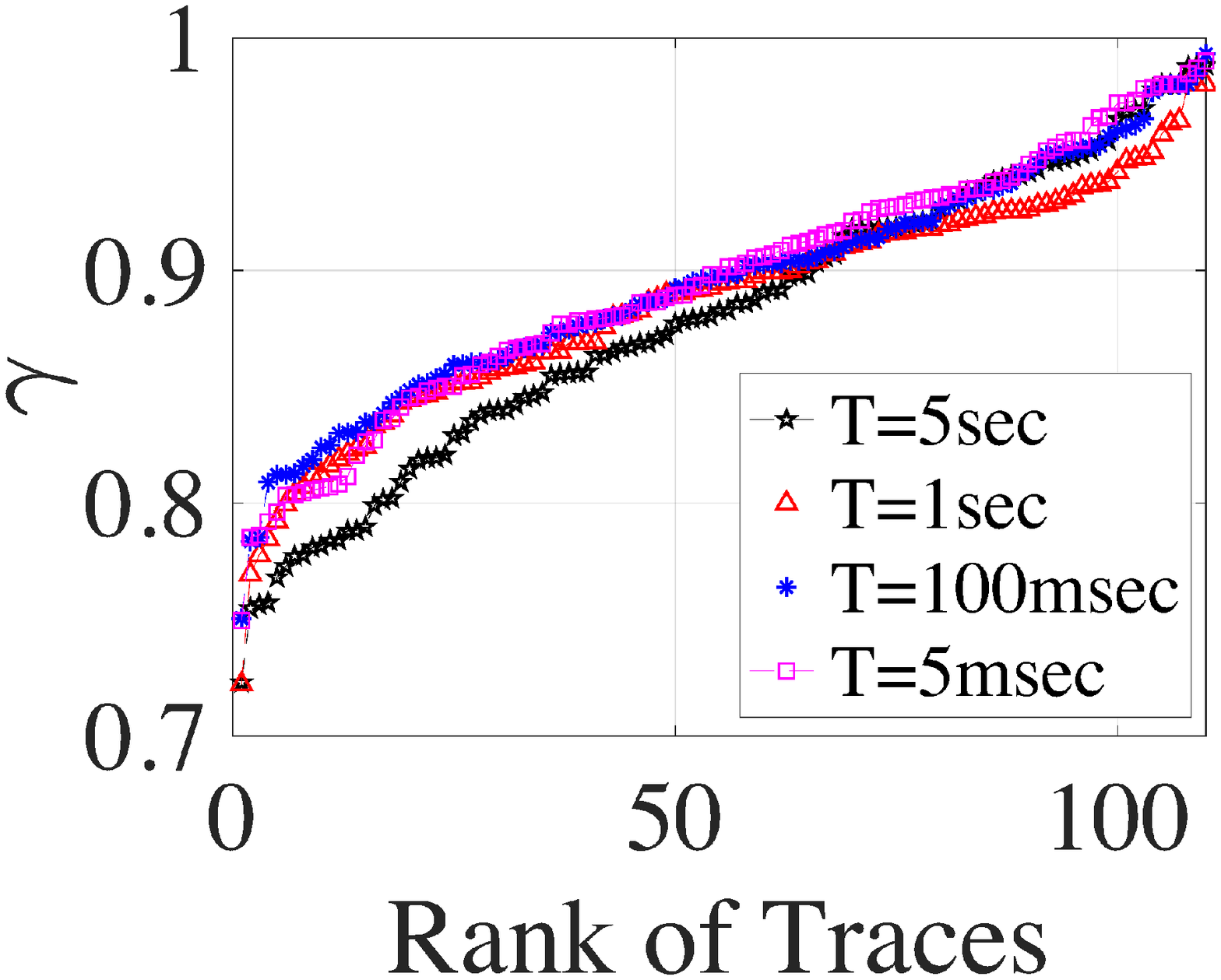}}	
		\subcaptionbox{CAIDA traces}[0.187\linewidth][c]{%
			\includegraphics[scale=0.184 ]{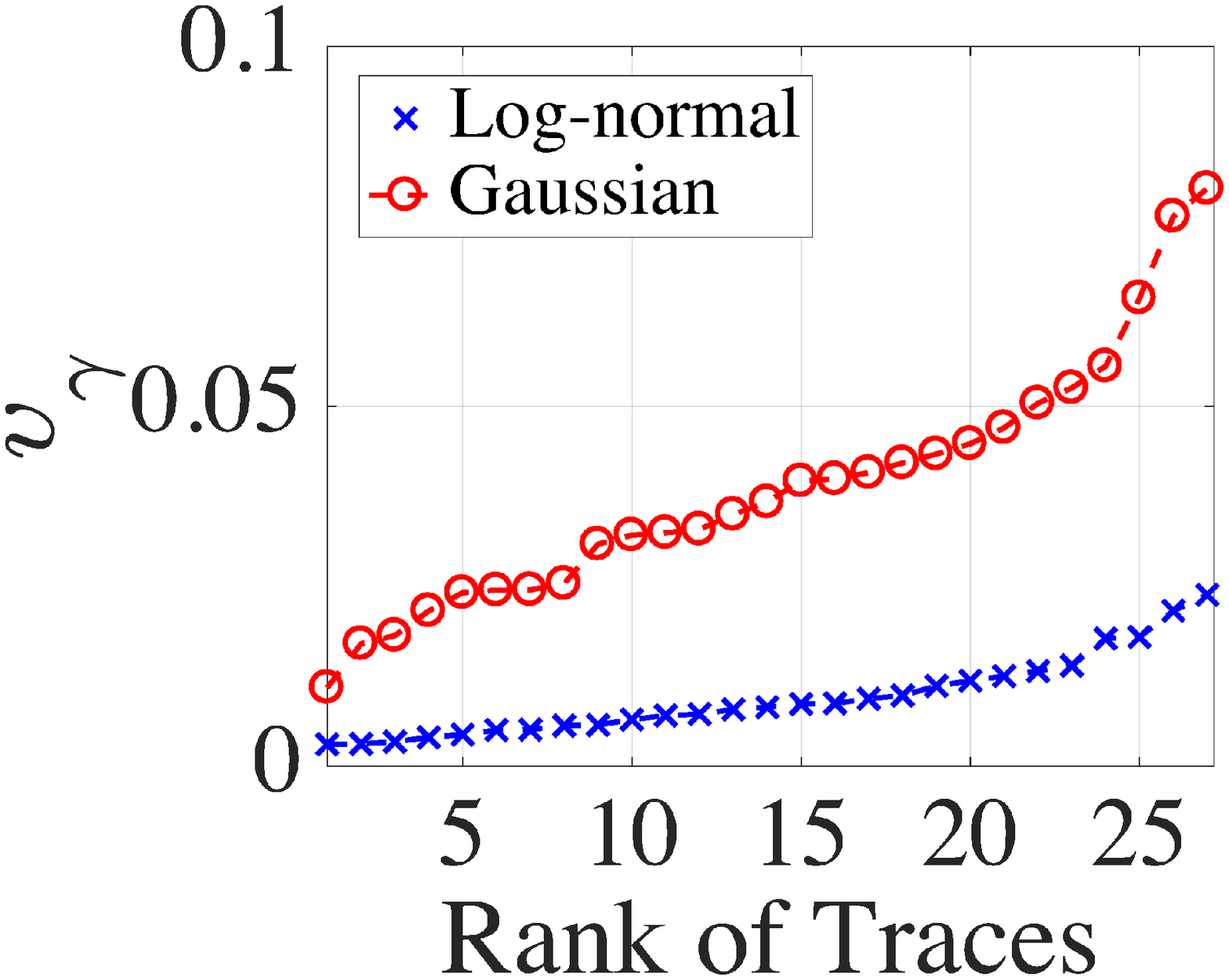}}\quad
		\subcaptionbox{Waikato traces}[0.187\linewidth][c]{%
			\includegraphics[scale=0.184 ]{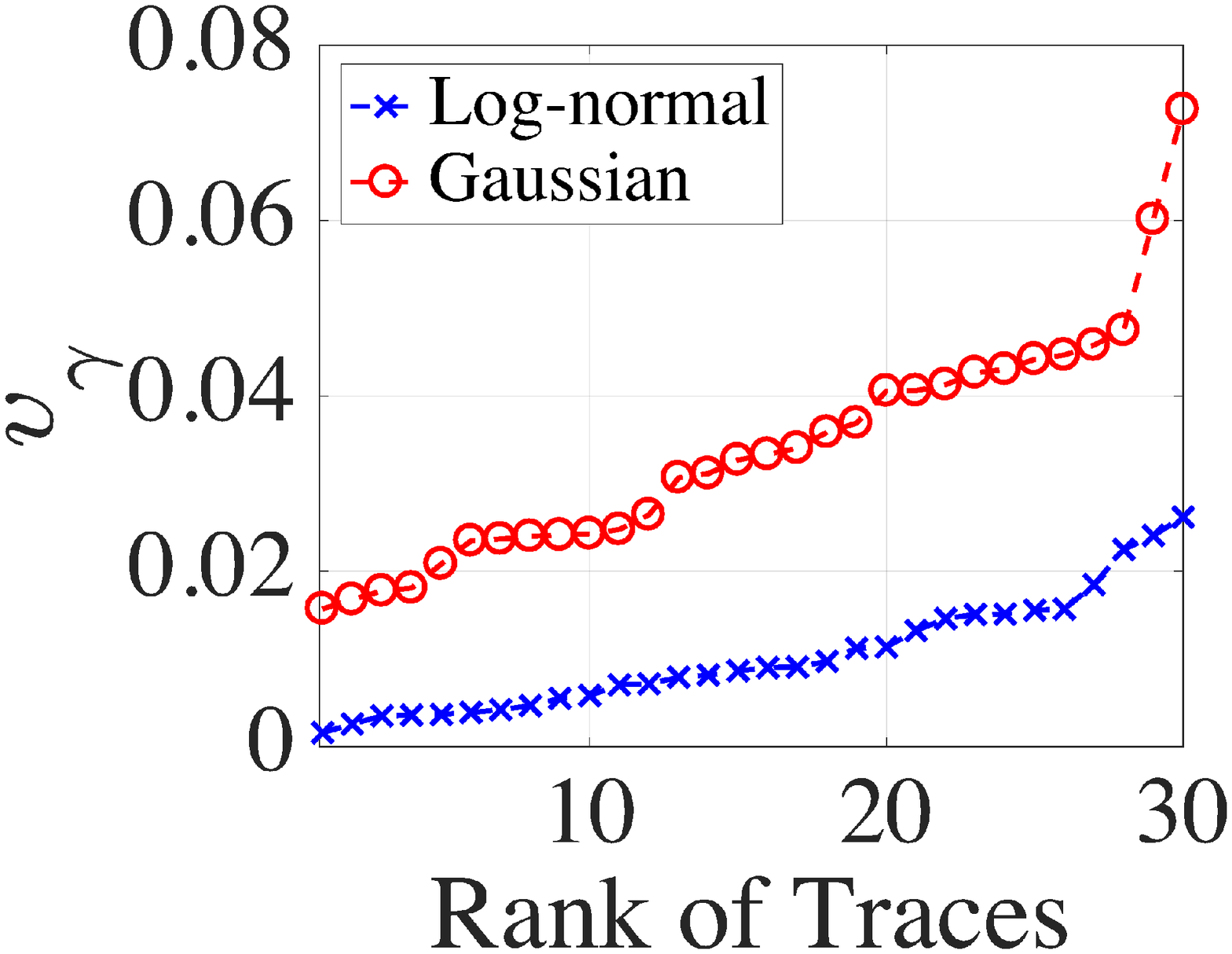}}\quad
		\subcaptionbox{Auckland traces}[0.187\linewidth][c]{%
			\includegraphics[scale=0.184 ]{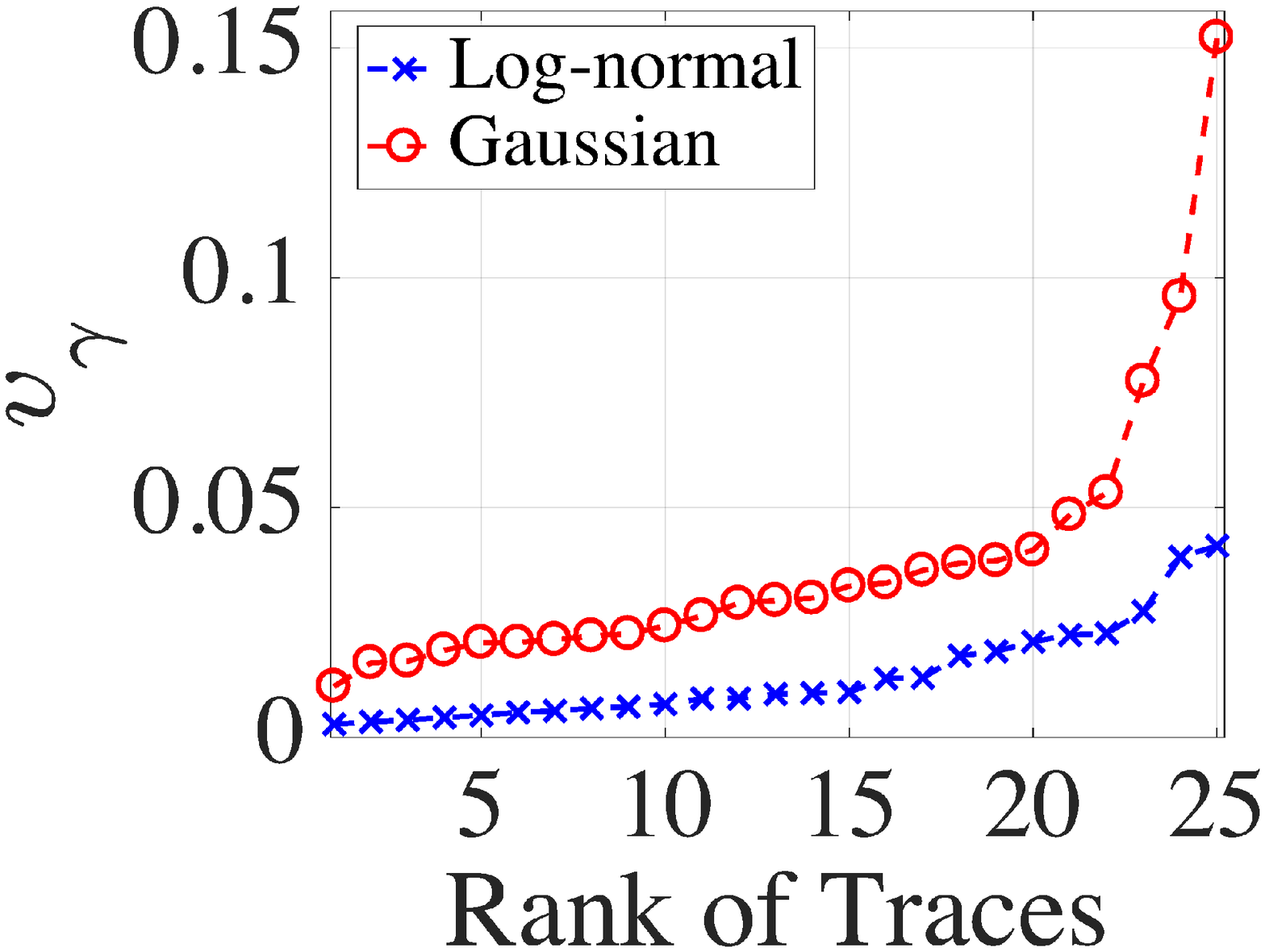}}
		\subcaptionbox{Twente traces}[0.187\linewidth][c]{%
			\includegraphics[scale=0.184 ]{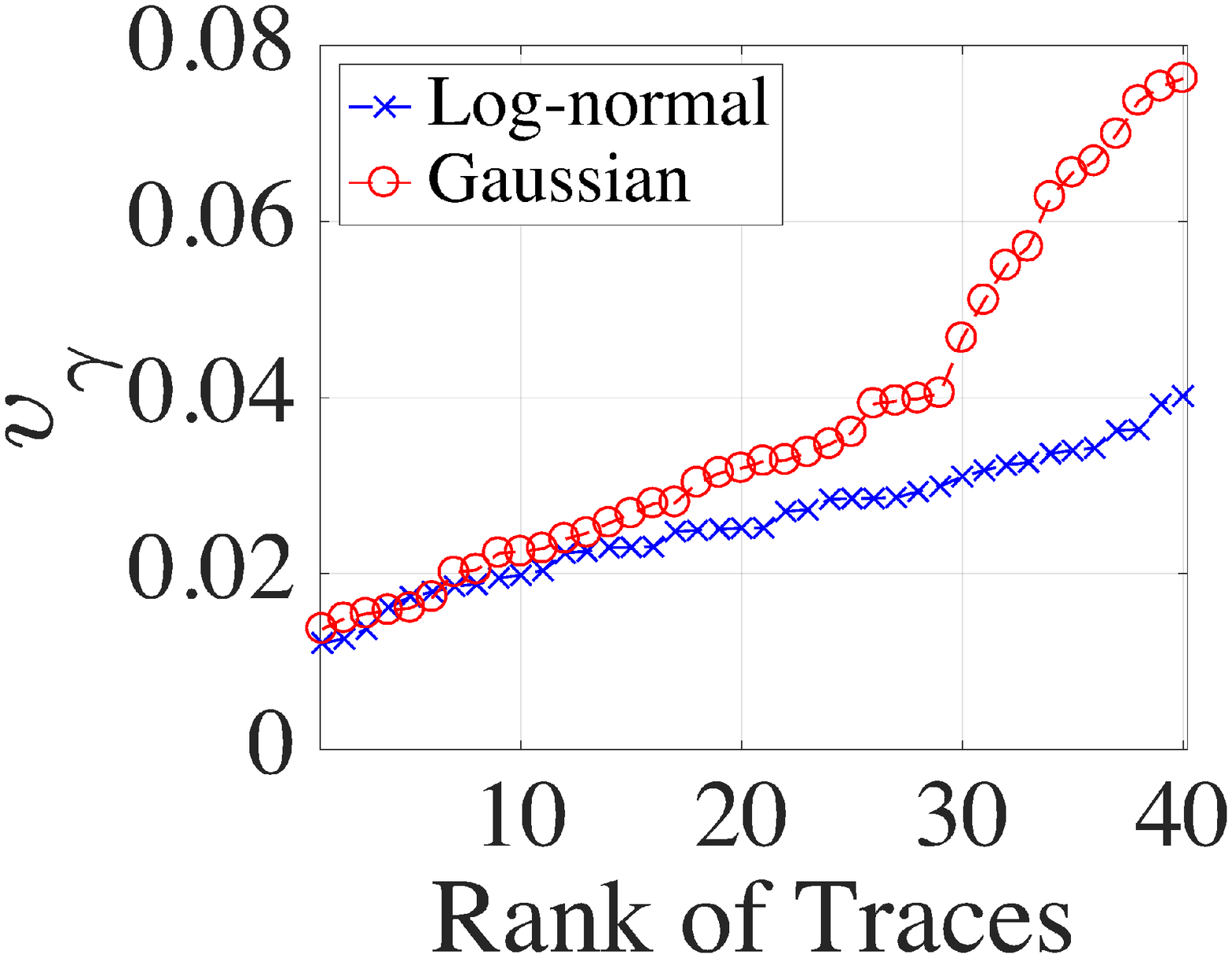}}\quad
		\subcaptionbox{MAWI traces}[0.1875\linewidth][c]{%
			\includegraphics[scale=0.184 ]{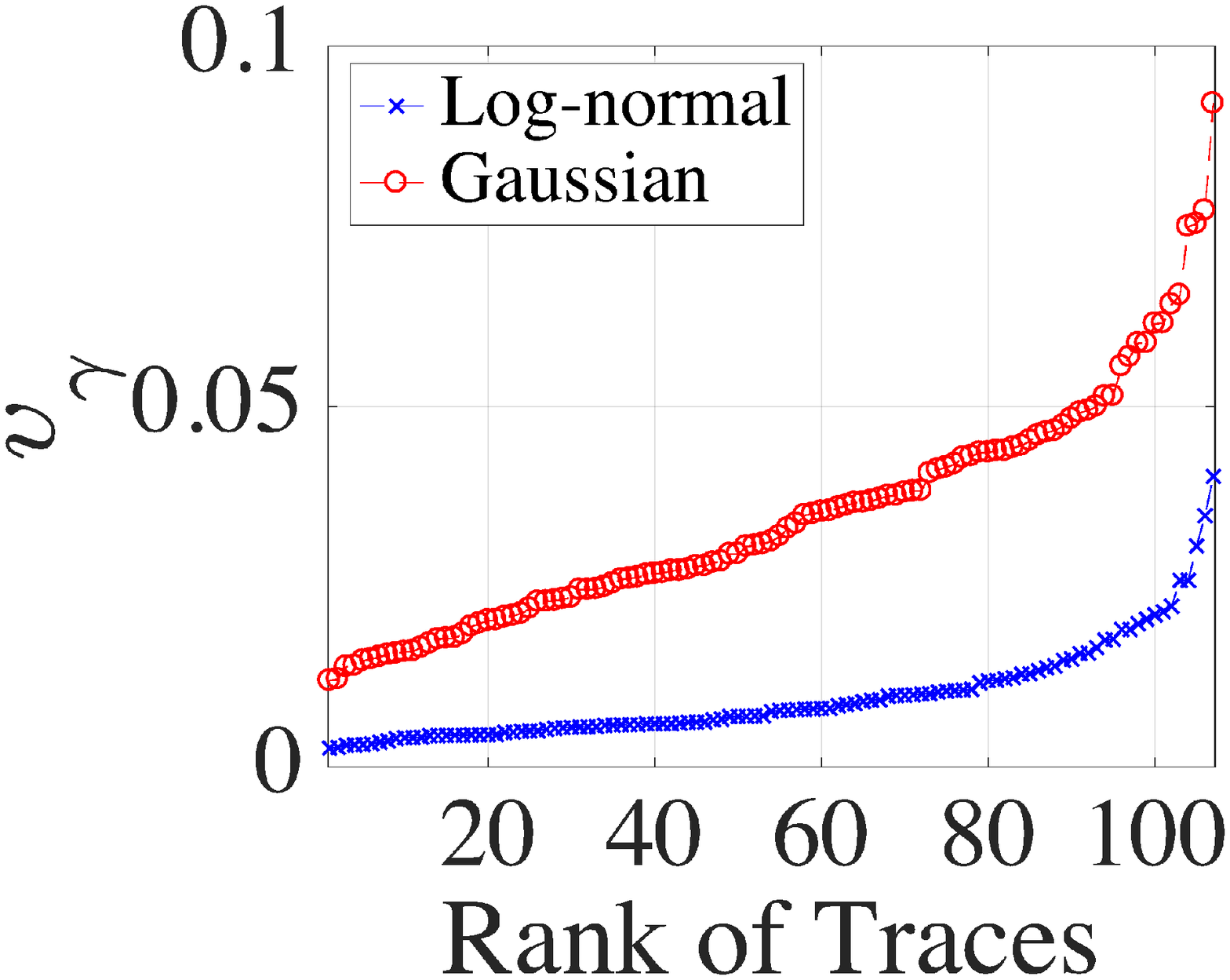}}\quad	
		\subcaptionbox{CAIDA traces }[.186\linewidth][c]{%
			\includegraphics[height=3cm, width=3.4cm ]{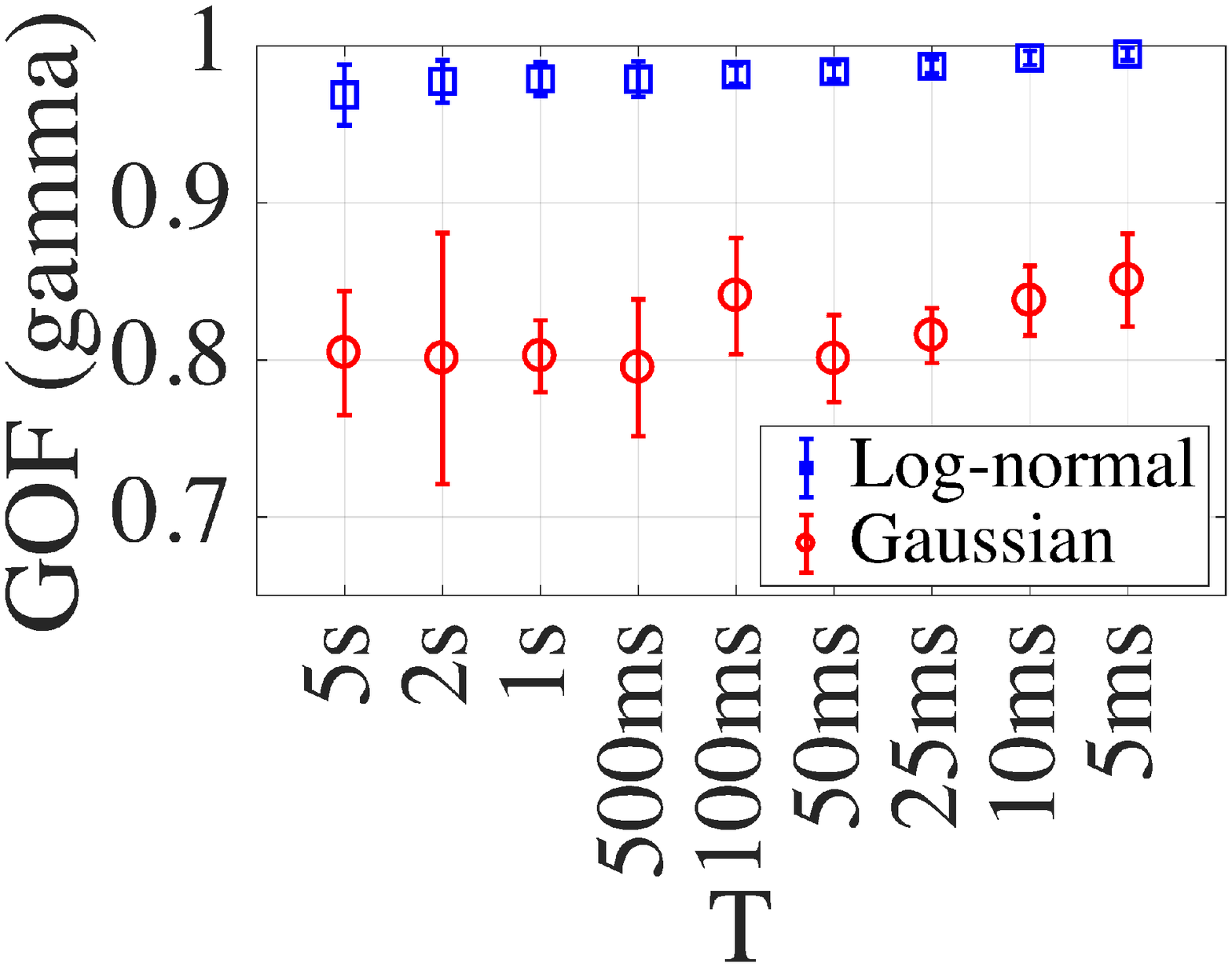}}\quad
		\subcaptionbox{Waikato traces}[.186\linewidth][c]{%
			\includegraphics[height=3cm, width=3.4cm ]{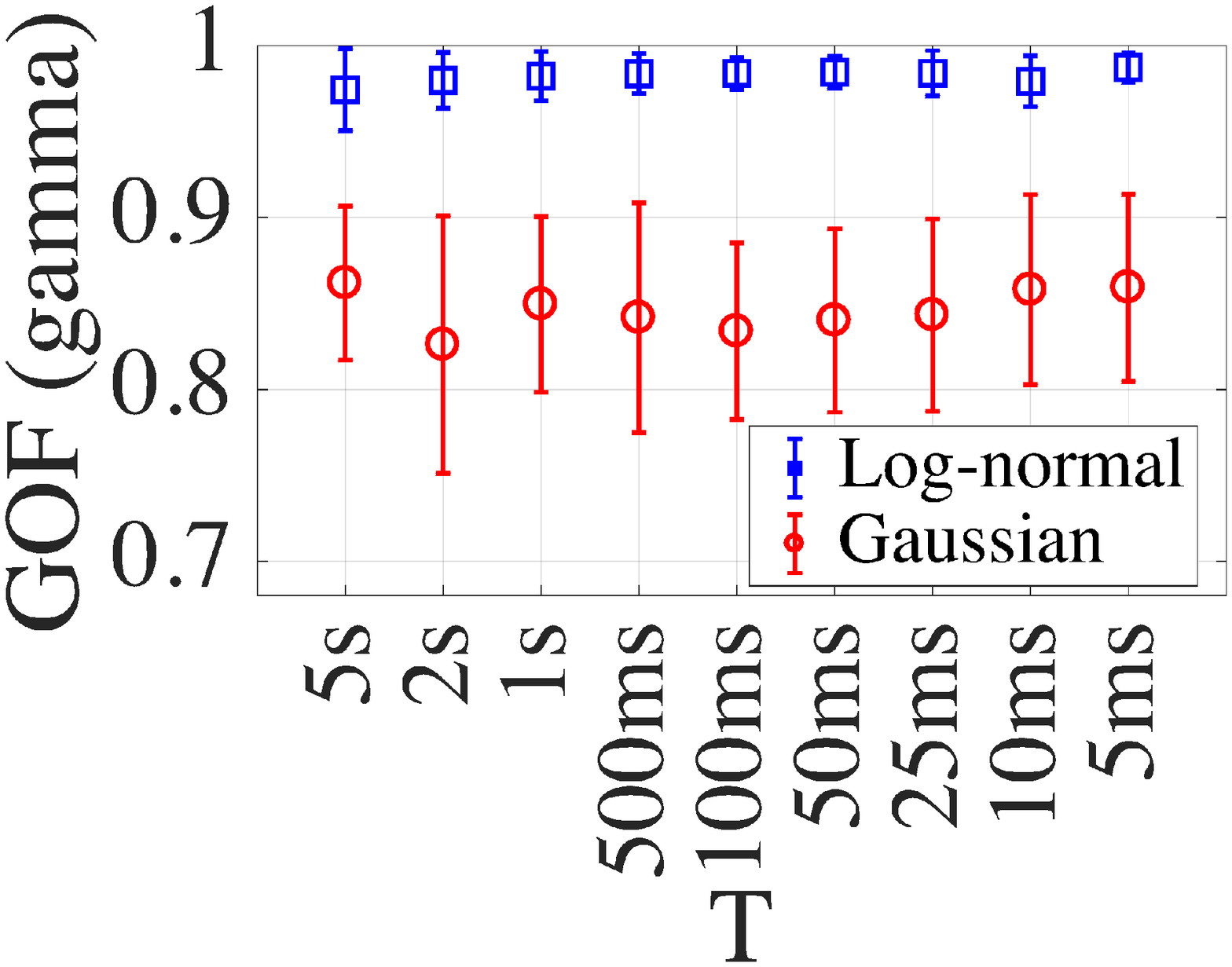}}\quad
		\subcaptionbox{Auckland traces}[.188\linewidth][c]{%
			\includegraphics[height=3cm, width=3.4cm ]{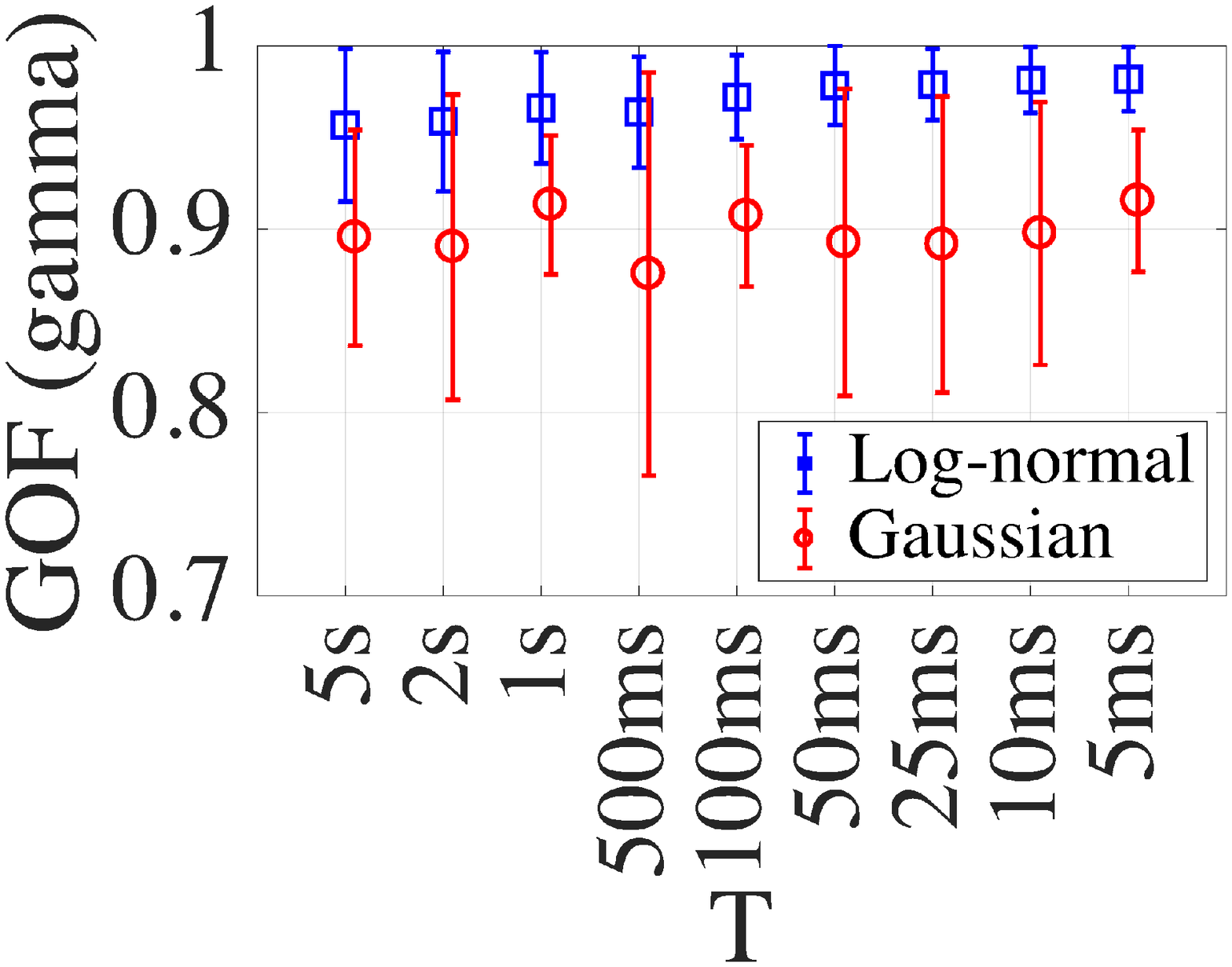}}
		\subcaptionbox{Twente traces}[.186\linewidth][c]{%
			\includegraphics[height=3cm, width=3.4cm ]{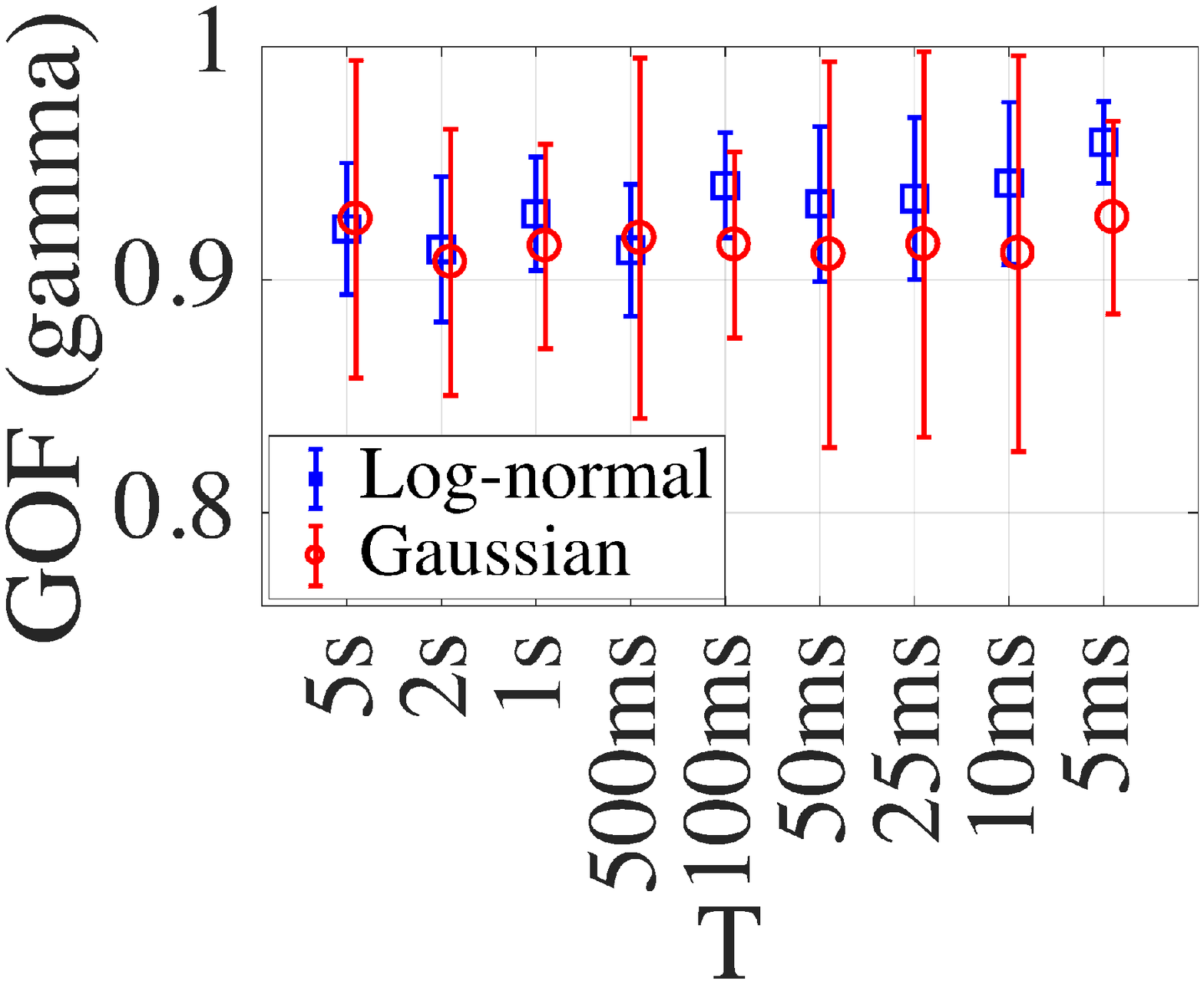}}\quad
		\subcaptionbox{MAWI traces}[.186\linewidth][c]{%
			\includegraphics[height=3cm, width=3.4cm ]{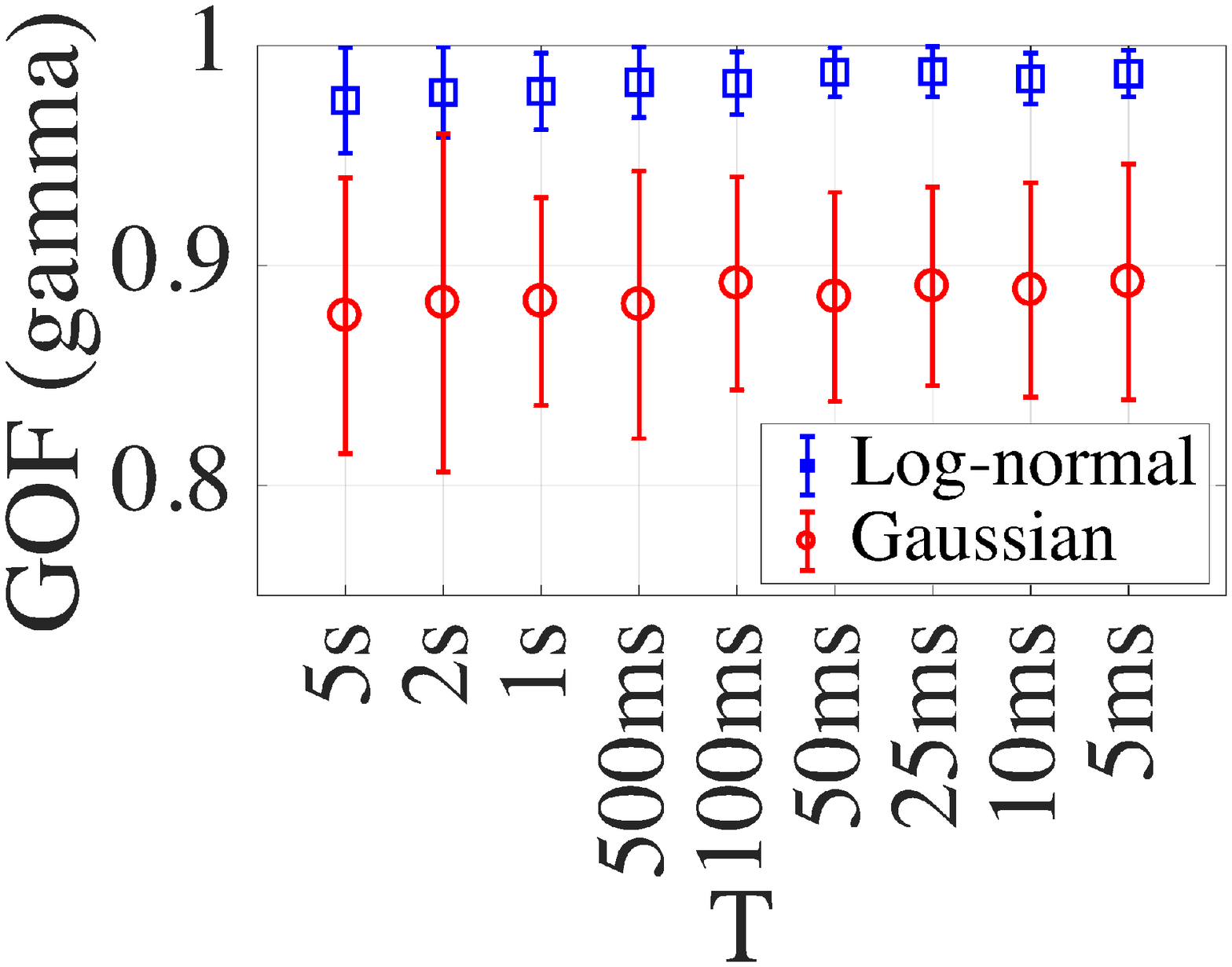}}\quad	
		\caption{\textcolor{black}{Correlation coefficient $\gamma$ test results for all studied traces and different timescales for the  log-normal (a-e) and Gaussian  (f-j) distributions. The variation $\upsilon_{\gamma}$ results (k-o). Goodness of fit (GOF)  results (p-t).}}
		\label{gammaRes} 	
	\end{figure*}
	
	\begin{figure*}[ht]
		\setlength{\belowcaptionskip}{-2.5pt}
		\centering
		\subcaptionbox{ADF test: CAIDA}[.20\linewidth][c]{%
			\includegraphics[width=1\linewidth]{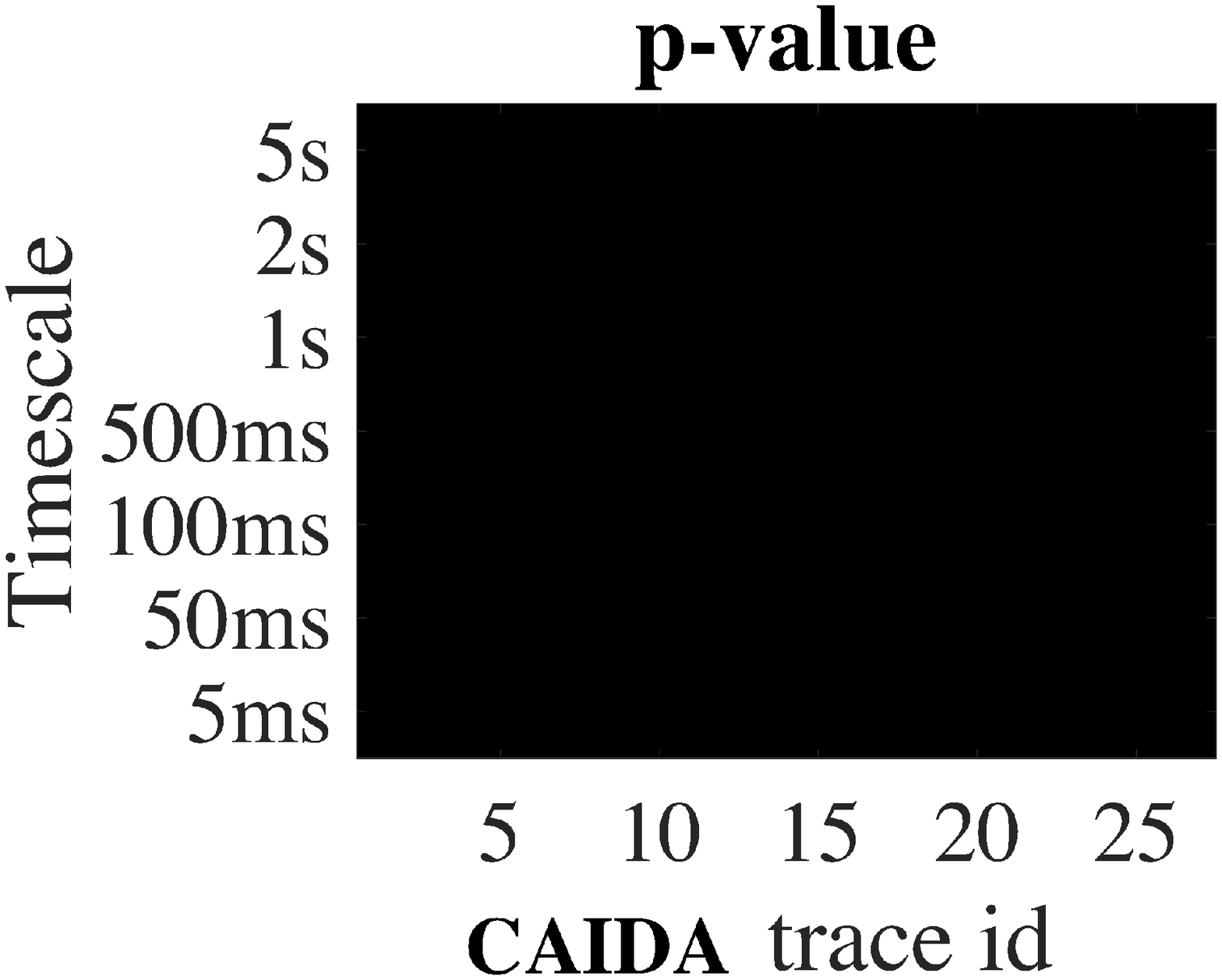}} 
		\hspace{-0.3cm}
		\subcaptionbox{ADF test: Waikato}[.20\linewidth][c]{%
			\includegraphics[width=1\linewidth]{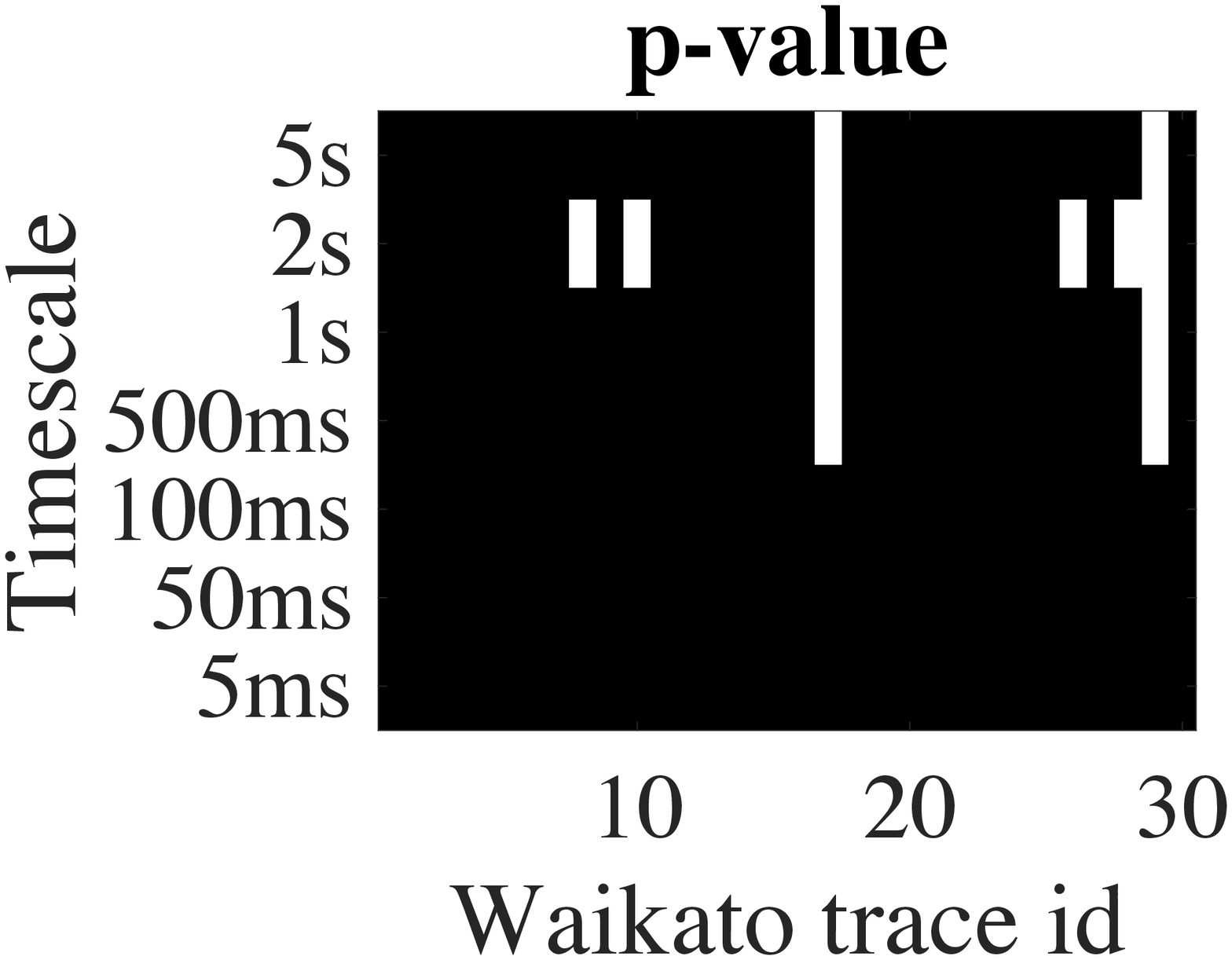}} 
		\hspace{-0.25cm}
		\subcaptionbox{ADF test: Auckland}[.20\linewidth][c]{%
			\includegraphics[width=1\linewidth]{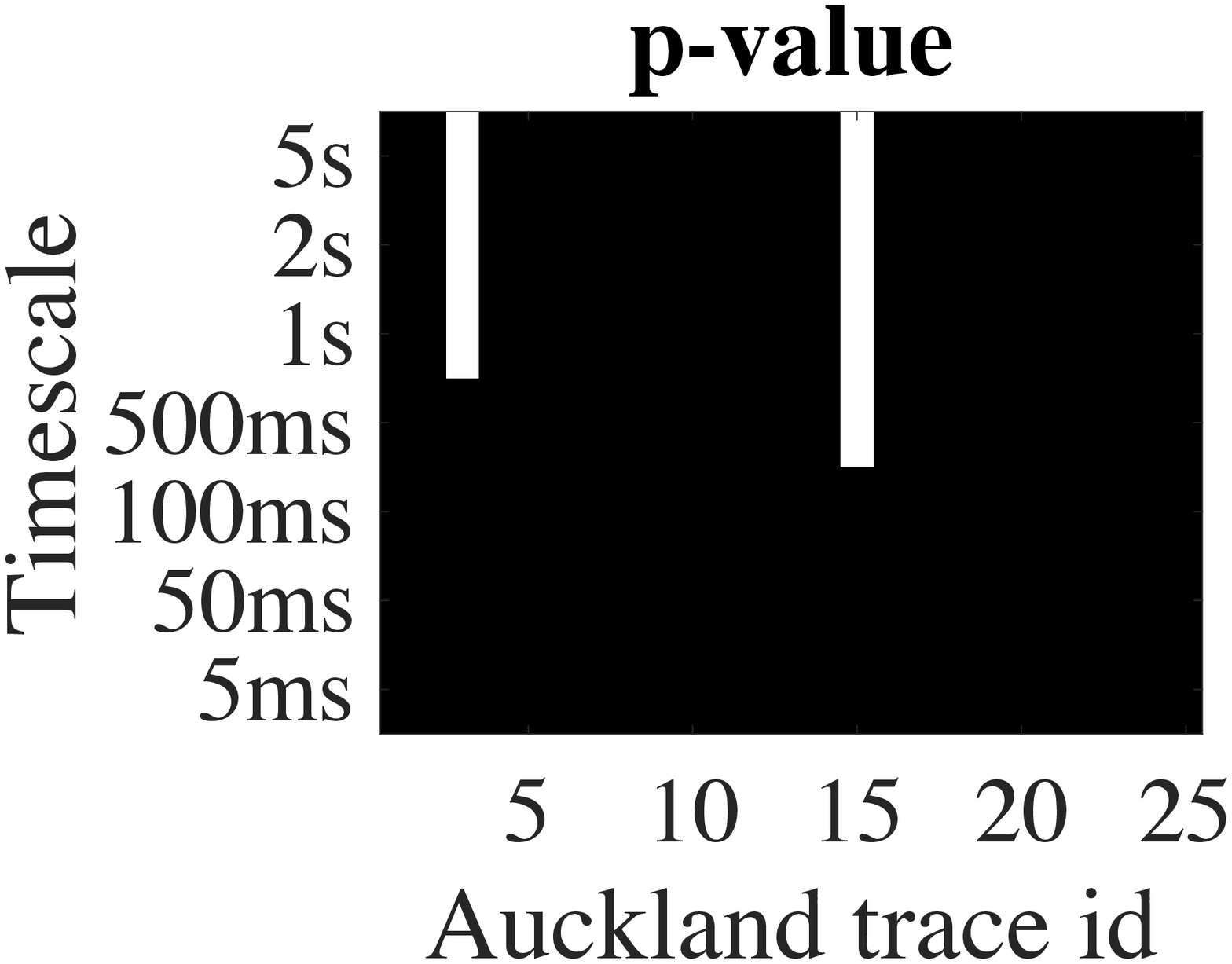}}
		\hspace{-0.25cm}
		\subcaptionbox{ADF test: Twente}[.20\linewidth][c]{%
			\includegraphics[width=1\linewidth]{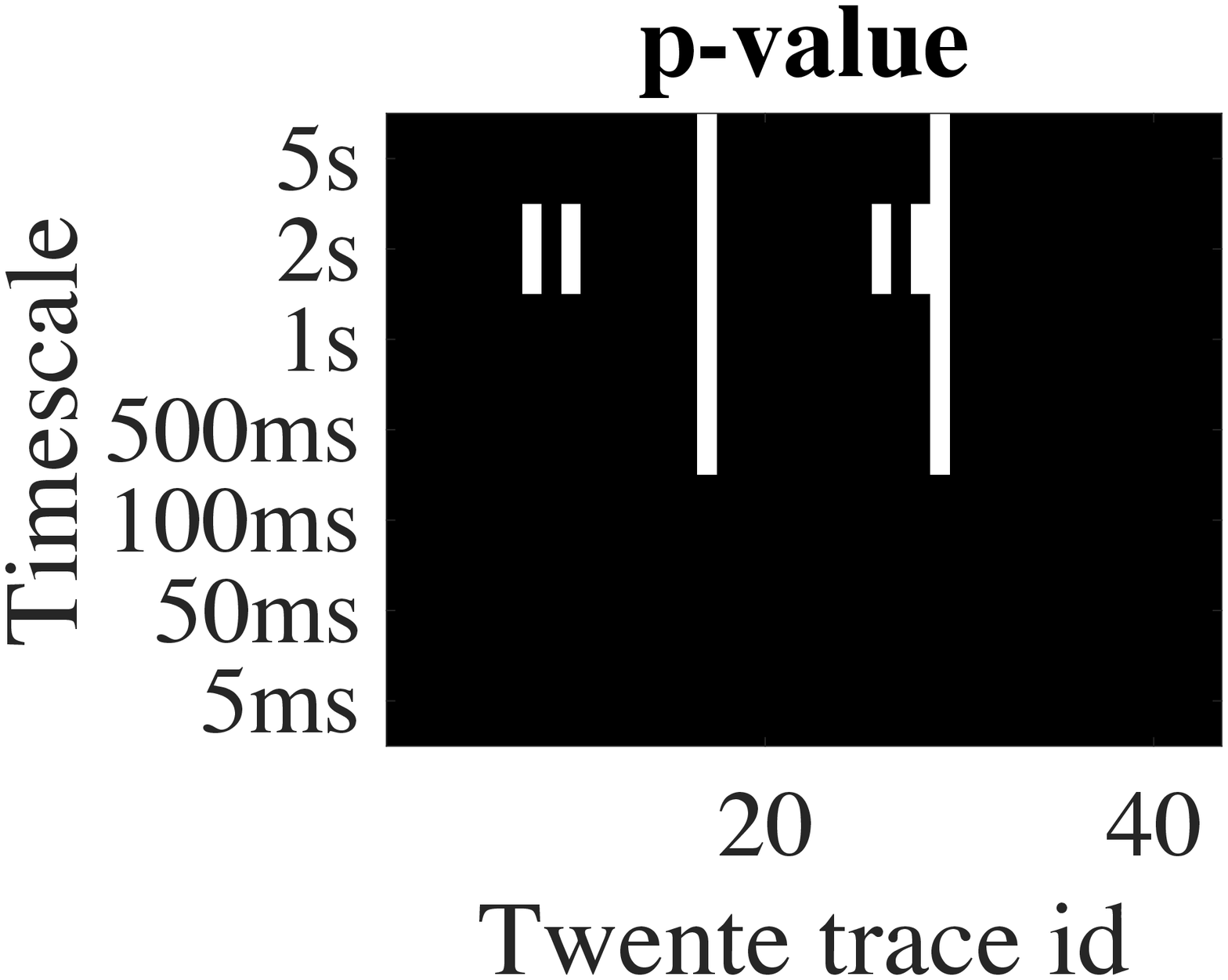}}
		\hspace{-0.3cm}
		\subcaptionbox{ADF test: MAWI}[.20\linewidth][c]{%
			\includegraphics[width=1\linewidth]{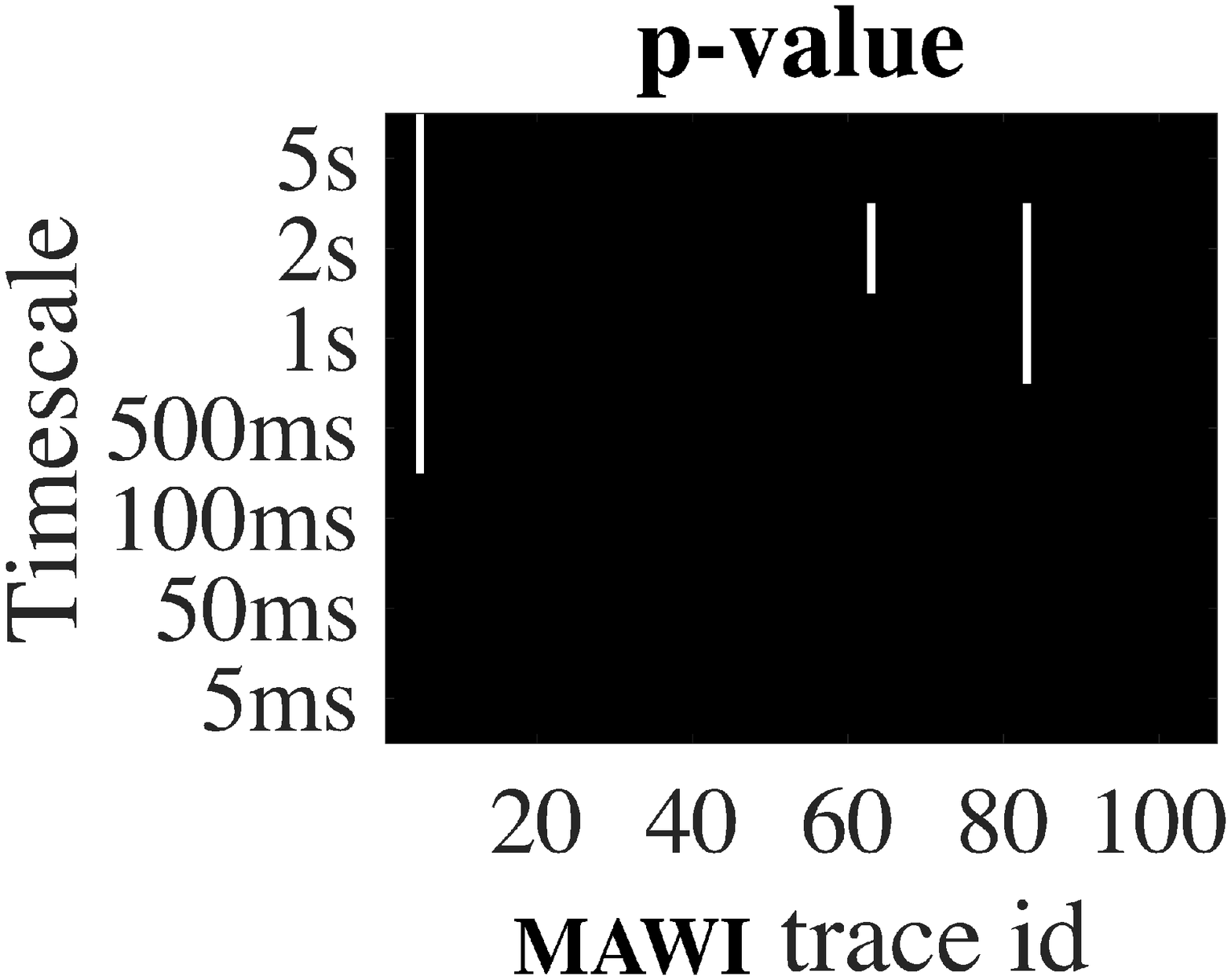}}
		\bigskip
		\\
		\setlength{\belowcaptionskip}{-2.5pt}
		\subcaptionbox{PP test: CAIDA}[.20\linewidth][c]{%
			\includegraphics[width=1\linewidth]{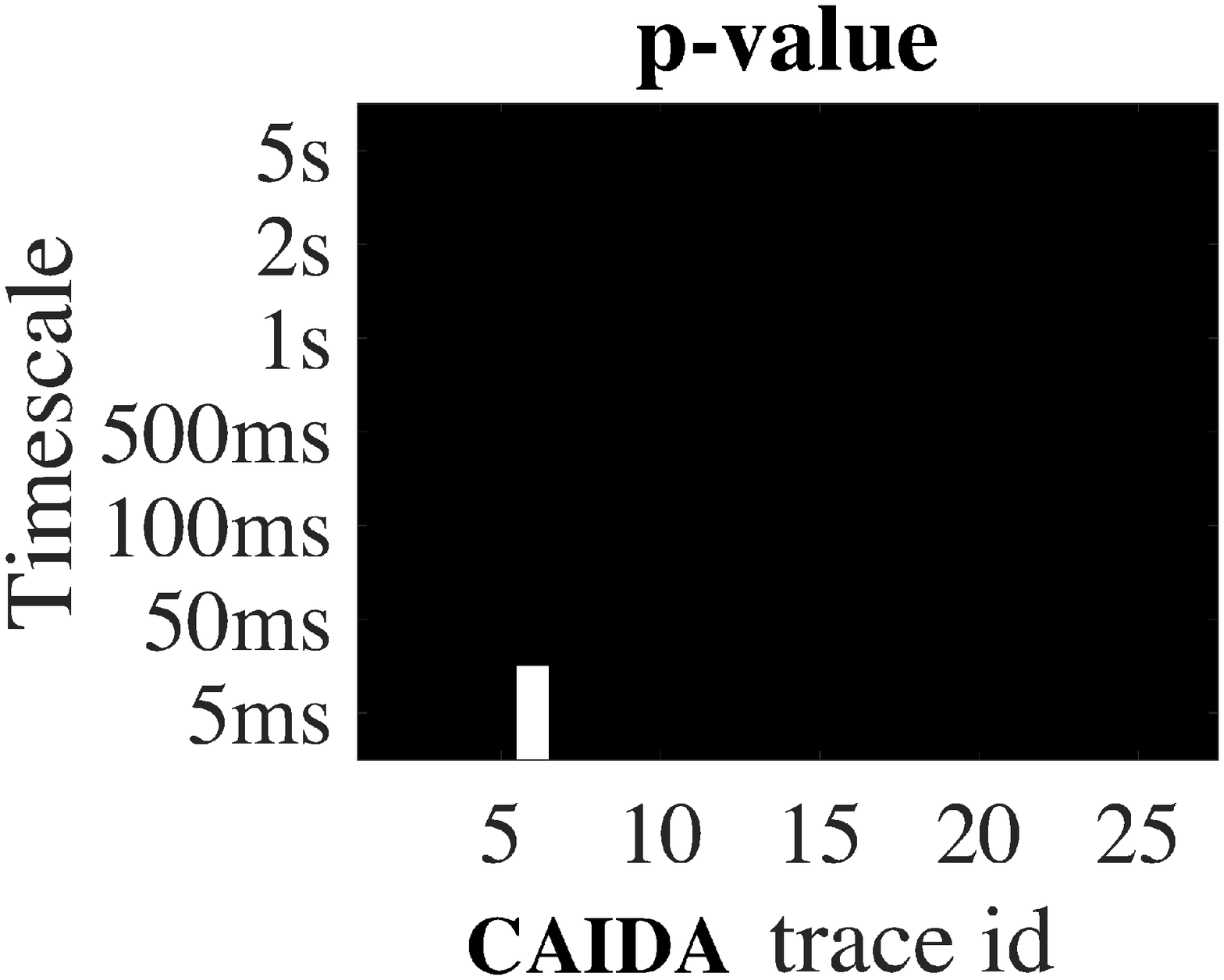}} 
		\hspace{-0.4cm}
		\subcaptionbox{PP test: Waikato}[.20\linewidth][c]{%
			\includegraphics[width=1\linewidth]{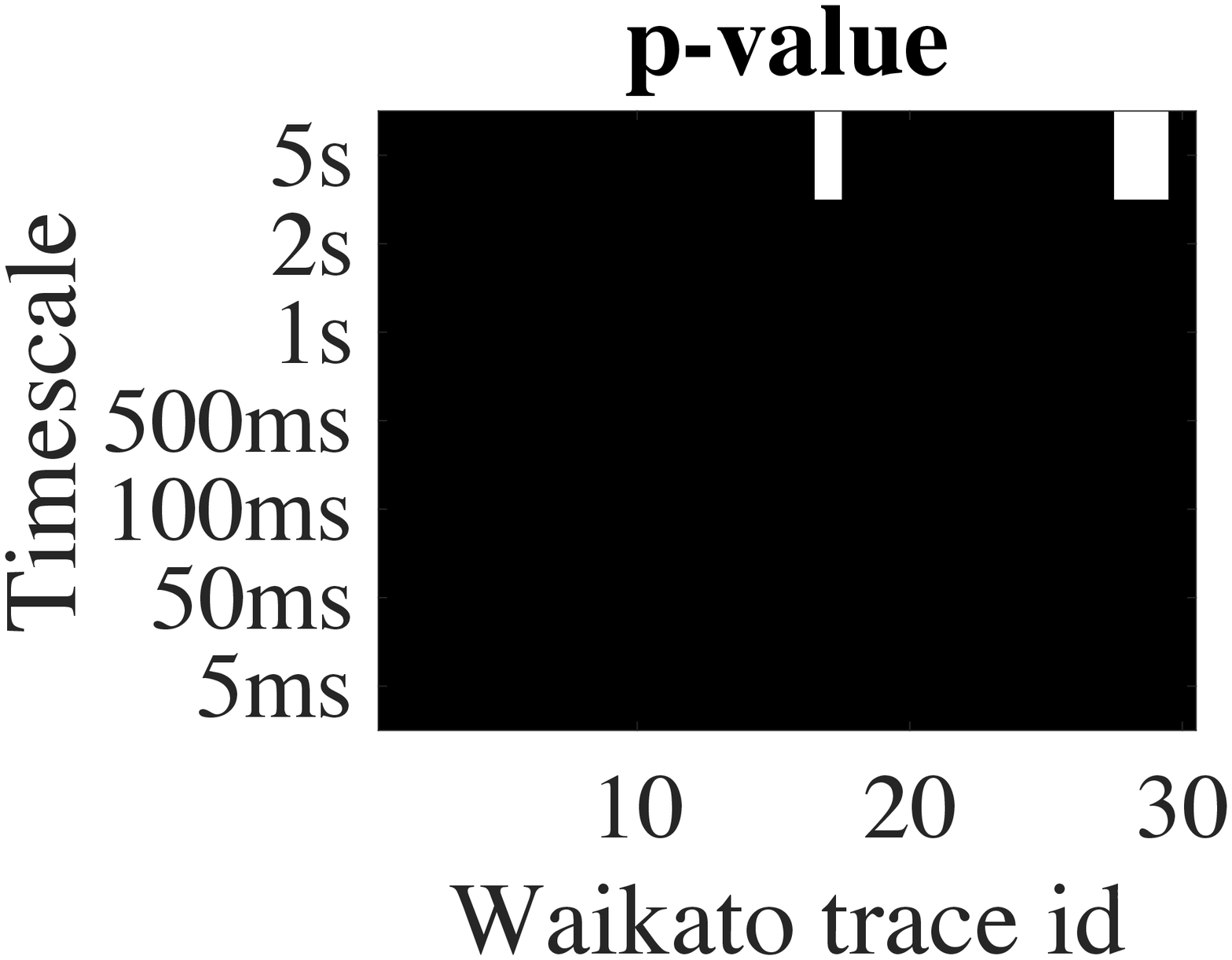}} 
		\hspace{-0.27cm}
		\subcaptionbox{PP test: Auckland}[.20\linewidth][c]{%
			\includegraphics[width=1\linewidth]{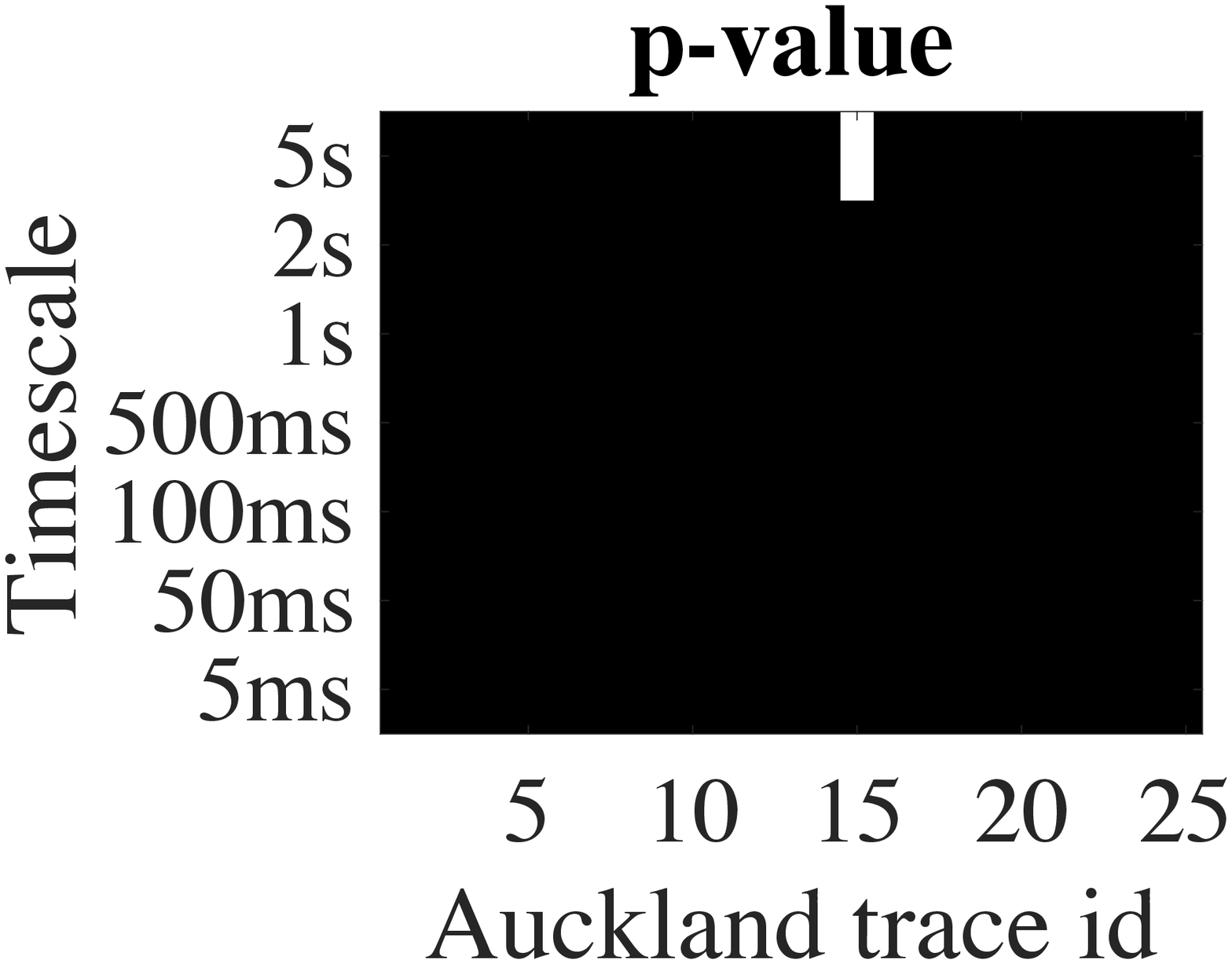}}
		\hspace{-0.27cm}
		\subcaptionbox{PP test: Twente}[.20\linewidth][c]{%
			\includegraphics[width=1\linewidth]{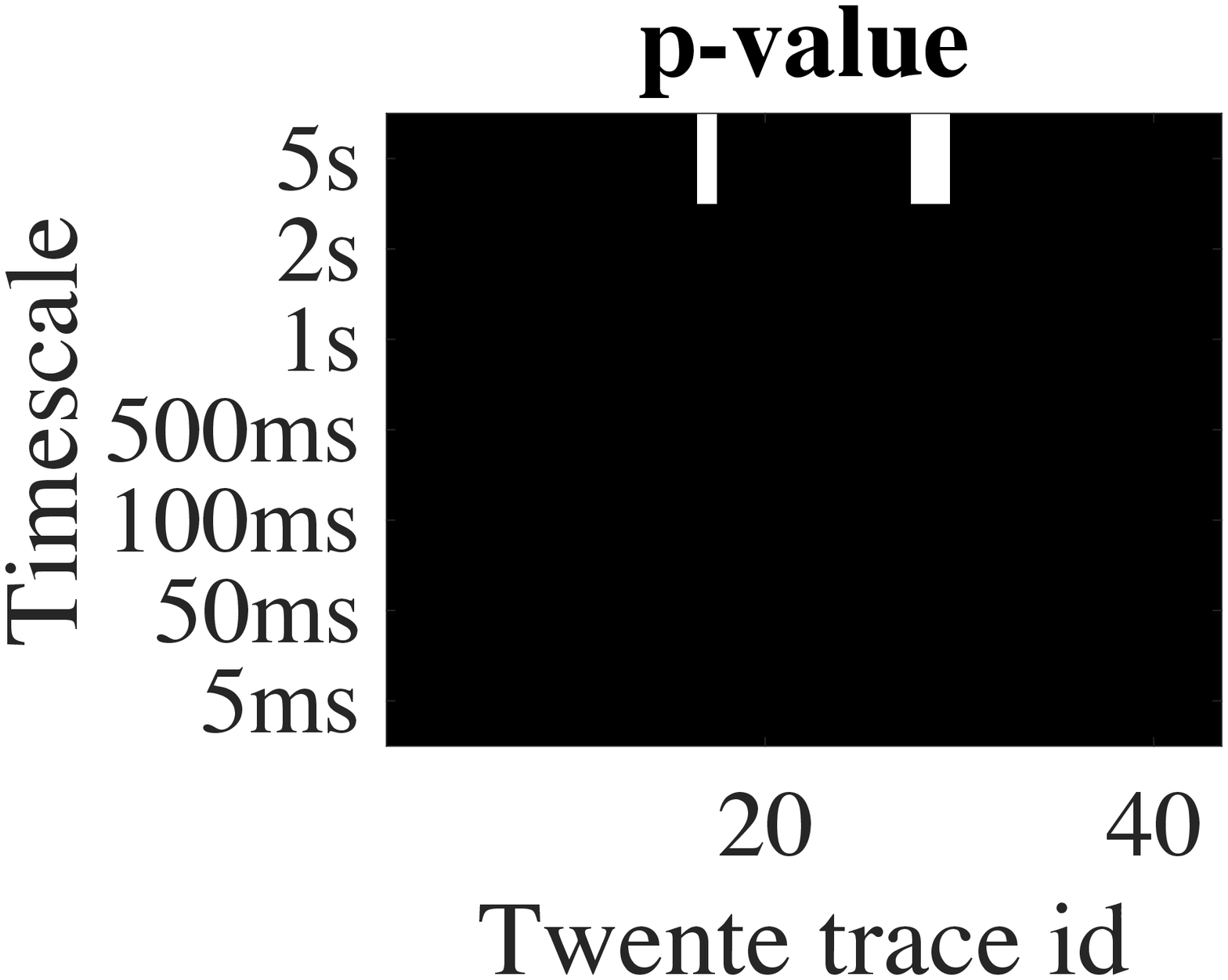}}
		\hspace{-0.3cm}
		\subcaptionbox{PP test: MAWI}[.20\linewidth][c]{%
			\includegraphics[width=1\linewidth]{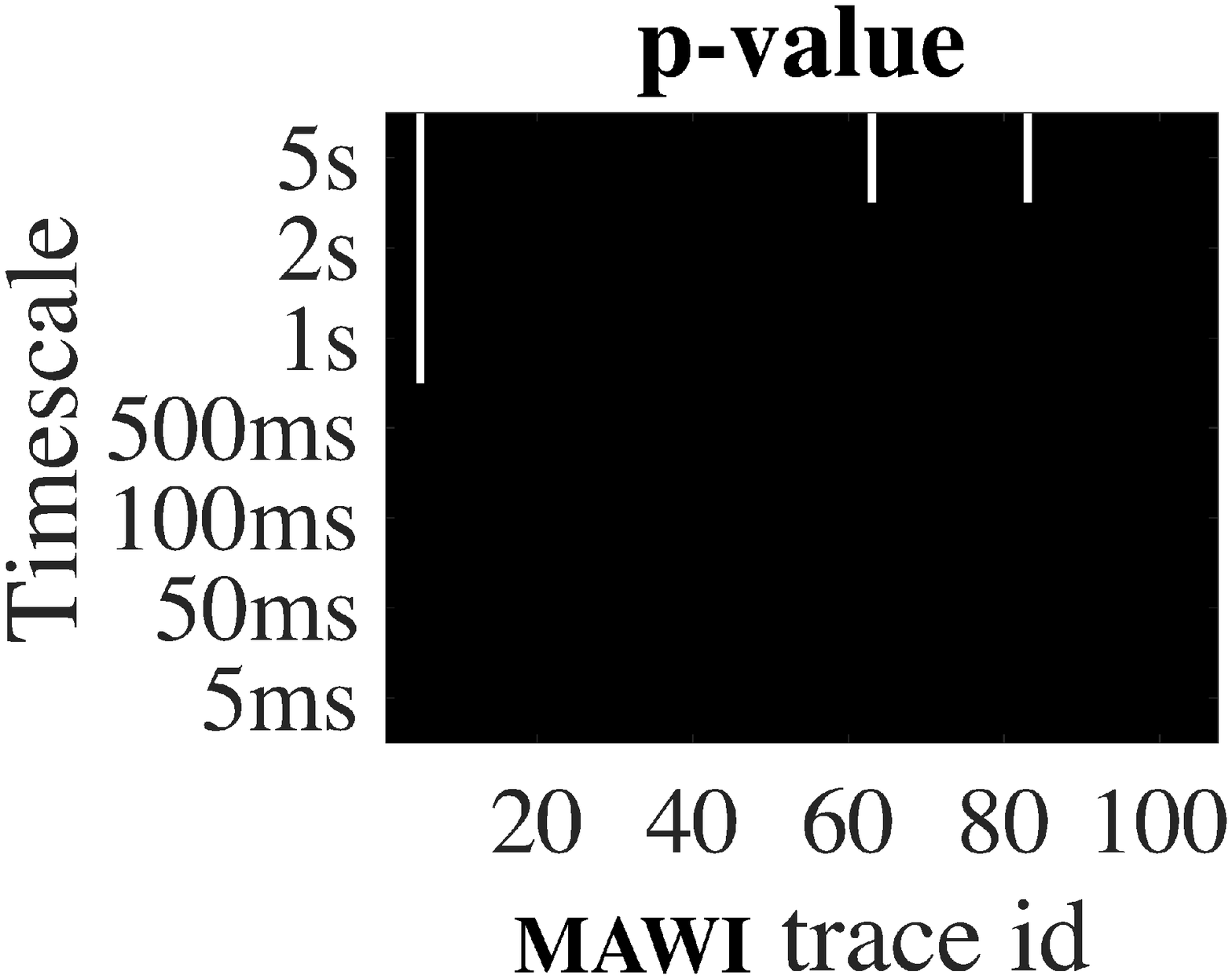}}
		\bigskip
		\\
		\setlength{\belowcaptionskip}{-2.5pt}
		\subcaptionbox{KPSS test: CAIDA}[.20\linewidth][c]{%
			\includegraphics[width=1\linewidth]{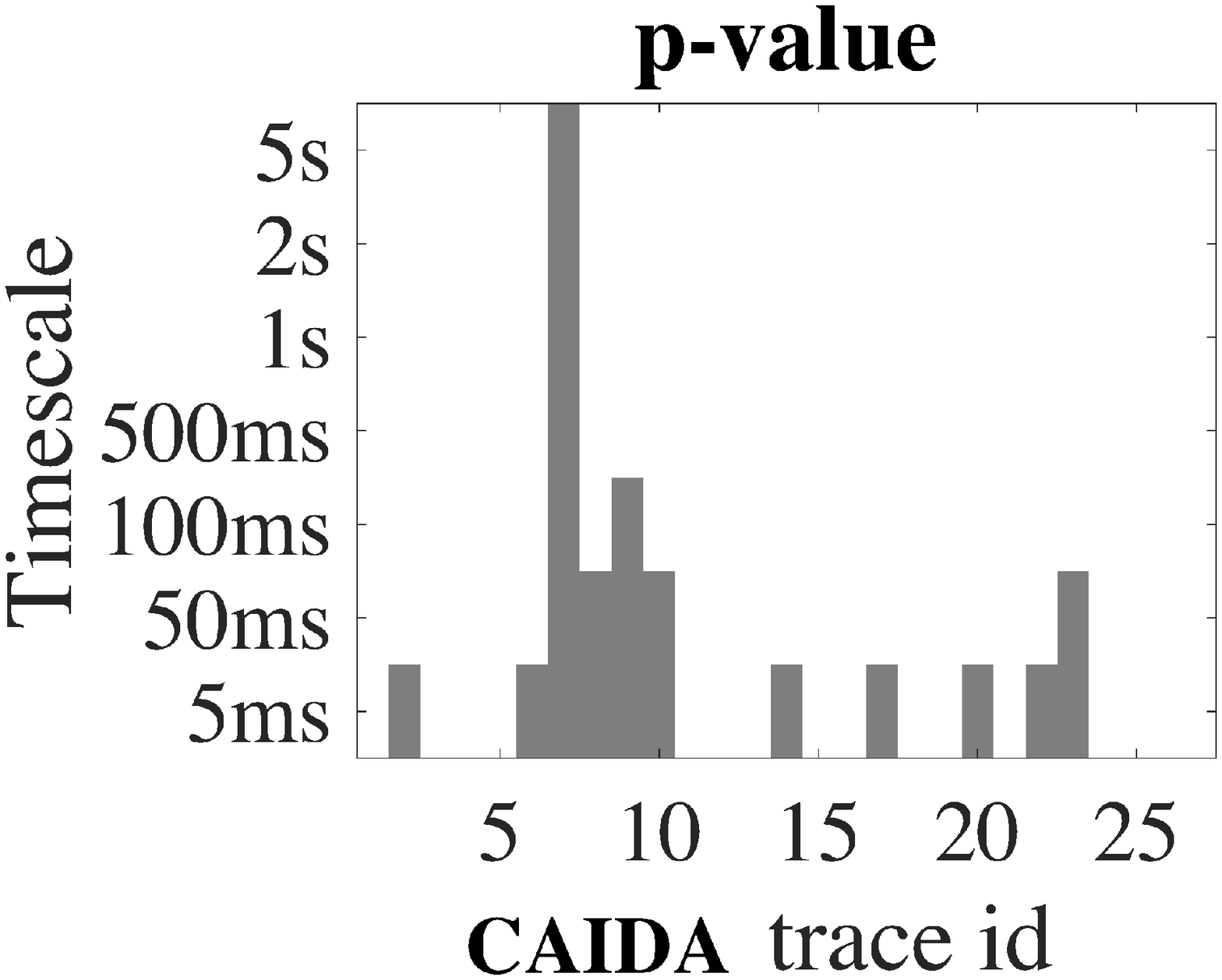}} 
		\hspace{-0.2cm}
		\subcaptionbox{KPSS test: Waikato}[.20\linewidth][c]{%
			\includegraphics[width=1\linewidth]{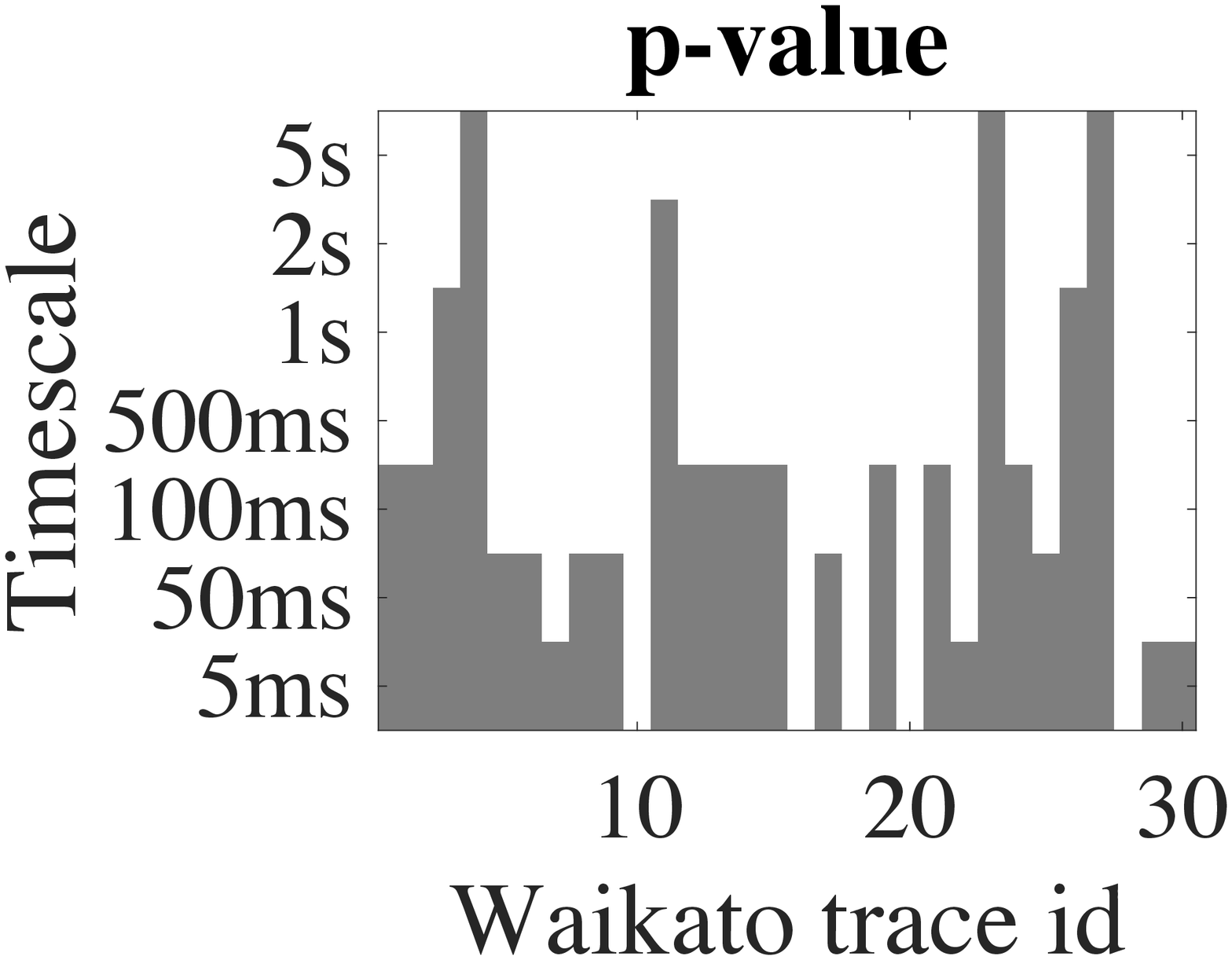}} 
		\hspace{-0.25cm}
		\subcaptionbox{KPSS test: Auckland}[.20\linewidth][c]{%
			\includegraphics[width=1\linewidth]{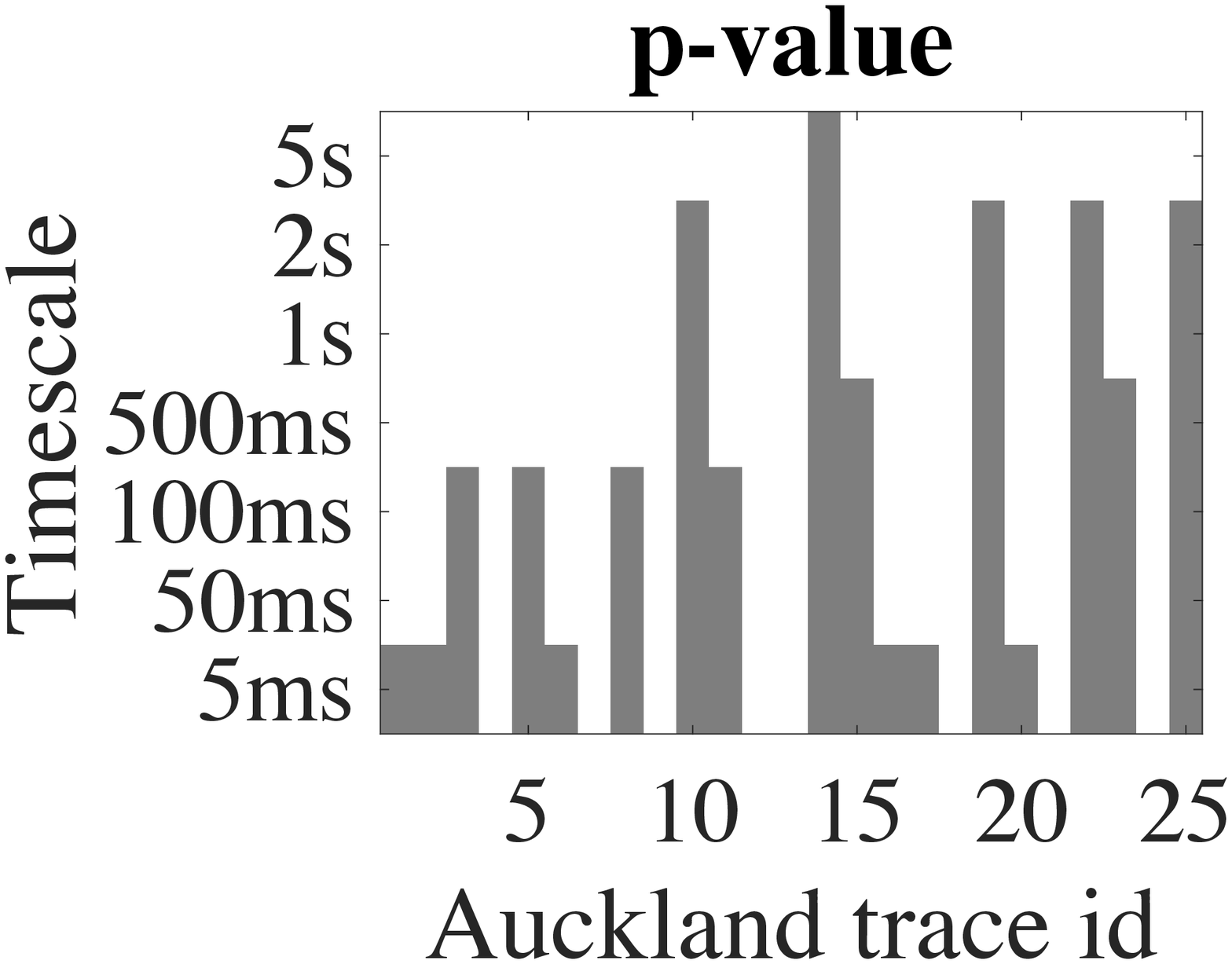}}
		\hspace{-0.25cm}
		\subcaptionbox{KPSS test: Twente}[.20\linewidth][c]{%
			\includegraphics[width=1\linewidth]{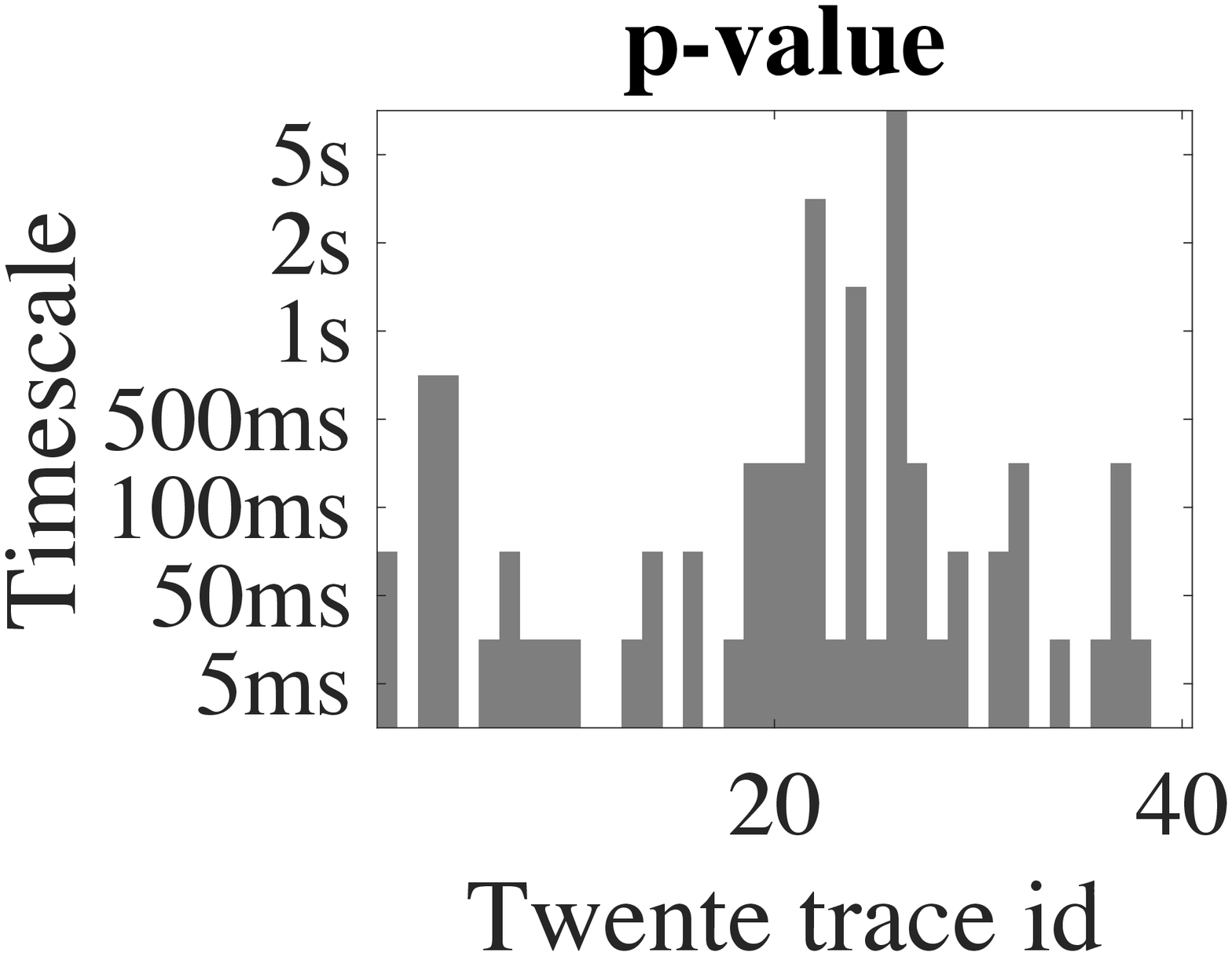}}
		\hspace{-0.25cm}
		\subcaptionbox{KPSS test: MAWI}[.20\linewidth][c]{%
			\includegraphics[width=1\linewidth]{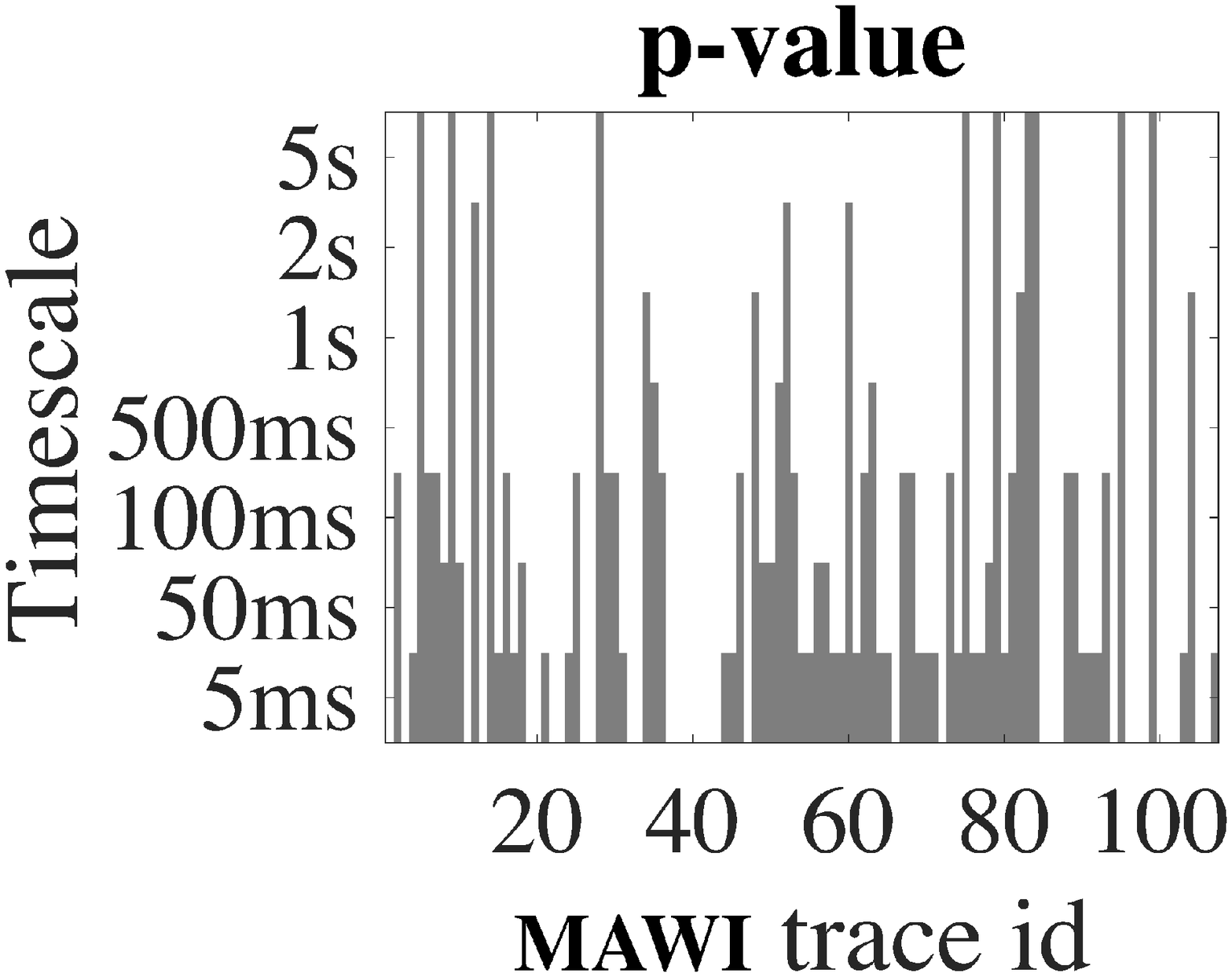}}
		\bigskip
		\\
		\subcaptionbox{KPSS: 1\textsuperscript{st} order diff.}[.20\linewidth][c]{%
			\includegraphics[width=1\linewidth]{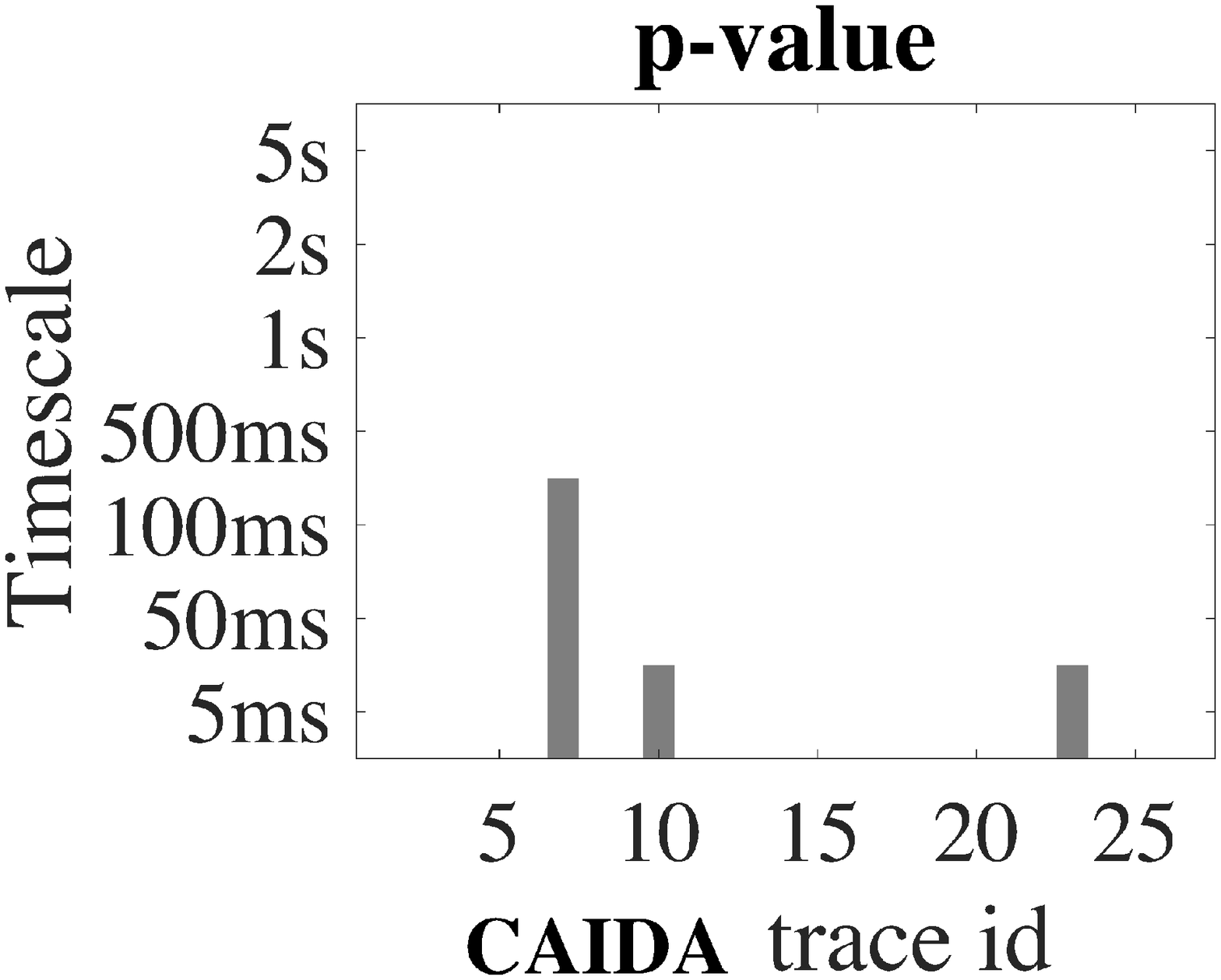}} 
		\hspace{-0.2cm}
		\subcaptionbox{KPSS: 1\textsuperscript{st} order diff.}[.20\linewidth][c]{%
			\includegraphics[width=1\linewidth]{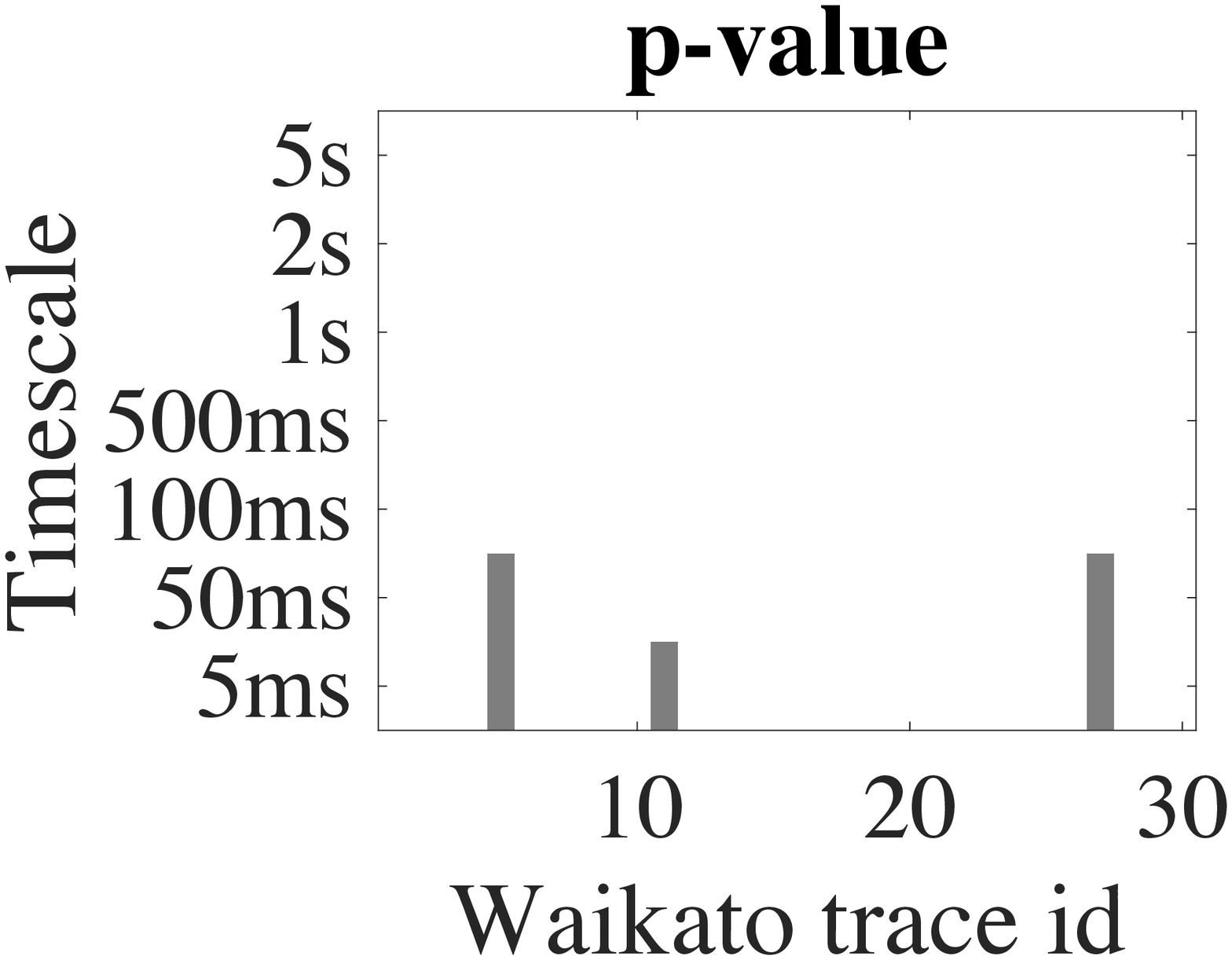}} 
		\hspace{-0.25cm}
		\subcaptionbox{KPSS: 1\textsuperscript{st} order diff.}[.20\linewidth][c]{%
			\includegraphics[width=1\linewidth]{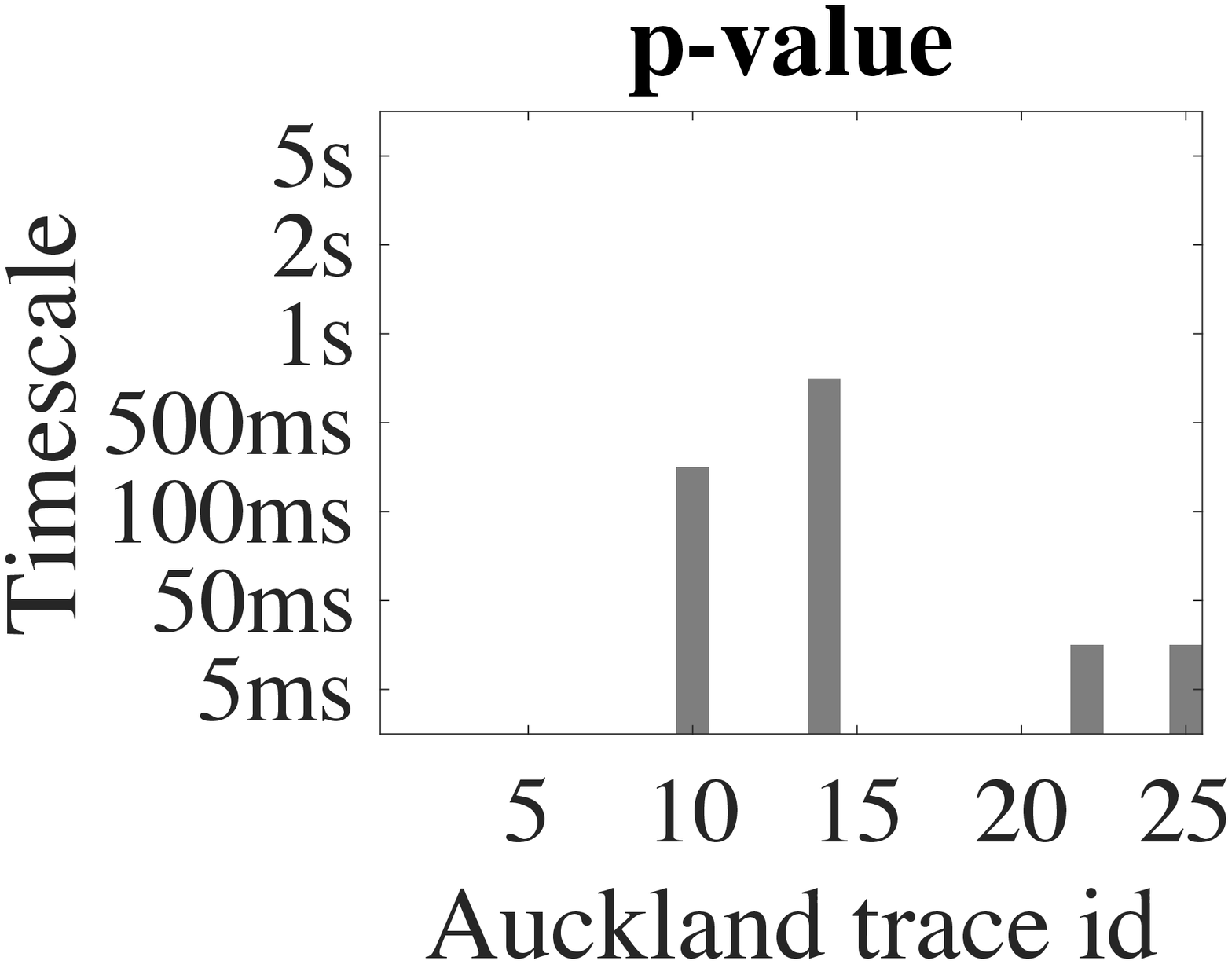}}
		\hspace{-0.25cm}
		\subcaptionbox{KPSS: 1\textsuperscript{st} order diff.}[.20\linewidth][c]{%
			\includegraphics[width=1\linewidth]{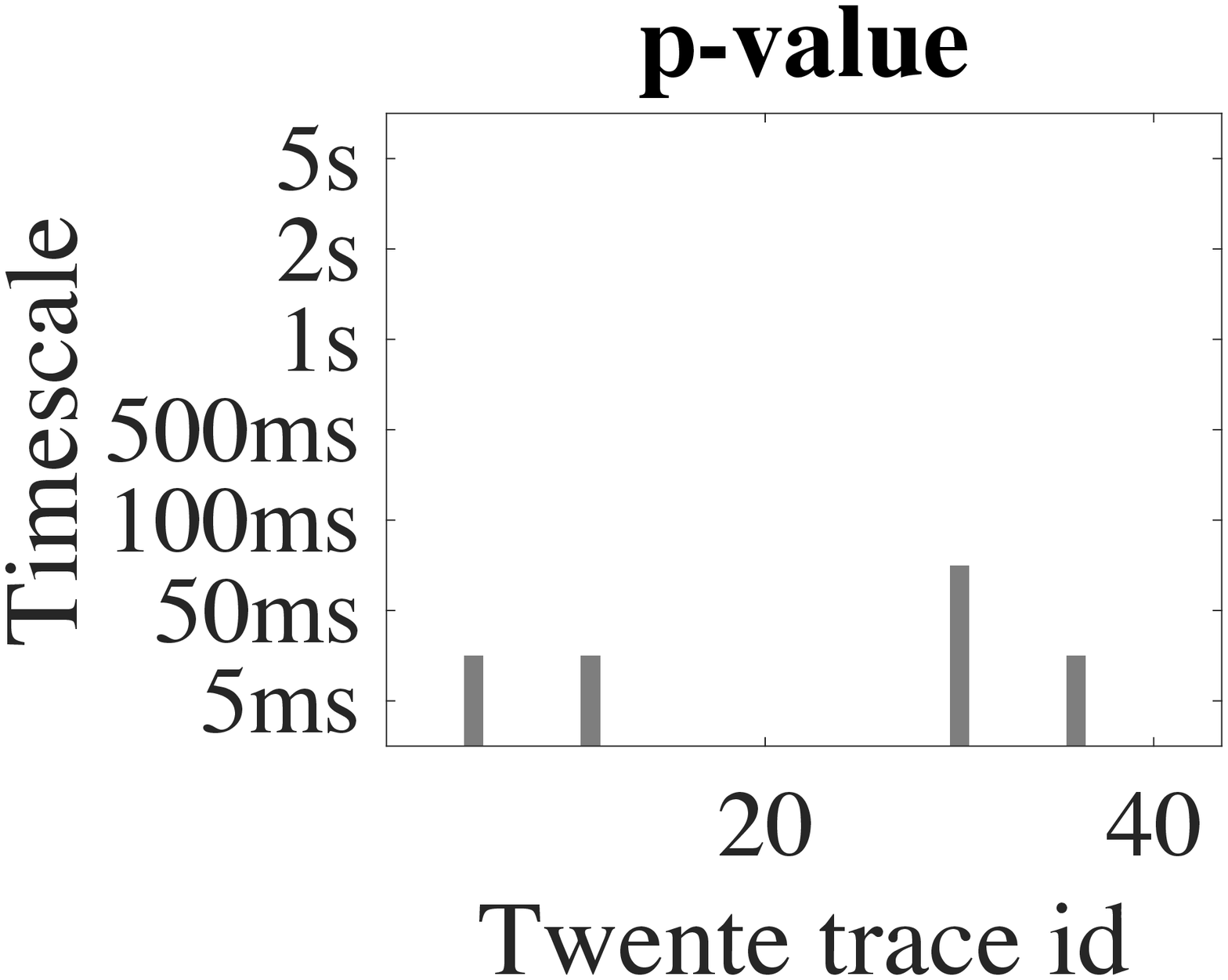}}
		\hspace{-0.2cm}
		\subcaptionbox{KPSS: 1\textsuperscript{st} order diff.}[.20\linewidth][c]{%
			\includegraphics[width=1\linewidth]{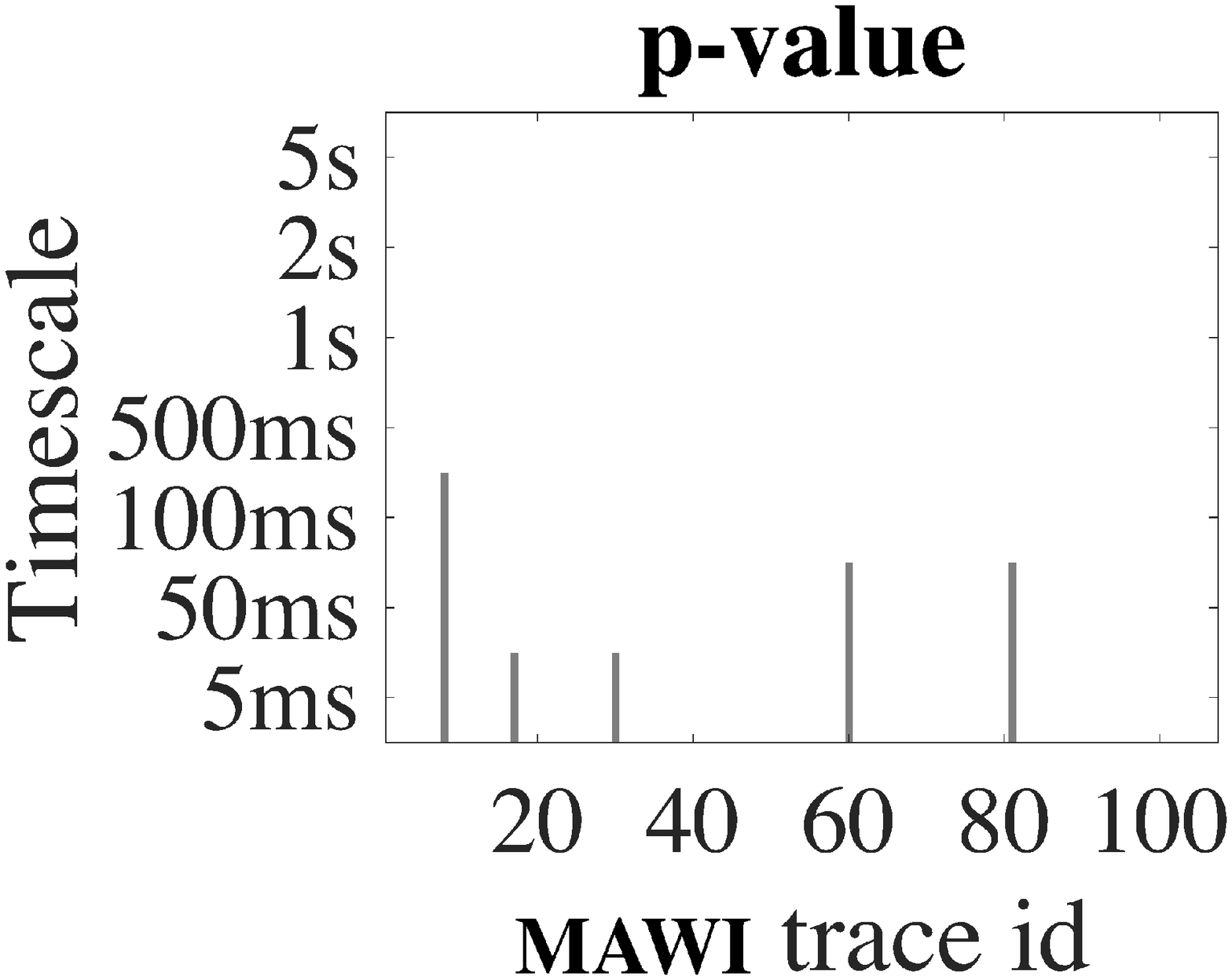}}
		\caption{Stationarity tests' results of the 15-minute long traces. Black: stationary, grey: non-stationary, white: inconclusive. In ADF \textcolor{black}{(a-e)} and PP \textcolor{black}{(f-j)} tests, the black areas represent $p$-value $\leq$ 0.05 (stationary results), while the white areas represent $p$-value $>$ 0.05 (inconclusive results). In KPSS test \textcolor{black}{(k-o)} and \textcolor{black}{KPSS first-order differencing results (p-t)}, the grey areas represent $p$-value $\leq$ 0.05  (non-stationary results), while the white areas represent $p$-value $>$ 0.05 (inconclusive result) (see Table~\ref{adf-pp-kpss-tests-table}). }
		\label{StationarityResults} 	
\end{figure*}
	
	The linear correlation coefficient test has been widely used to assess the fit of a distribution to empirical data. To reinforce the results of Section~\ref{fit-log-normal}, we assess the fit of the log-normal and Gaussian distributions to all studied traces. We use the linear correlation coefficient as defined in~\cite{TestingGaussianApproximation}:
	\begin{equation}
	\gamma = \frac{\sum_{i=1}^{n} \left ( S_{(i)}- \hat{\mu} \right )\left ( x_{i}-\hat{x} \right )}{\sqrt{\sum_{i=1}^{n}\left ( S_{(i)}-\hat{\mu} \right )^{2}.\sum_{i=1}^{n}\left ( x_{i}-\hat{x} \right )^{2}}} 
	\end{equation}
	where $S_{(i)}$ is the observed sample $i$,  $\hat{\mu}=\frac{1}{n}\sum_{i=1}^{n}S_{(i)}$ is the samples' mean value, and $x_{i}$ is sample $i$ from the reference distribution (log-normal in our case), which can be calculated from the inverse cumulative distribution function (CDF) of the reference random variable $x_{i} =F^{-1}\left ( \frac{i}{n+1} \right )$ and $\hat{x}=\frac{1}{n}\sum_{i=1}^{n}x_{i}$ is the respective mean value. The value of the correlation coefficient can vary between $-1\leq \gamma \leq 1$, with a $1$, $0$ and $-1$ indicating perfect correlation, no correlation and perfect anti-correlation, respectively. Strong goodness-of-fit (GOF) is assumed to exist when the value of $\gamma$ is greater than $0.95$~\cite{ResDimension}.
	
	We measure the linear correlation coefficient for all datasets at four different aggregation timescales (ranging from 5 msec to 5 sec) and plot the results in Figures~\ref{gammaRes}(a-e) for the log-normal distribution and Figures~\ref{gammaRes}(f-j) for the Gaussian distribution. Traces are ordered by the value of $\gamma$ for the given timescale. It can be clearly seen that $\gamma>0.95$ for most traces when employing the test for the log-normal distribution, but this is not the case for the Gaussian distribution. $\gamma$ is larger for smaller aggregation timescales indicating that the log-normal distribution is an even better fit as the aggregation gets finer. For very small values of $T$, i.e. lower than 1 msec, data samples exhibit binary behaviour, where either a packet is transmitted or not during each examined time frame~\cite{transaction2015}. We have examined $\gamma$ for very short (and large) aggregation timescales, and can confirm the absence of a model describing the data (for brevity, we have omitted the relevant figures).
	
	Next, we calculate $\upsilon_{\gamma}$ (the variation of $\gamma$) for each dataset. $\upsilon_{\gamma}$ gives an indication of the stability of $\gamma$ for each dataset, for all timescales tested. This metric is defined as:
	\begin{equation}
	\upsilon_{\gamma} = \sqrt{  var(\gamma _{T_{1}},\gamma _{T_{2}}, \gamma _{T_{3}},\gamma _{T_{4}} )}  
	\end{equation}
	where $ {T_{1}}= 5$ sec, $ {T_{2}}= 1$ sec, $ {T_{3}}= 100$ msec and $ {T_{4}}= 5$ msec. Figures~\ref{gammaRes}(k-o) show the results for each dataset with the traces ranked by $\upsilon_{\gamma}$. For log-normal model, $\upsilon_{\gamma}$ is very small (below $0.045$) for all traces, therefore we can conclude that $\gamma$ is almost constant for all studied aggregation timescales. While $\upsilon_{\gamma}$ is higher for the Gaussian model. Furthermore, the error bars in Figures~\ref{gammaRes}(p-t) represent the standard deviation of the correlation coefficient at different timescales (see x-axis). This again shows that for the log-normal model $\gamma$ is larger than $0.95$ (at different T values) for most CAIDA and MAWI traces, while it is larger than $0.9$ for all other datasets. This is not the case with the Gaussian model, where most ${\gamma}$ values are less than $0.9$. 
	
	Overall, the correlation coefficient test reinforces the results extracted in Section~\ref{fit-log-normal}, providing strong evidence that the log-normal distribution is the best fit for all studied traces. The superior
	performance of our model can also be seen from comparison of our
	results for correlation coefficient with those in \cite{ieee-network-2009} where the Gaussian model was used.


\section{Stationarity Testing}
\label{sec:stationarity}
\textcolor{black}{In the previous section we showed that the log-normal distribution is the best fit among all studied distributions for the vast majority of traces, for all studied aggregation timescales. In this section, we investigate whether these results are applicable on any traffic sample and we need to find what is the aggregation timescale and time period levels (i.e. length of captured trace) at which these results remain valid. This can be done by applying time series analysis on the traffic by using stationarity tests. From a probabilistic point of view, stationarity means that the distribution remains unchanged when shifted in time. Stationarity plays an important role in time series analysis \cite{Lauks2011, stationaryModelingLS, NonstationarityIP}. It is important to mention that the study of Internet traffic as a time series depends on two factors. Firstly, a time period over which to study the traffic. Obviously, over long time periods (hours and days) the data is not stationary as it is subject to daily and weekly variations related to human activity. Secondly we use a timescale that is used to aggregate the traffic over a specific time period. If the aggregation timescale is very small then the traffic volume will really be a product of exactly how many packets are classified as arriving within that period leading to very noisy measurements. If the timescale is longer, our measured time period will contain very few samples and the statistics calculated will lack power to reject hypotheses, producing instead inconclusive results simply because they have insufficient data. Hence we want to establish whether we can reliably say that a sample of 15-minute or 1-hour long of the captured data is typically stationary in the data set and at which aggregation timescale. Our stationarity test results show that over a 15-minute and 1-hour periods the data is stationary when aggregated at timescales of $0.5$ sec to $5$ sec.}

\begin{table}[]
	\setlength{\belowcaptionskip}{-6pt}
	\scriptsize
	\caption{ADF, PP and KPSS tests}
	\centering
	\begin{tabular}{l|l|l|}
		\cline{2-3}
		& \multicolumn{1}{c|}{\cellcolor[HTML]{C0C0C0}ADF and PP tests}                   & \multicolumn{1}{c|}{\cellcolor[HTML]{C0C0C0}KPSS test}                       \\ \hline
		\multicolumn{1}{|l|}{\cellcolor[HTML]{C0C0C0}\begin{tabular}[c]{@{}l@{}}null hypothesis (H0)\end{tabular}}      & \begin{tabular}[c]{@{}l@{}}unit root is present\end{tabular}         & \begin{tabular}[c]{@{}l@{}}series is stationary\end{tabular}          \\ \hline
		\multicolumn{1}{|l|}{\cellcolor[HTML]{C0C0C0}\begin{tabular}[c]{@{}l@{}}alternative hypothesis (H1)\end{tabular}} & \begin{tabular}[c]{@{}l@{}}series is stationary\end{tabular}             & \begin{tabular}[c]{@{}l@{}}unit root is present\end{tabular}      \\ \hline
		\multicolumn{1}{|l|}{\cellcolor[HTML]{C0C0C0}$p$-value $>$ 0.05}                                              & \begin{tabular}[c]{@{}l@{}}H0 is not rejected: \\result is inconclusive \end{tabular} & \begin{tabular}[c]{@{}l@{}}H0 is not rejected: \\ result is inconclusive\end{tabular} \\ \hline
		\multicolumn{1}{|l|}{\cellcolor[HTML]{C0C0C0}$p$-value $\leq$	 0.05}                                                 & \begin{tabular}[c]{@{}l@{}} H0 is rejected: \\series is stationary\end{tabular}            & \begin{tabular}[c]{@{}l@{}}H0 is rejected: \\series is non-stationary \\(due to a unit root)\end{tabular}     \\ \hline
	\end{tabular}
	\label{adf-pp-kpss-tests-table} 	
\end{table}

We examine traffic stationarity using the traces discussed in Section \ref{sec:dataset} at different timescales using three tests commonly used for stationarity testing, namely the Augmented Dickey-Fuller (ADF)~\cite{adf-test}, Phillips-Perron (PP)~\cite{pp-test} and Kwiatkowski-Phillips-Schmidt-Shin (KPSS)~\cite{kpss-test} tests. In the ADF and PP tests, the \emph{null} hypothesis is that a \emph{unit root} is present and the alternate hypothesis is \emph{stationarity}. In the KPSS test, the \emph{null} hypothesis is that the time series is \emph{stationary} and the alternate hypothesis is that a \emph{unit root} is present. Table~\ref{adf-pp-kpss-tests-table} summarises the hypotheses of the three tests and outlines the outcome of each test according to the $p$-value.

\subsection{Stationarity tests of 15-minute long traces}

We begin by conducting the stationarity tests on the $232$ 15-minute long traces (described in Section~\ref{sec:dataset}). \textcolor{black}{The numbers of data points used in the stationarity tests for each dataset are $180000$, $18000$, $9000$, $1800$, $900$, $450$ and $180$ for aggregation timescales of $5$ msec, $50$ msec, $100$ msec, $500$ msec, $1$ sec,  $2$ sec and $5$ sec, respectively.} Figure~\ref{StationarityResults} shows the results for the ADF, PP and KPSS tests at all studied aggregation timescales.

According to the ADF and PP tests (Figures~\ref{StationarityResults}(a-j)), the majority of time series are stationary for all aggregation timescales; the $p$-value is less than $0.05$, therefore the \emph{null} hypothesis is rejected and there is enough evidence to support the alternative hypothesis. These traces are shown as black areas in the figures. There are a few traces for which the $p$-value is greater than $0.05$ at some aggregation timescales. These are illustrated as white areas in the figures. For these traces the null hypothesis cannot be rejected. These are the \textit{anomalous traces} discussed in Section~\ref{anomalous}. Below, we employ the KPSS test to provide evidence that these series are non-stationary; i.e. to show that for the studied traffic traces, where the log-normal was not a good fit for a specific trace, the underlying time series was not stationary. As shown in Figures~\ref{anomalousTraces}a\&c, said traces appear with a bi-modal distribution.
\begin{figure}
	\setlength{\belowcaptionskip}{-5pt}
	\centering
	\subcaptionbox{Trace with trends}[.48\linewidth][c]{%
		\includegraphics[width=1\linewidth]{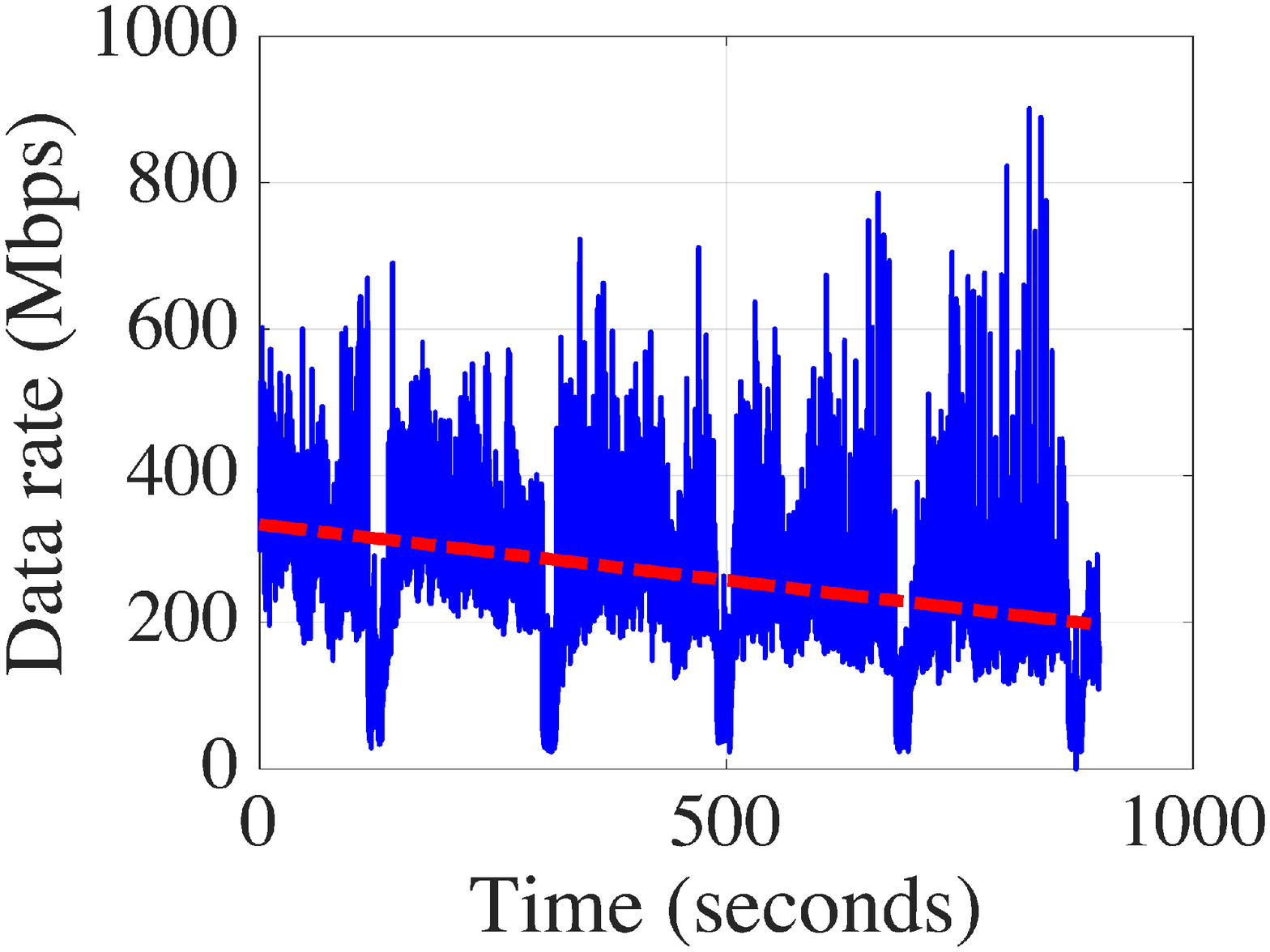}}\quad
	\subcaptionbox{Differenced trace}[.48\linewidth][c]{%
		\includegraphics[width=1\linewidth]{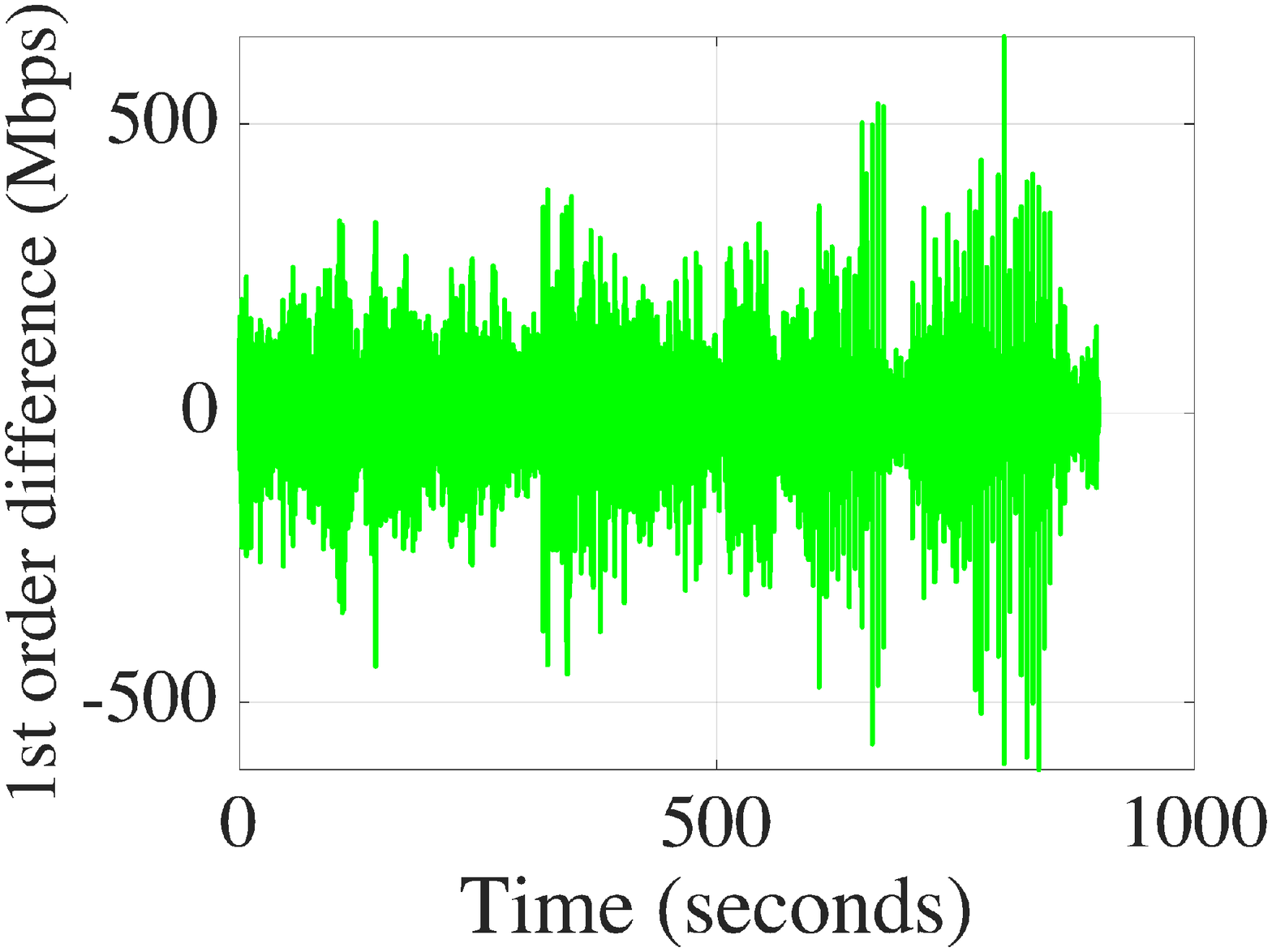}}\quad
	\caption{First-order differencing of a MAWI trace with trends.}
	\label{kpss-trend} 
\end{figure}
\begin{figure*}[t]
	\setlength{\belowcaptionskip}{-4pt}
	\centering
	\subcaptionbox{24-hour long MAWI trace: data rate plot over 24 hours at different timescales}[.7\linewidth][]{%
		\includegraphics[width=1.05\linewidth]{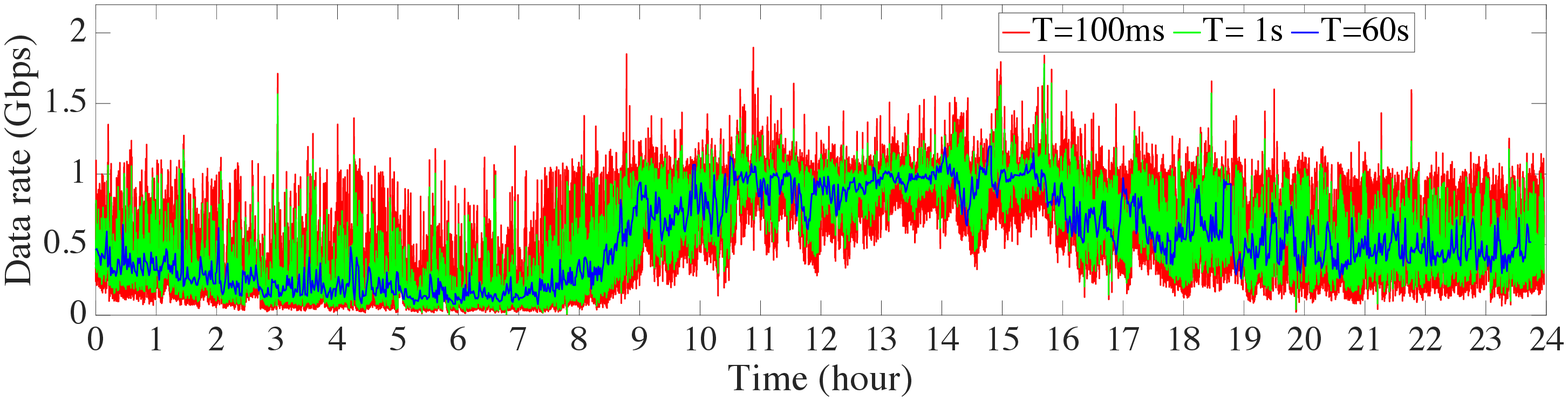}}
	\hspace{0.6cm}  
	\subcaptionbox{24-hour MAWI trace: PDF}[.25\linewidth][]{%
		\includegraphics[scale=0.227]{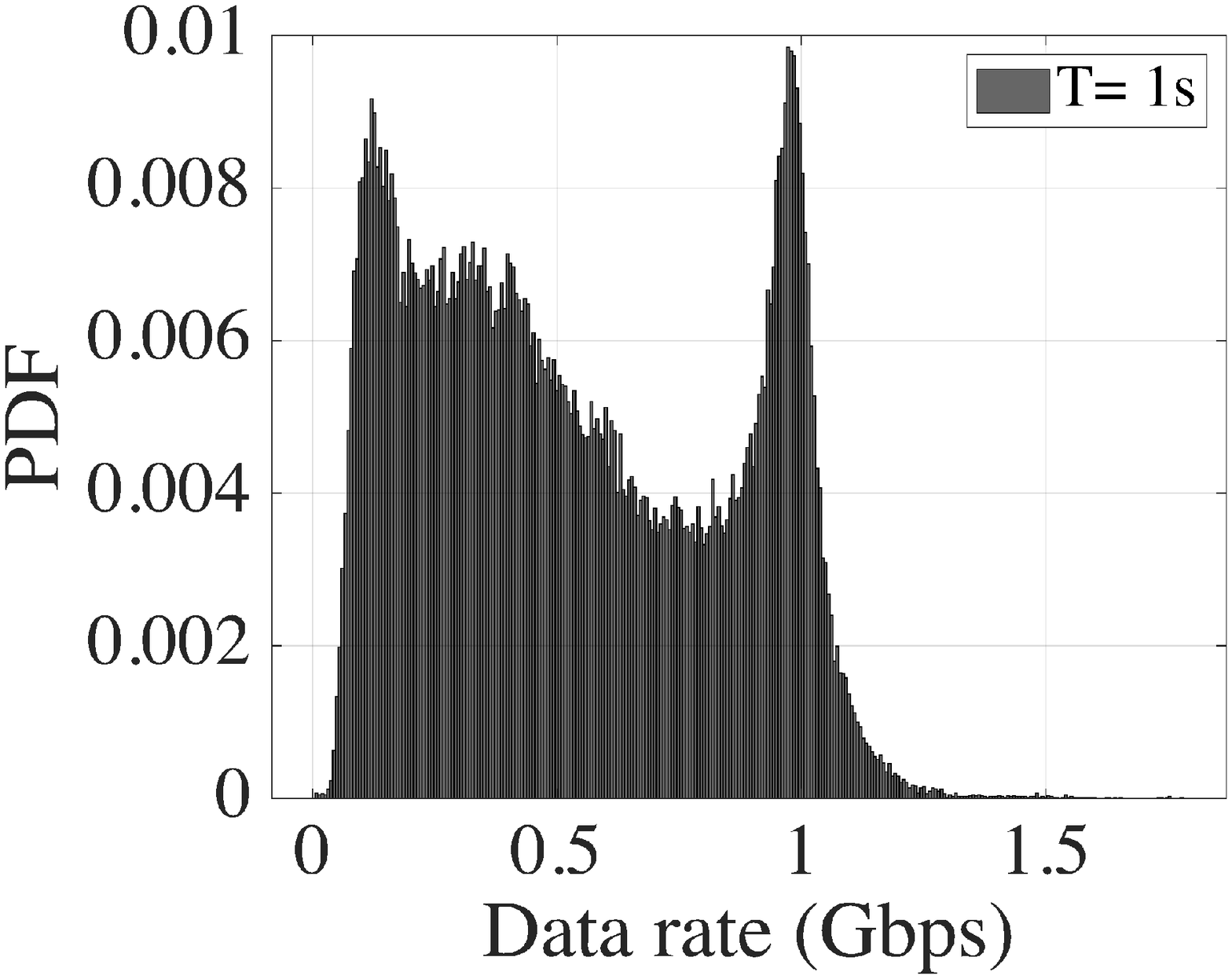}}\quad
	\caption{24-hour long MAWI trace.}
	\label{mawi-trace-24-series} 
\end{figure*} 

\begin{figure*}[h!]
	\setlength{\belowcaptionskip}{-5pt}
	\centering
	\subcaptionbox{ADF test}{%
		\includegraphics[scale=0.172]{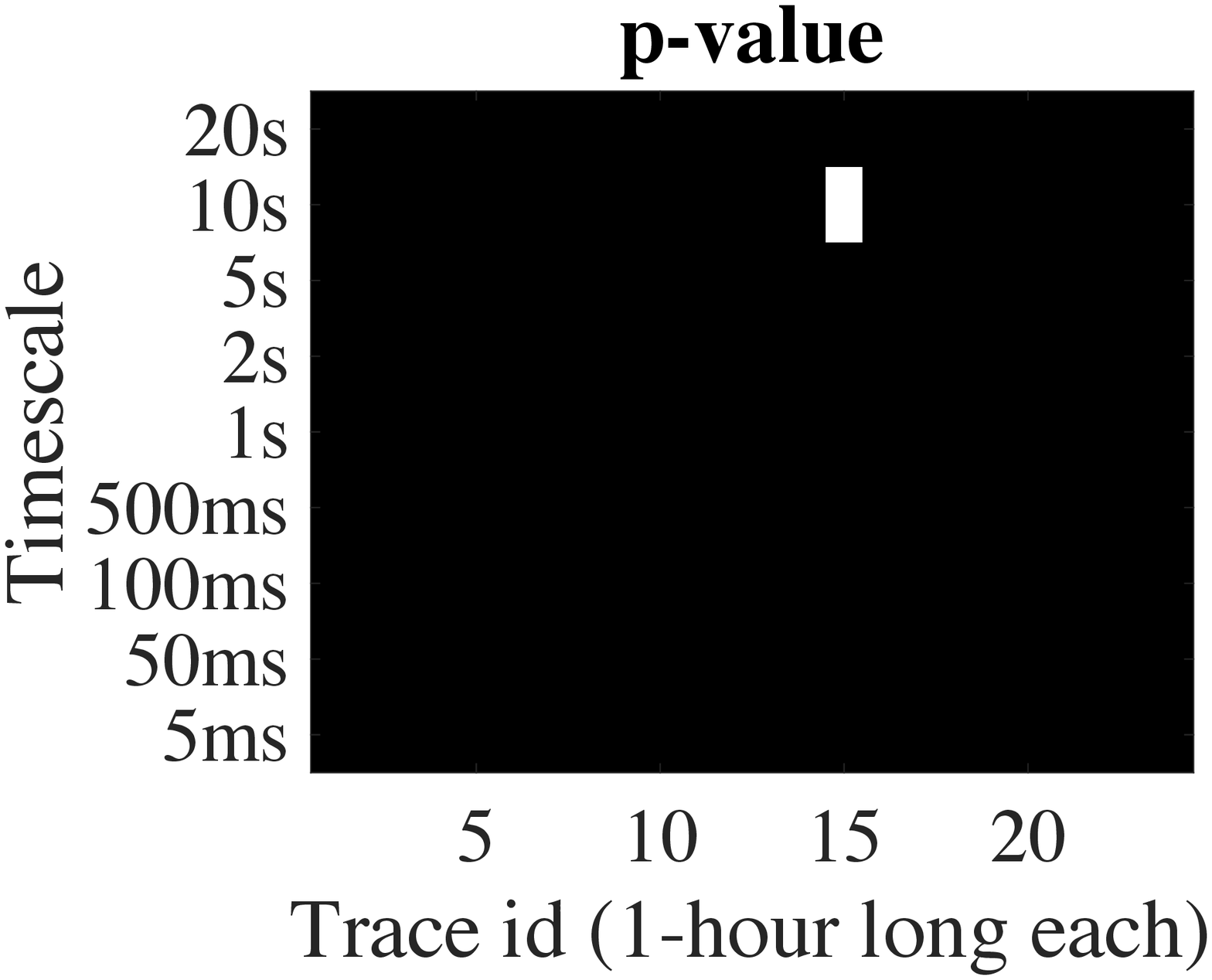}} \quad
	\subcaptionbox{PP test}{%
		\includegraphics[scale=0.172]{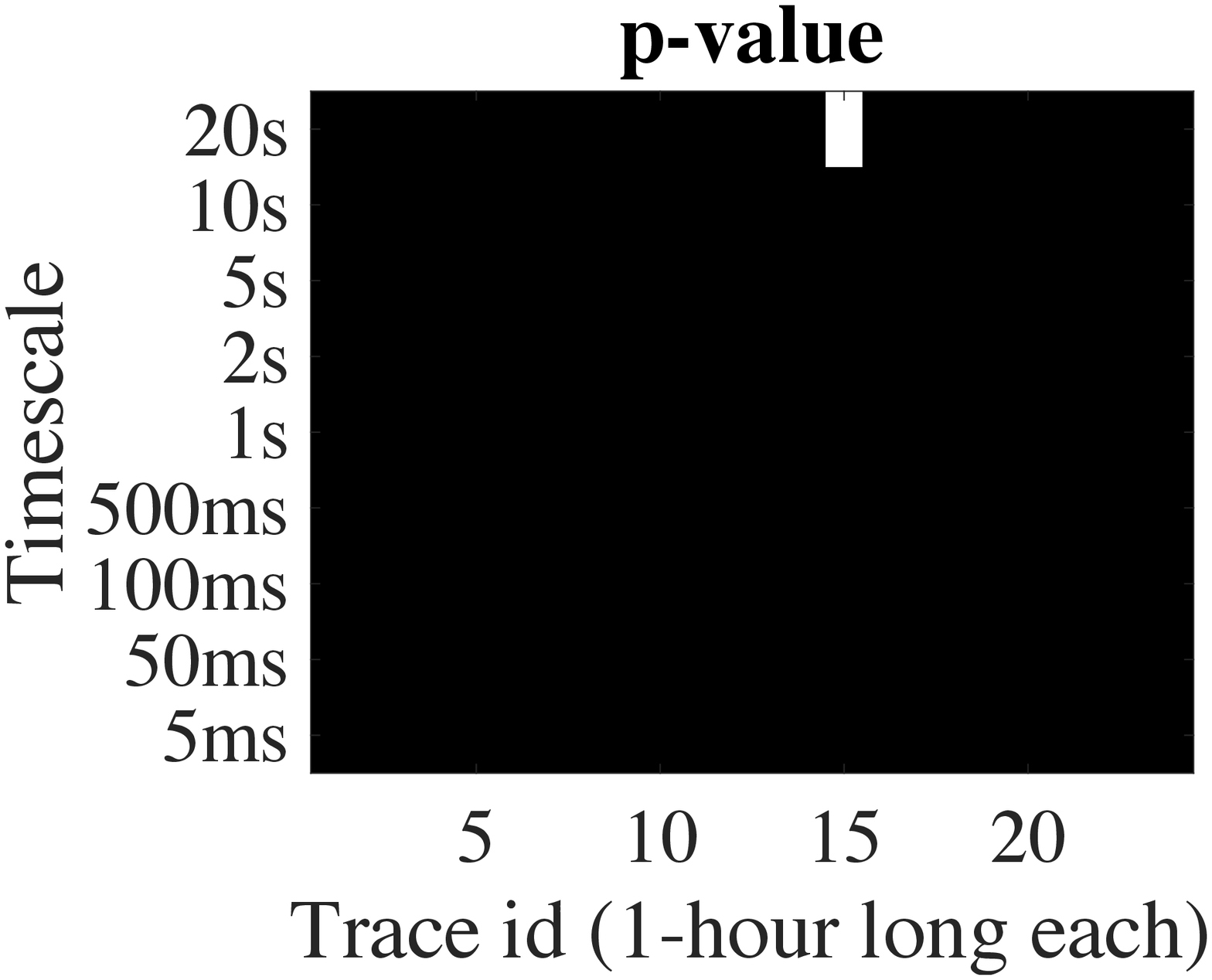}} \quad
	\subcaptionbox{KPSS test}{%
		\includegraphics[scale=0.172]{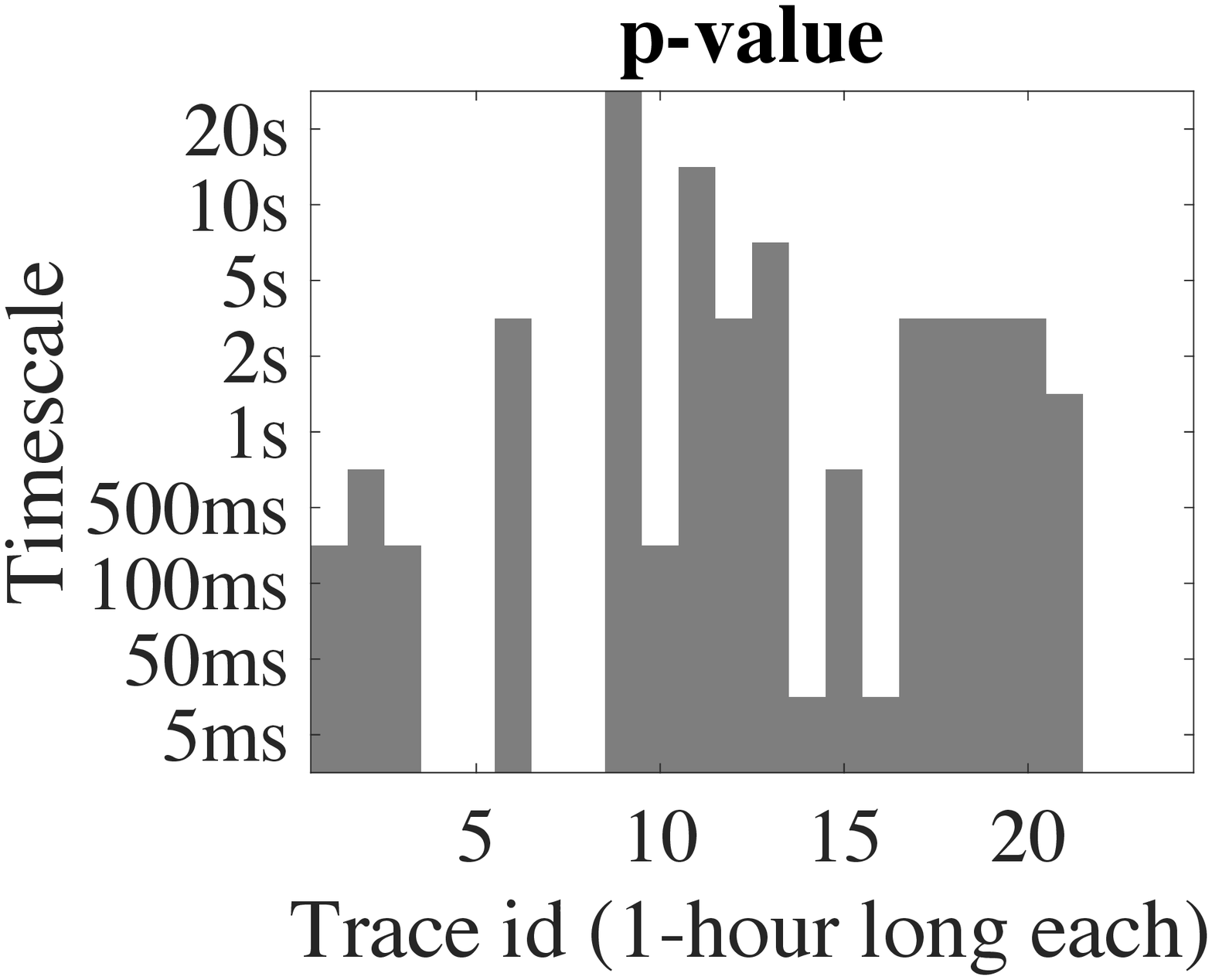}}\quad
	\subcaptionbox{KPSS: 1\textsuperscript{st} order difference}{%
		\includegraphics[scale=0.172]{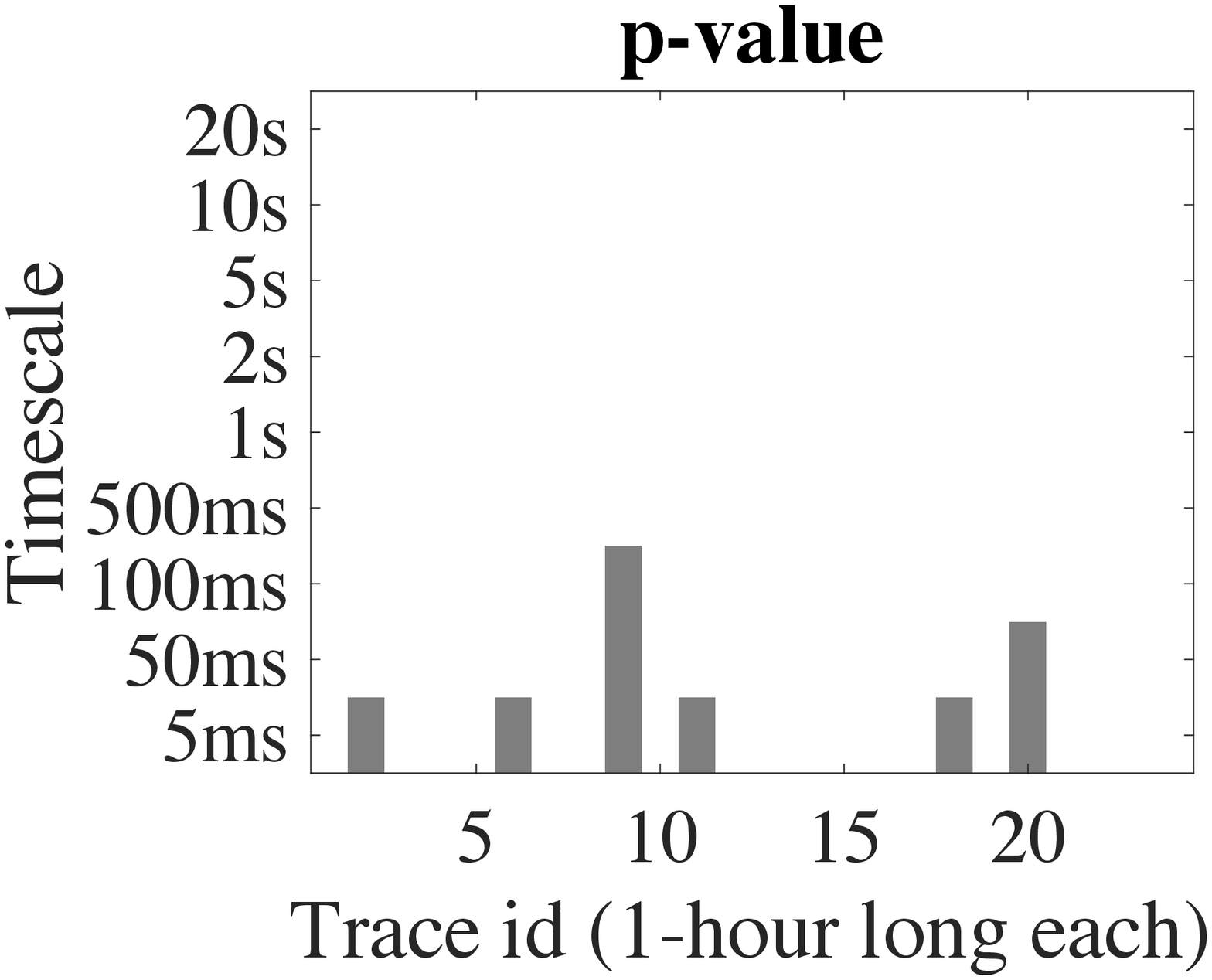}}\quad
	\subcaptionbox{24-hour MAWI trace stationarity results}{%
		\includegraphics[scale=0.15]{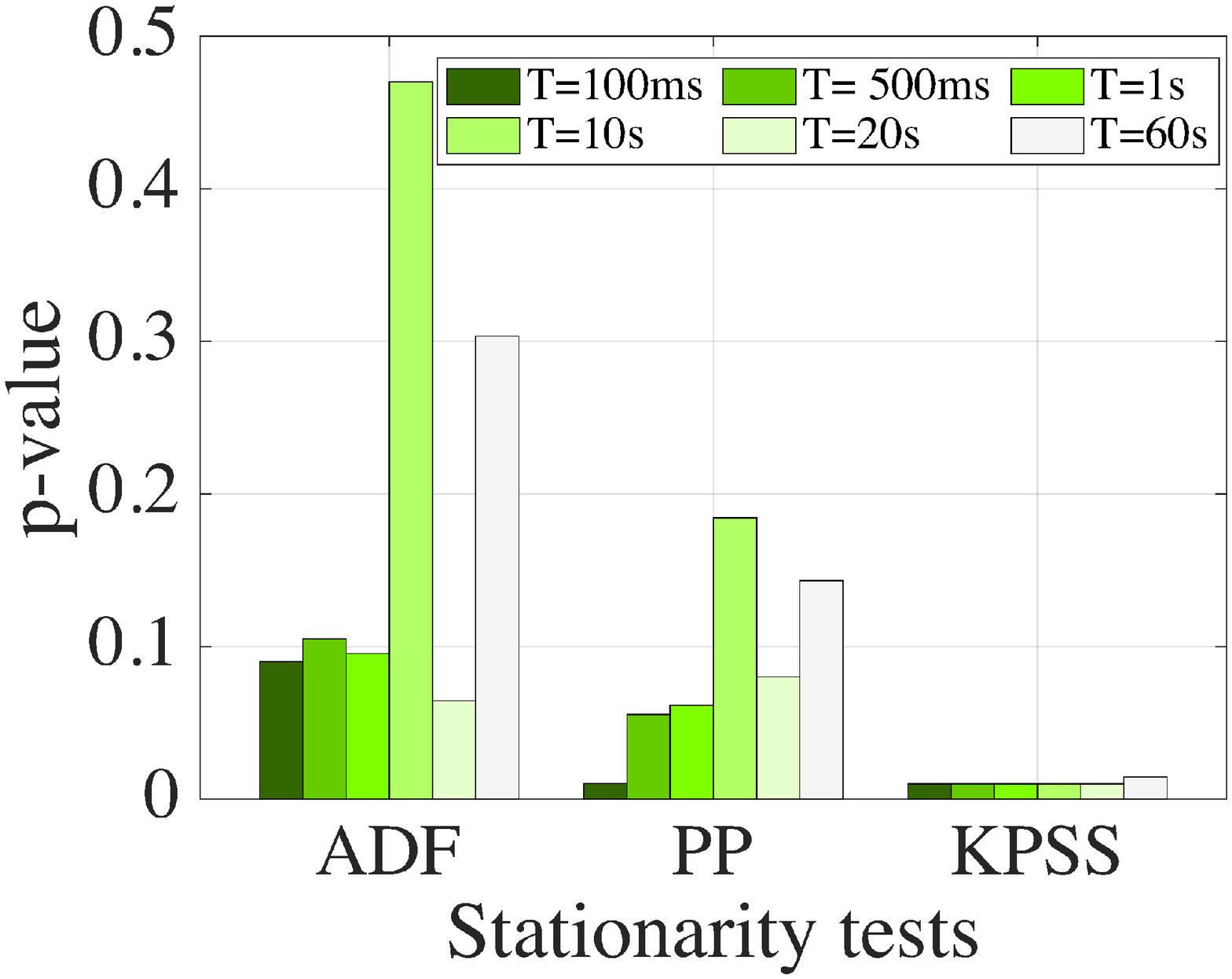}}
	\caption{(a-d) Stationarity tests' results of 24 1-hour long subtraces from the 24-hour long MAWI trace. Black: stationary, grey: non-stationary, white: inconclusive. In ADF and PP tests, the black areas represent $p$-value $\leq$ 0.05 (stationary results), while the white areas represent $p$-value $>$ 0.05 (inconclusive results). In KPSS test, the grey areas represent $p$-value $\leq$ 0.05  (non-stationary results), while the white areas represent $p$-value $>$ 0.05 (inconclusive result) (see Table~\ref{adf-pp-kpss-tests-table}).}
	\label{stationarity-1-hour-groups} 	 	
\end{figure*}
\textcolor{black}{In KPSS test results (Figures~\ref{StationarityResults}(k-o)), we fail to reject the null hypothesis for most traces, as the $p$-value is larger than $0.05$. These traces with inconclusive results are shown as white  areas in the figures. For some traces (commonly at small aggregation timescales), the null hypothesis is rejected, and therefore there is evidence that the series is non-stationary. These are shown as grey areas in the figures. }\\
\textcolor{black}{\noindent\textbf{De-trending by differencing in KPSS.} The KPSS test is known to be vulnerable to mistaking trends for non-stationarity i.e. the KPSS test is sensitive to trends~\cite{statsmodels}. This problem appears for small aggregation timescales because fluctuations appear, and these are mistaken as trends within the time series. If KPSS indicates non-stationarity and ADF indicates stationarity, then the series is difference stationary~\cite{statsmodels} (that is the case with the grey areas results in Figures~\ref{StationarityResults}(k-o)). In order to further explore this observation, we re-ran the analysis by first de-trending the series. De-trending is carried out by using differencing which can be used to remove the series' dependence on time, including structures like trends and seasonality~\cite{firstOrderDiff}. The differenced series is checked for stationarity as not all non-stationary time series are difference stationary. Figure~\ref{kpss-trend}(b) shows the first-order difference of the trace that is shown in Figure~\ref{kpss-trend}(a). This gives a $p$-value of $0.1$, i.e. we fail to reject the stationarity null hypothesis. Figures~\ref{StationarityResults}(p-t) show that we always fail to reject the stationarity null hypothesis for most traces (at the tested timescales) in our datasets when applying the KPSS test on the first-order difference sequences (i.e. the results are inconclusive). The results shown in Figure~\ref{StationarityResults}(k-t) are consistent with the conclusion that the data is stationary at larger aggregation timescales: $0.5 - 5$ sec and first-difference stationary at smaller aggregation timescales: $5 - 100$ msec. }

\subsection{Stationarity tests of an hour long samples within a 24-hour trace}

In this section we consider the hour-long samples within a 24-hour MAWI trace (described in Section~\ref{sec:dataset}). This 24-hour long trace is used to see if the assumption of stationarity holds for periods longer than 15 minutes. We conclude that these are also stationary.\\ Figures~\ref{mawi-trace-24-series}(a-b) show the data rate plots as PDF and as a time series, respectively. It is obvious that the 24-hour long series is not stationary; for example, the average data rate in this series between 12:00 am to 05:00 am is $0.252$ Gbps, while the average in the time period between 09:00 am to 17:00 pm is $0.875$ Gbps.

We run the stationarity tests for this traffic trace and for different aggregation timescales at different sampling times. We start by applying ADF, PP and KPSS tests on 1-hour long groups (subtraces) in this trace (the time at each group or subtrace starts at the beginning of each hour). The stationarity tests results are shown in Figures~\ref{stationarity-1-hour-groups}(a-d). It is clear from the first two tests' results that the majority of the 1-hour long groups are stationary (black areas in the figure) as their null hypothesis is rejected. The KPSS test shows many subtraces (white areas) where we failed to reject the null hypothesis i.e. inconclusive results. We ran the KPSS test on the first-order difference series (white areas in Figure~\ref{stationarity-1-hour-groups}d) and the number of contradictory results was greatly reduced as the KPSS test failed to reject the null-hypothesis of trend stationarity. 

It is worth mentioning that these results might slightly change for some groups if we use 1-hour long groups that do not start at the beginning of each hour (e.g. when using a 1-hour long group that starts at 08:30 am, as a jump in the captured data rate will appear at the second half of this group causing it to be non-stationary). Figure~\ref{stationarity-1-hour-groups}e shows the stationarity tests results of the 24-hour long MAWI trace. As expected, and based on the stationarity tests results, this 24-hour long trace is non-stationary.

\textcolor{black}{In the next two sections (\ref{sec:provision}\&\ref{sec:pricing}) we present the impact of traffic distribution on two sample traffic engineering problems: link dimensioning and traffic billing. We do not intend our examples to be fully worked systems for practical deployment. We wish to demonstrate using motivational examples that the improved predictions made possible by these models could in the future have practical utility. }

\section{Bandwidth Provisioning}
\label{sec:provision}

It has been previously suggested that network link provisioning could be based on fitted traffic models instead of relying on straightforward empirical rules~\cite{ieee-network-2009}. In this way, over- or under-provisioning can be mitigated or eliminated even in the presence of strong traffic fluctuations. Such approaches rely on having a statistical model that accurately describes the network traffic. This is therefore an excellent area for applying our findings on fitting the log-normal distribution to Internet traffic data. In the literature, the following inequality (the authors call it the ``link transparency formula") has been used for bandwidth provisioning~\cite{transaction2015}:
\begin{equation}
P\left ( A(T)\geq CT \right ) \leq \varepsilon.    \label{link-tran}  
\end{equation}

In words, this inequality states that the probability that the captured traffic $A(T)$ over a specific aggregation timescale  $T$ is larger than the link capacity has to be smaller than the value of a performance criterion $\varepsilon$. The value of $\varepsilon$ is chosen carefully by the network provider in order to meet a specific SLA~\cite{ieee-network-2009}. Likewise, the value of the aggregation time $T$ should be sufficiently small so that the fluctuations in the traffic can be modelled as well, taking into account the buffering capabilities of network switching devices\footnote{Large traffic fluctuations at very short aggregation timescales are smoothed by the presence of buffers at network routers and switches.}.

We compare bandwidth provisioning  using Meent's approximation formula~\cite{ieee-network-2009} (assuming Gaussian) and using a log-normal traffic model.

\begin{figure}[t]
	\setlength{\belowcaptionskip}{-7pt}
	\centering
	\includegraphics[width=1\linewidth, scale=0.23]{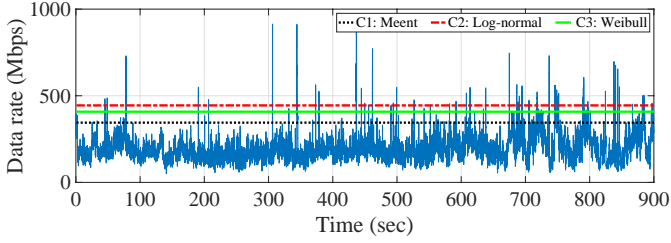}\quad
	\caption{Data rate of a MAWI trace ($T=100$ msec and $\varepsilon=0.01$). The horizontal lines represent the calculated link capacity based on different models.}
	\label{time-domain} 	
\end{figure}

\subsection{Bandwidth provisioning using Meent's formula} 
To find the minimum required link capacity, Meent et al.~\cite{ieee-network-2009} proposed a bandwidth provisioning approach that is based on the assumption that the traffic follows a Gaussian distribution. Meent's dimensioning formula is defined as follows~\cite{ieee-network-2009}:
\begin{equation}
\  C1=\mu +\frac{1}{T} \sqrt{-2log(\varepsilon) .\upsilon (T)}   \label{meents-euq}
\end{equation}
where $\mu$ is the average value of the traffic, $\upsilon (T)$ is the variance at timescale $T$ and $\varepsilon$ is the performance criterion. The link capacity is obtained by adding a safety margin value 
\[
\textrm{Safety margin = }\sqrt{-2log(\varepsilon)}  \textrm{ . } \sqrt{\frac{\upsilon (T)}{T^{2}} }
\]
to the average of the captured traffic (see Equation~\ref{meents-euq}). This safety margin value depends on $\varepsilon$ and the ratio $\sqrt{\upsilon (T)/T^{2}}$. As the value of  $\varepsilon$ decreases the safety margin increases. For example, when the value of  $\varepsilon$ decreases from $10^{-2}$ to $10^{-4}$, then  value of the safety margin increases by $40\%$. This is different from conventional link dimensioning methods, where the safety margin is fixed to be 30\% above the average of the presented traffic~\cite{cisco,ieee-network-2009,NetworkLinkDimensioning}. Traffic tails are represented using the Chernoff bound, as follows:
\begin{equation}
P  \left  ( A(T)\geq CT  \right )\leq  e^{-SCT}E\left [e^{SA(T)}  \right].
\end{equation}
Here \noindent${E\left [e^{SA(T)} \right ]}$ is the moment generation function (MGF) of the captured traffic $A(T)$.
\begin{figure*}[t]
	\centering
	\subcaptionbox{target $\varepsilon = 0.5$}[.27\linewidth][c]{%
		\includegraphics[scale=0.2]{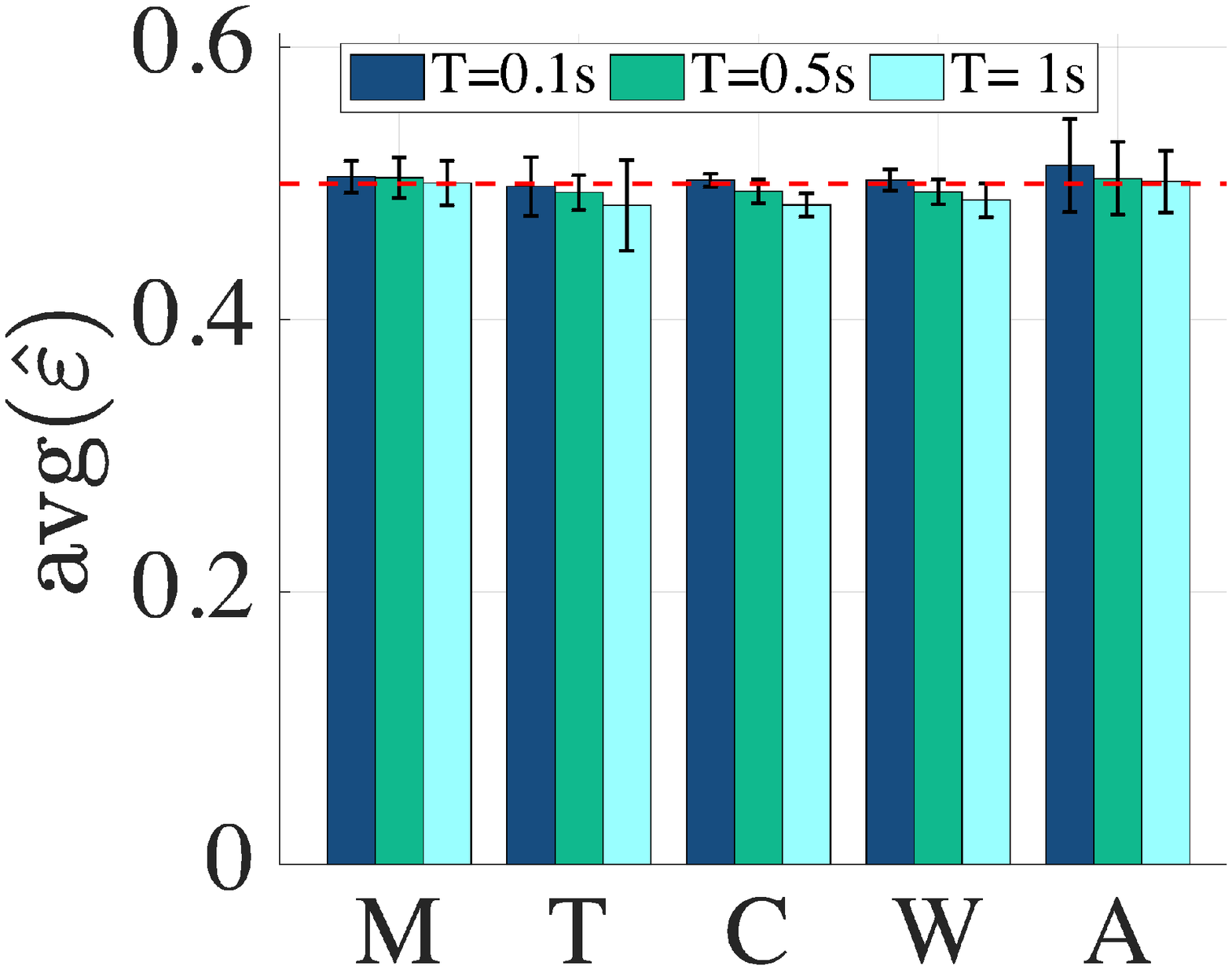}}
	\hspace{-0.8cm}  
	\subcaptionbox{target $\varepsilon = 0.1$}[.27\linewidth][c]{%
		\includegraphics[scale=0.2]{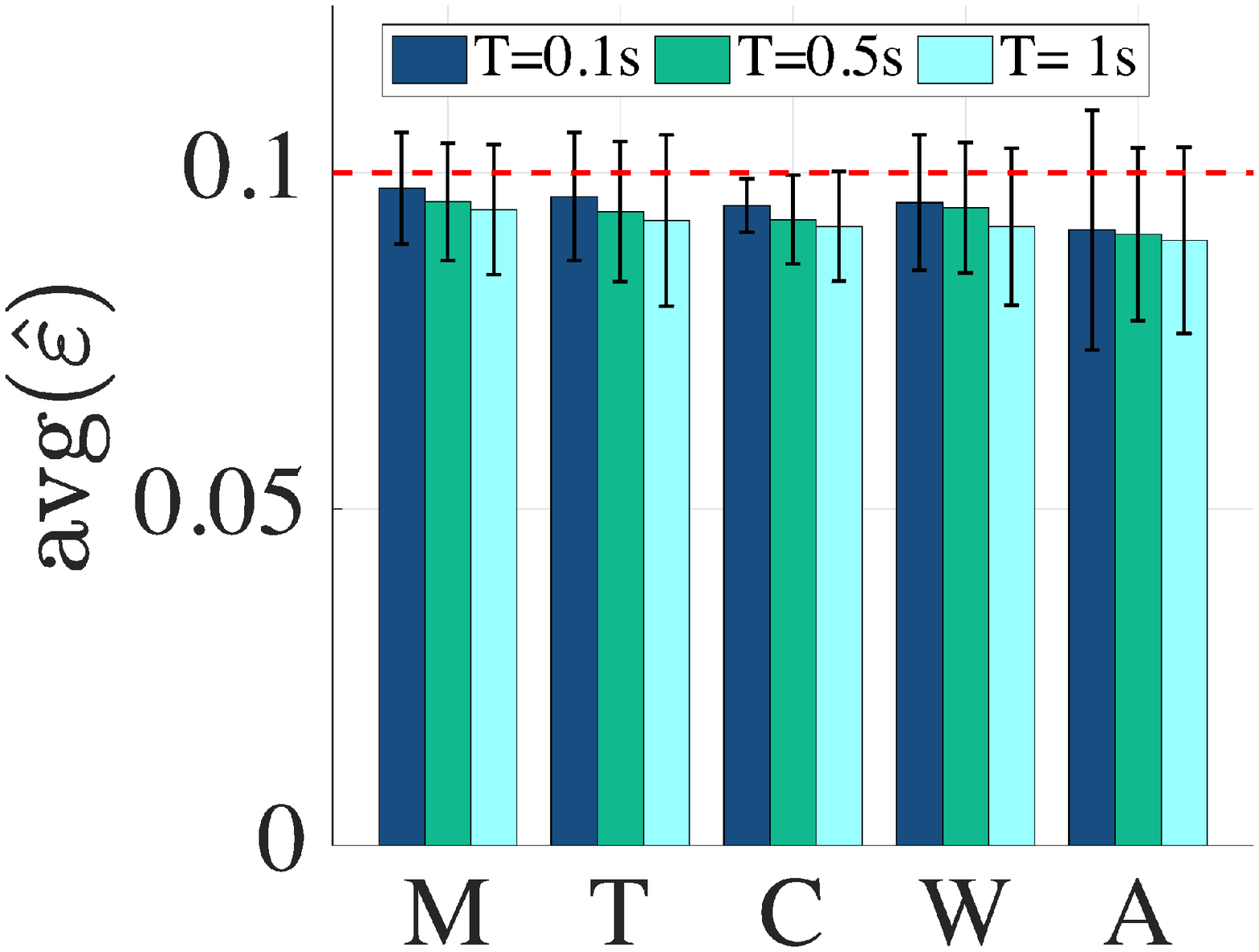}}
	\hspace{-0.8cm}
	\subcaptionbox{target $\varepsilon = 0.05$}[.27\linewidth][c]{%
		\includegraphics[scale=0.2]{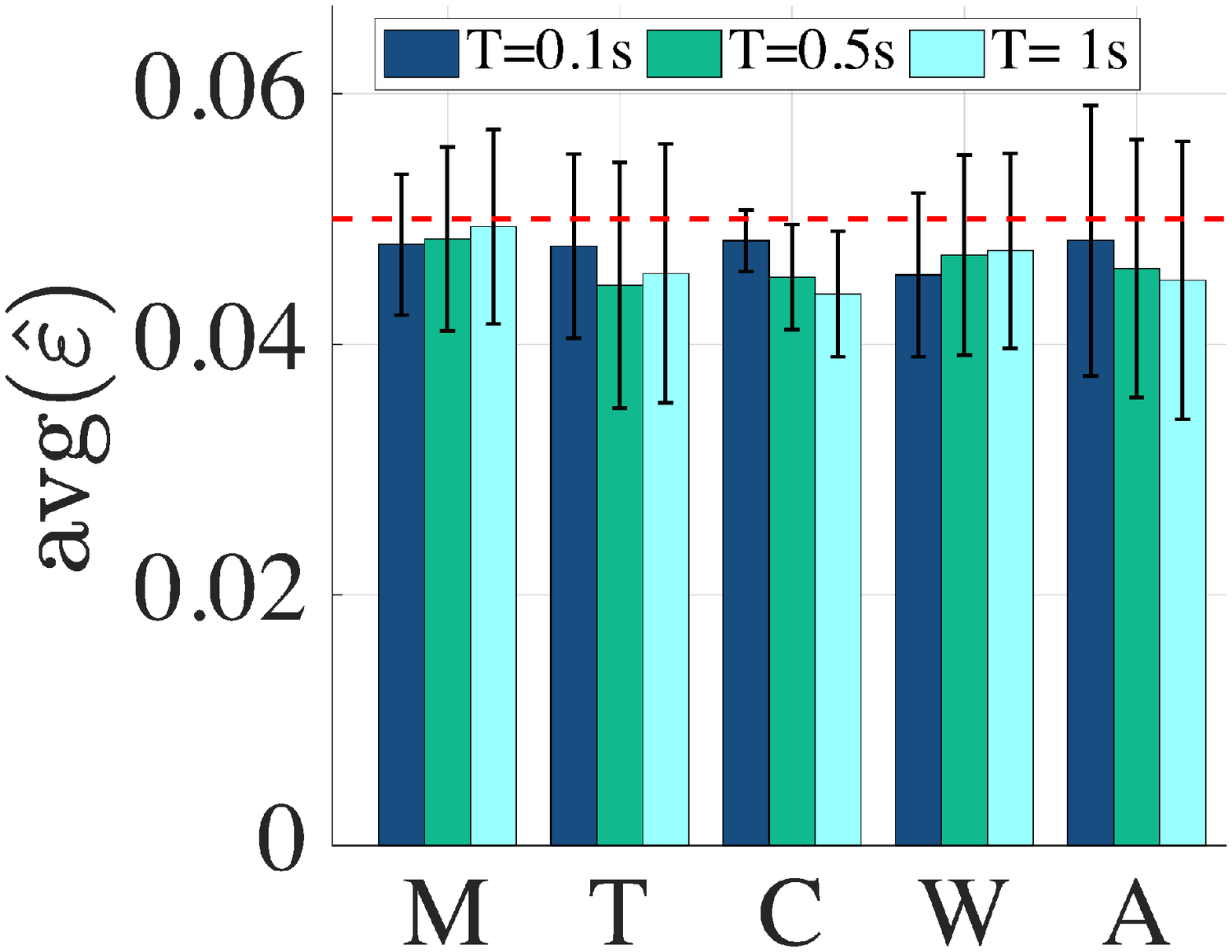}}
	\hspace{-0.8cm}
	\subcaptionbox{target $\varepsilon = 0.01$}[.27\linewidth][c]{%
		\includegraphics[scale=0.2]{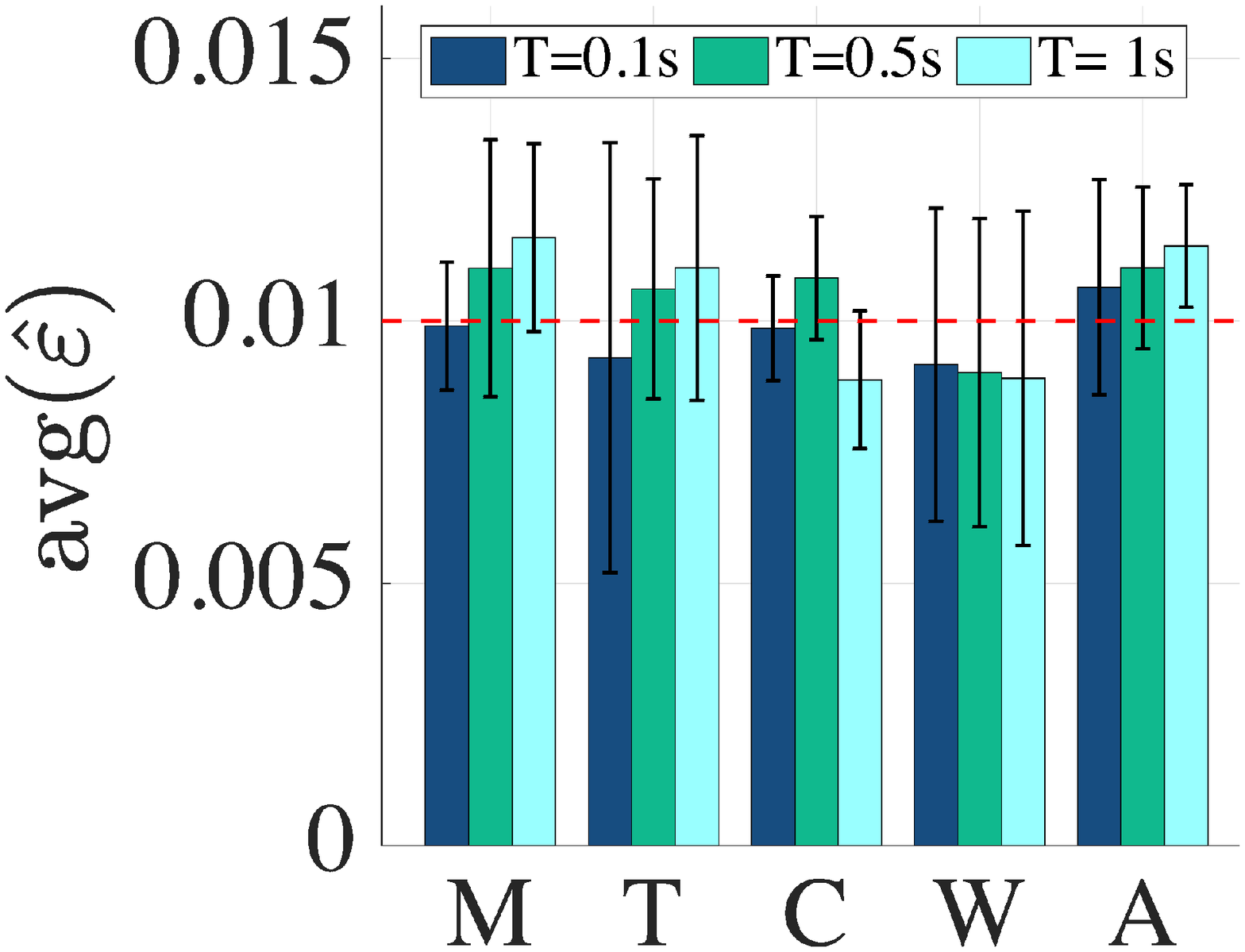}}
	
	
	\centering
	\subcaptionbox{target $\varepsilon = 0.5$}[.27\linewidth][c]{%
		\includegraphics[scale=0.2]{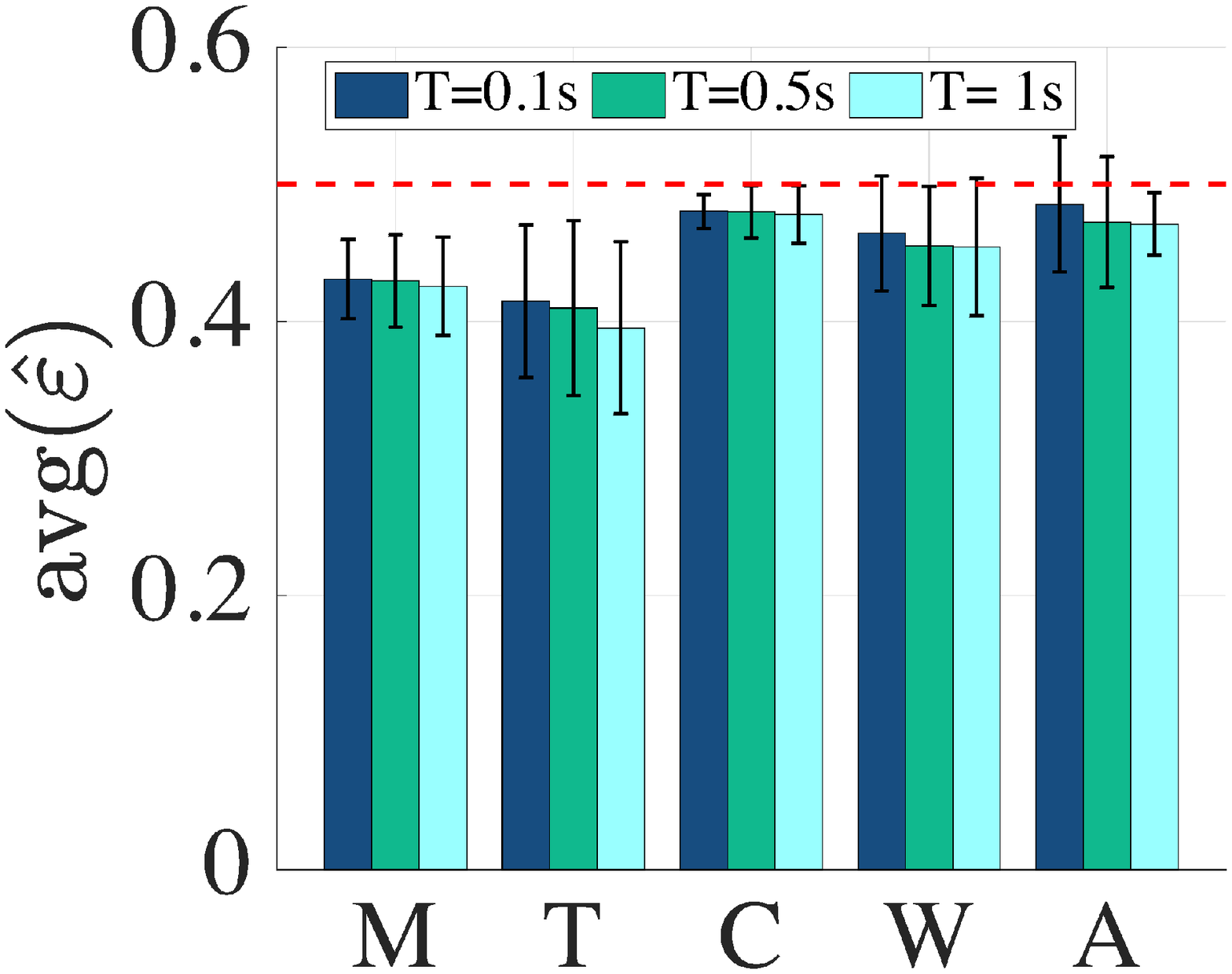}} 
	\hspace{-0.8cm}
	\subcaptionbox{target $\varepsilon = 0.1$}[.27\linewidth][c]{%
		\includegraphics[scale=0.2]{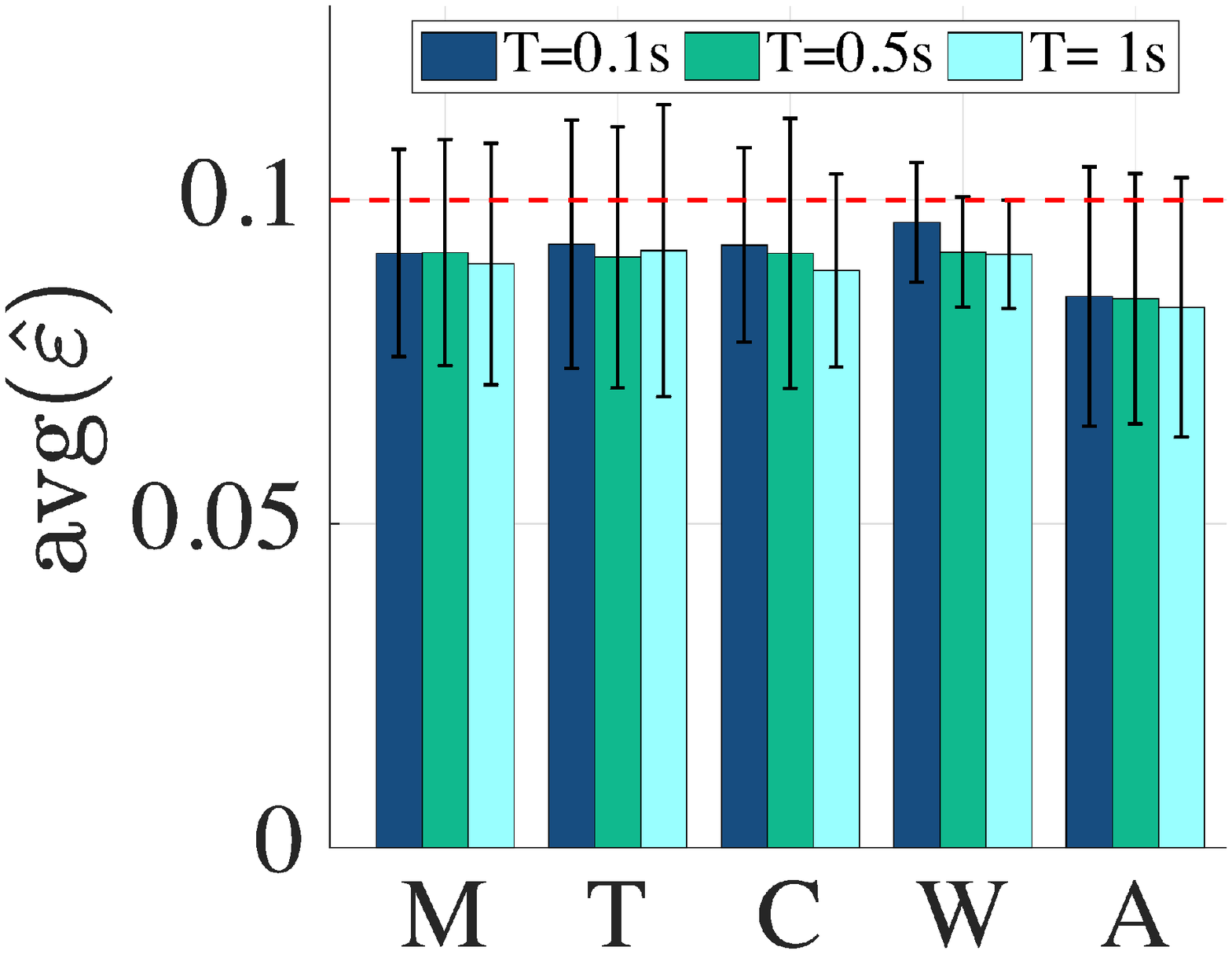}}
	\hspace{-0.8cm}
	\subcaptionbox{target $\varepsilon = 0.05$}[.27\linewidth][c]{%
		\includegraphics[scale=0.2]{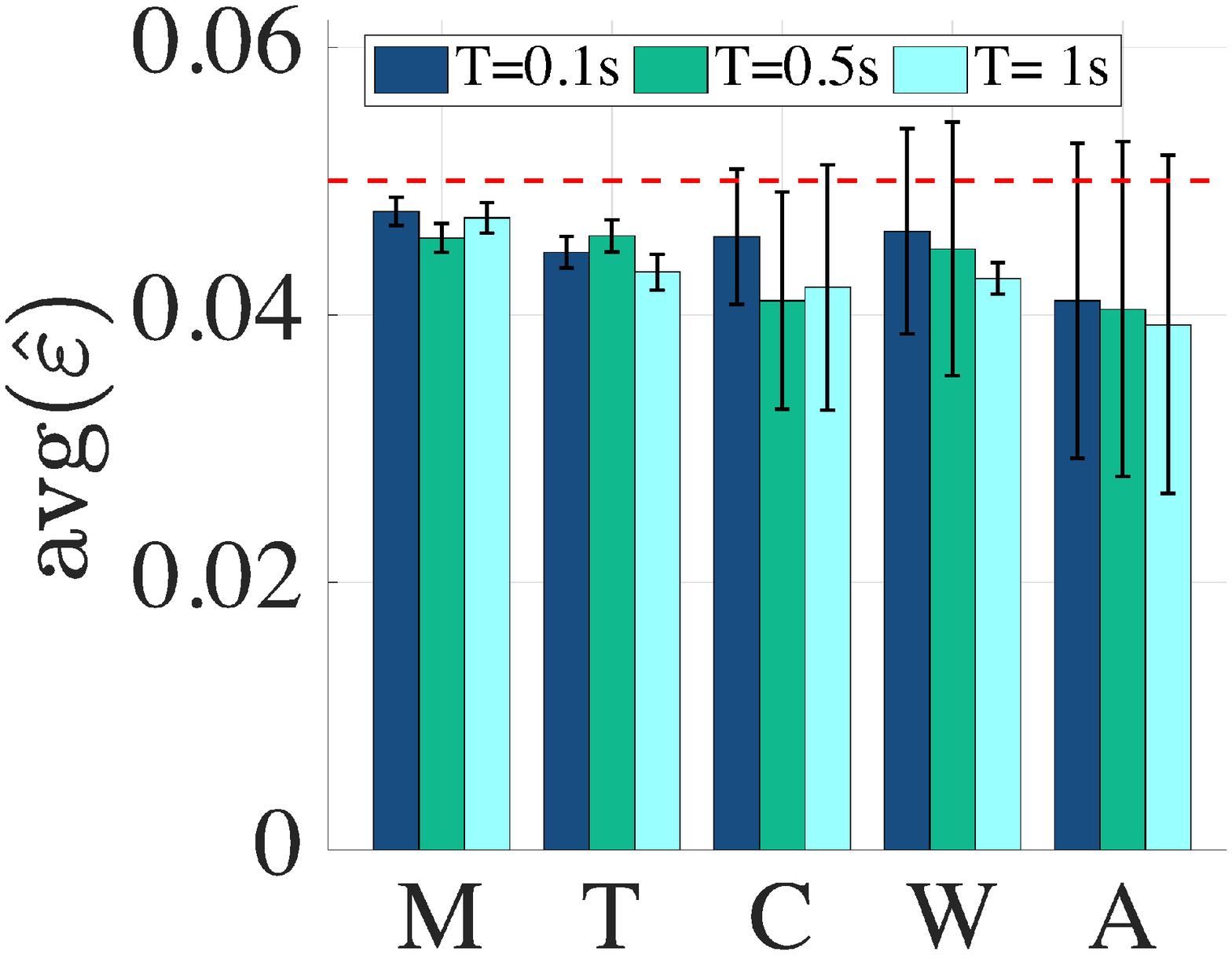}}
	\hspace{-0.8cm}
	\subcaptionbox{target $\varepsilon = 0.01$}[.27\linewidth][c]{%
		\includegraphics[scale=0.2]{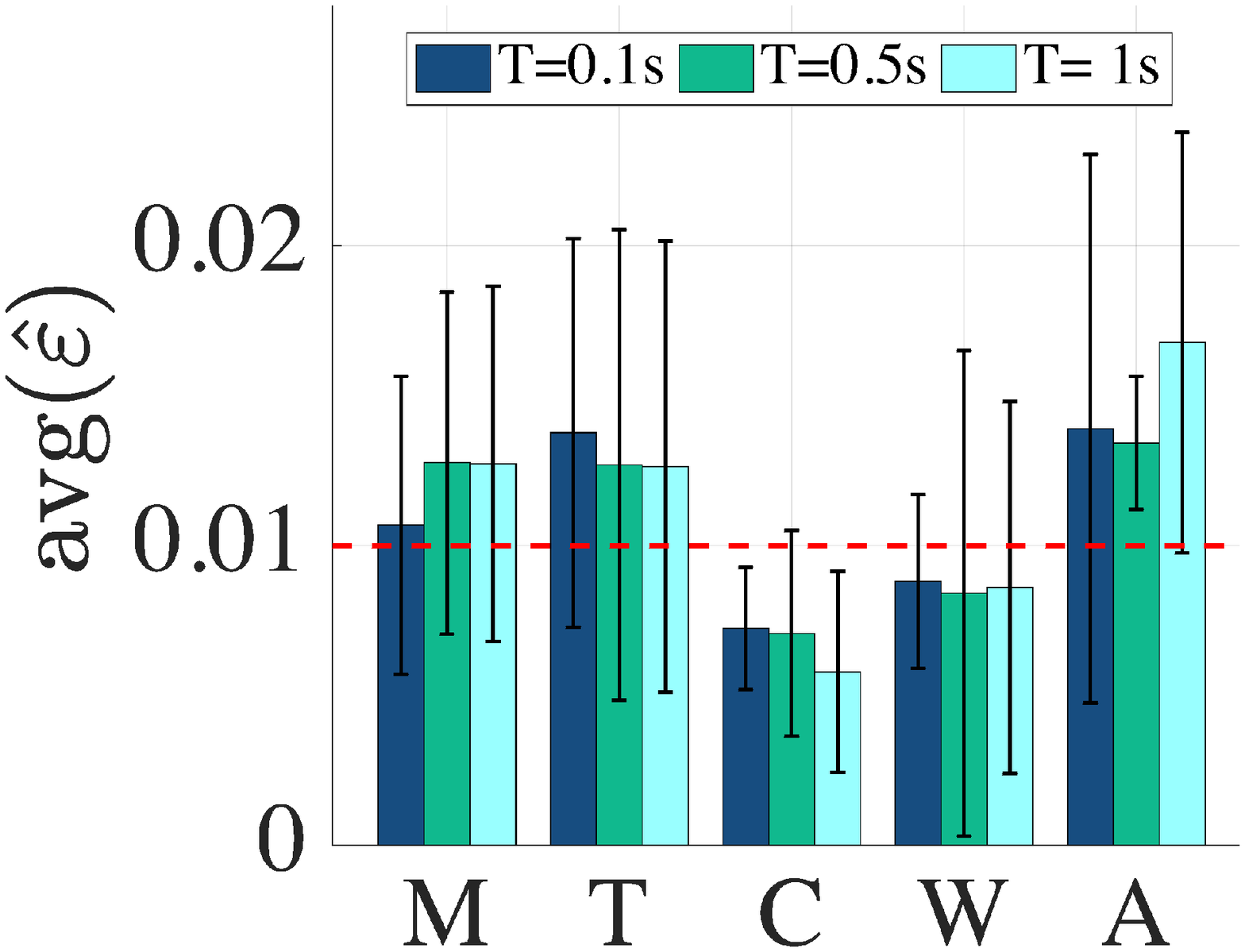}}
	
	\centering
	\setlength{\belowcaptionskip}{-3pt}
	\subcaptionbox{target $\varepsilon = 0.5$}[.27\linewidth][c]{%
		\includegraphics[scale=0.2]{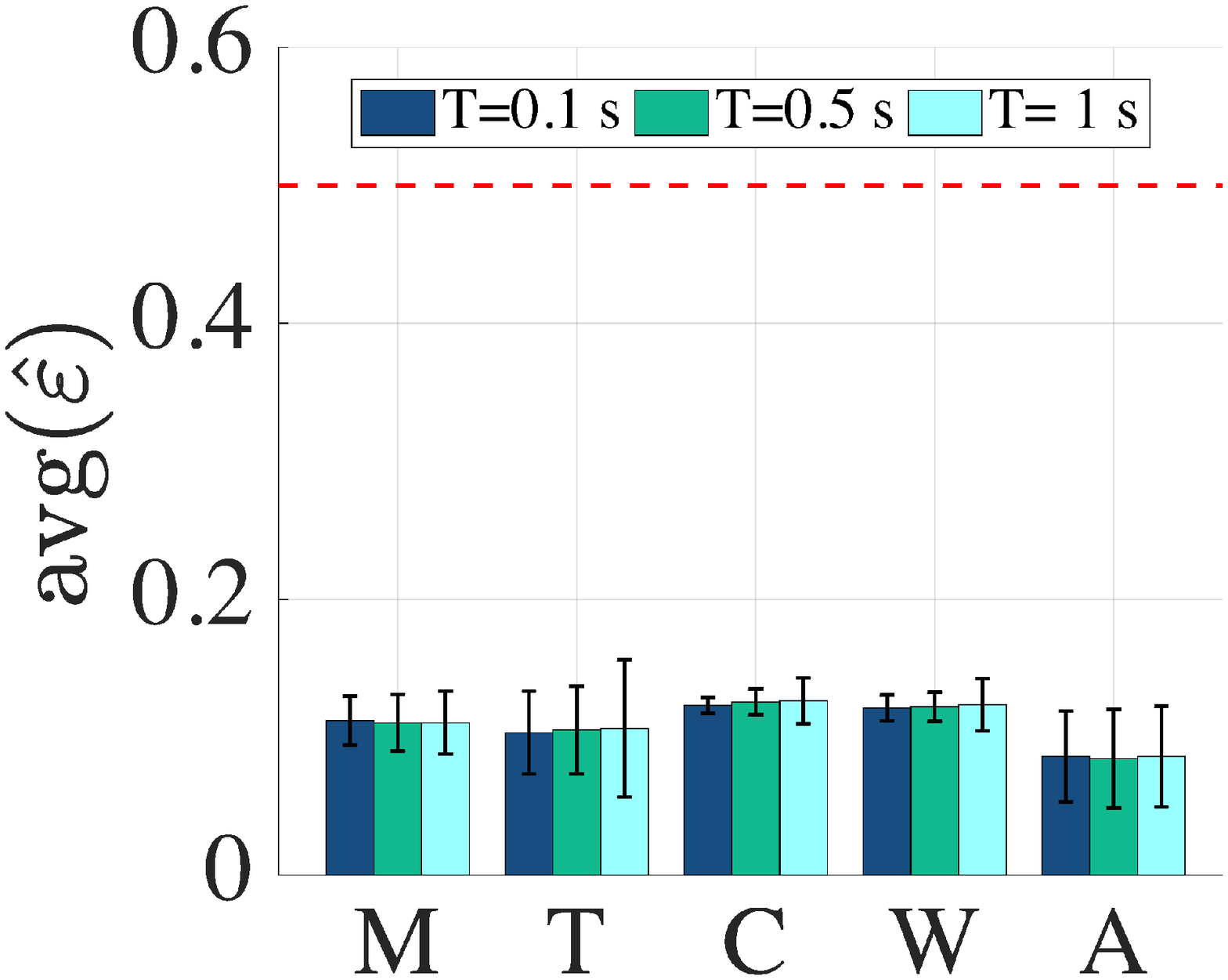}} \hspace{-0.8cm}
	\subcaptionbox{target $\varepsilon = 0.1$}[.27\linewidth][c]{%
		\includegraphics[scale=0.2]{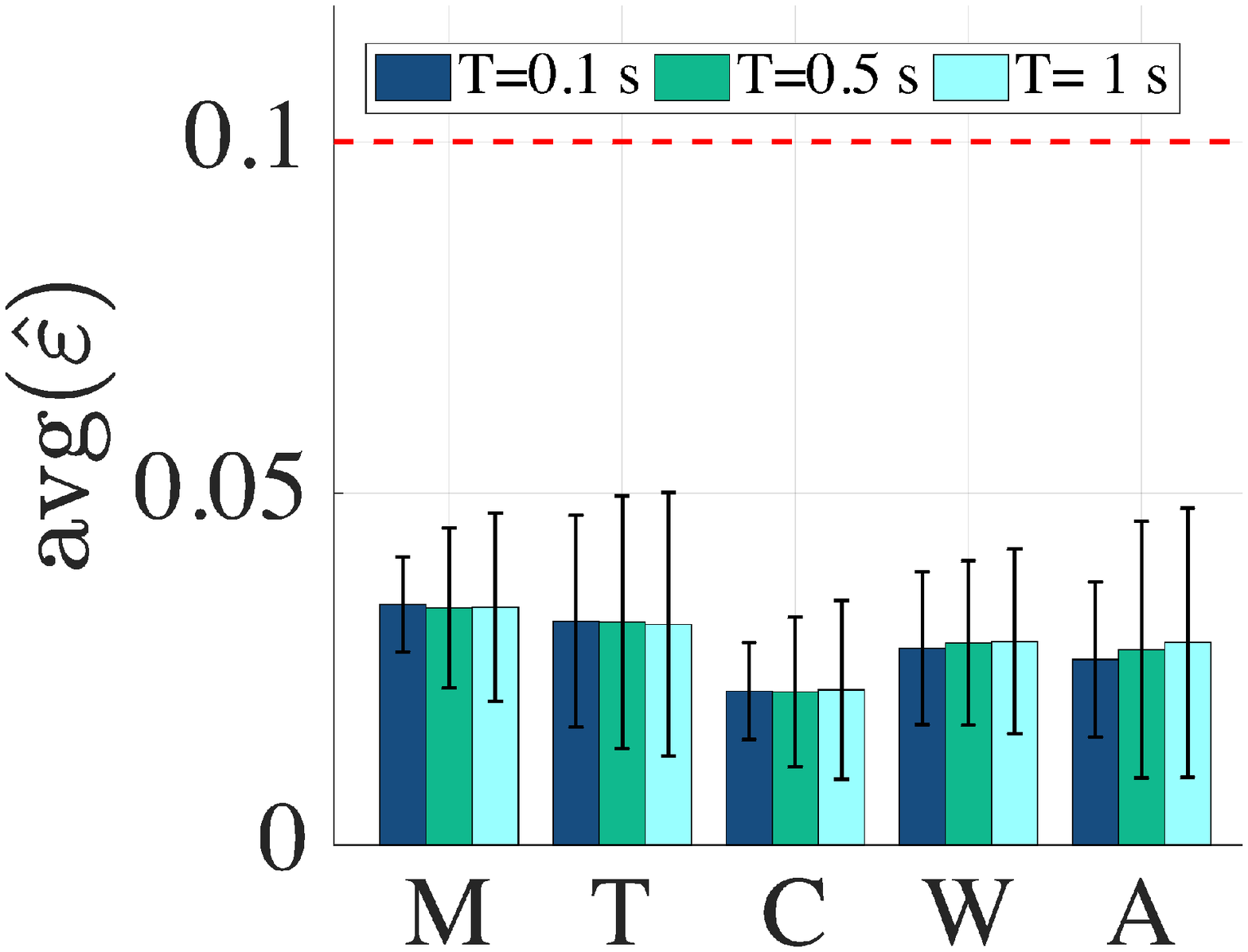}} \hspace{-0.8cm}
	\subcaptionbox{target $\varepsilon = 0.05$}[.27\linewidth][c]{%
		\includegraphics[scale=0.2]{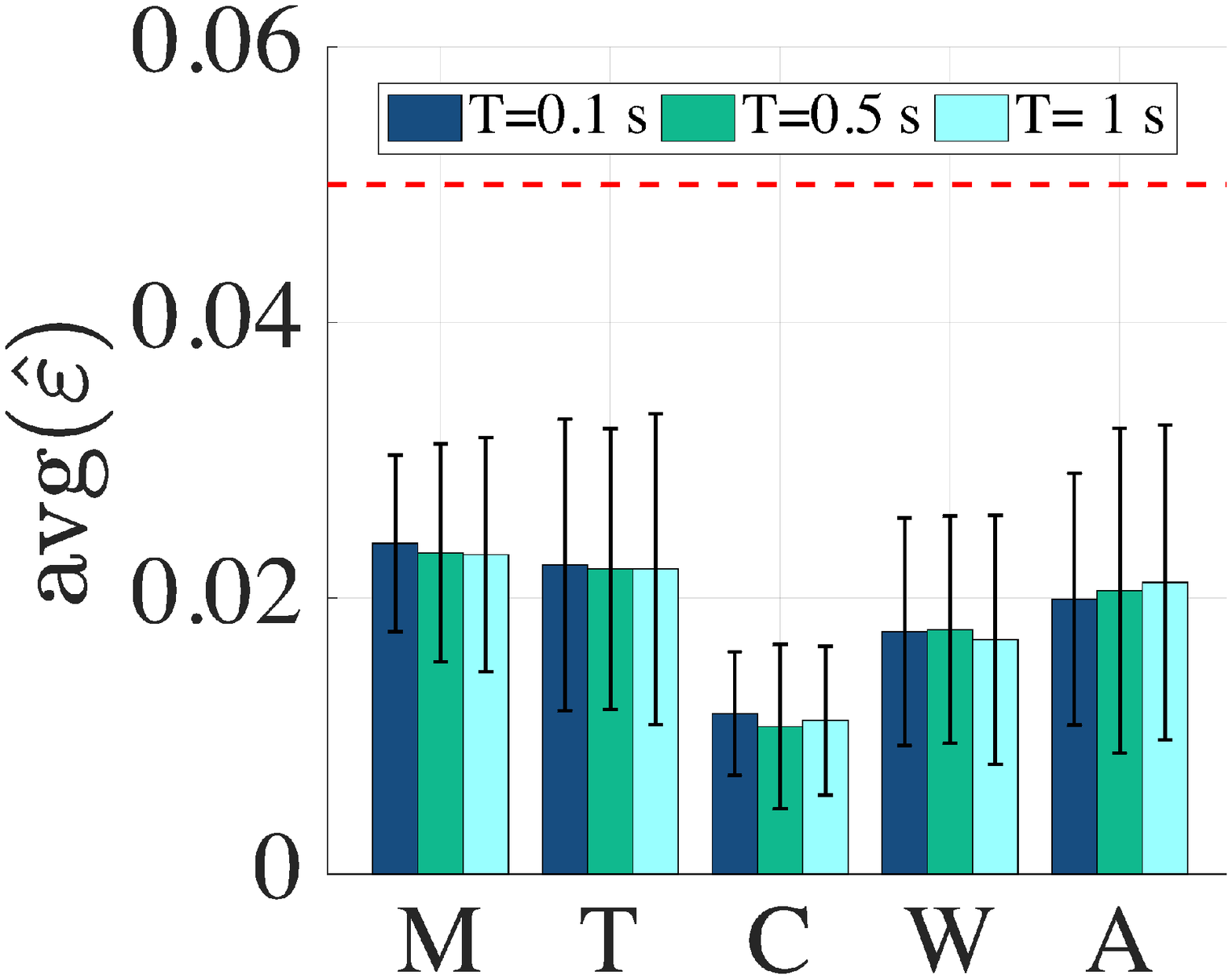}} \hspace{-0.8cm}
	\subcaptionbox{target $\varepsilon = 0.01$}[.27\linewidth][c]{%
		\includegraphics[scale=0.2]{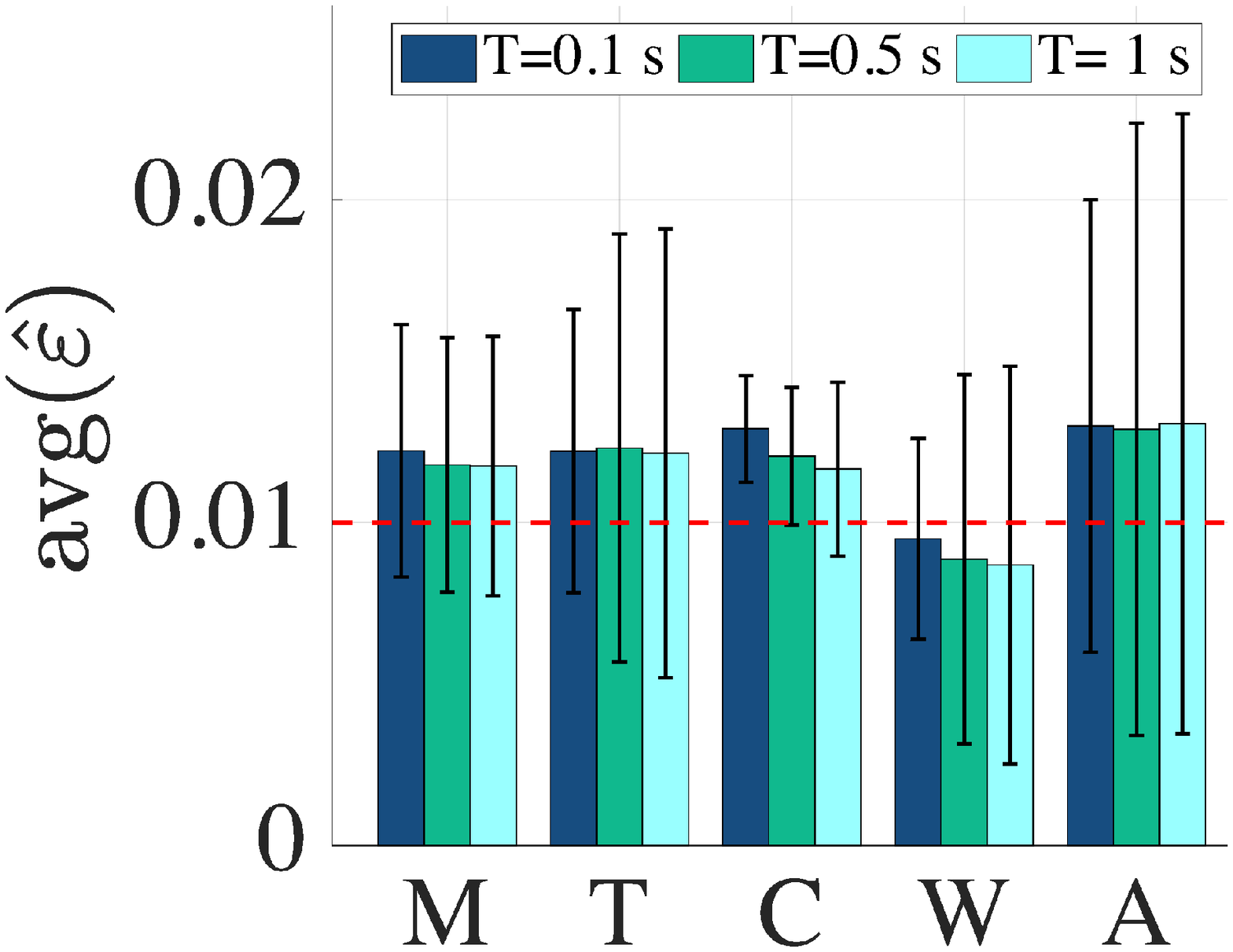}}
	\caption{Link dimensioning based on (a-d) log-normal model, (e-h) Weibull model and (i-l) Meent's formula: avg$(\hat{ \varepsilon})$ for different datasets (M: MAWI, T: Twente, C: CAIDA, W: Waikato, A: Auckland), aggregation timescales ($100$ msec, $500$ msec and $1$ s), and target values of $\varepsilon$ (0.5, 0.1, 0.05 and 0.01). Error bars represent stderr $|\varepsilon -\hat{\varepsilon}|$.}
	\label{empEpsiRes} 	
\end{figure*}

\subsection{Bandwidth provisioning based on the log-normal model}

Here we investigate whether we could achieve more reliable bandwidth provisioning by adopting the log-normal traffic model. We calculate the mean and variance from the captured trace and generate the respective log-normal model. Then, we use the CDF function ($F$) to solve the link transparency formula shown in Equation~\ref{link-tran}. Hence, $F$ is defined as $F(C) = P(A(T)/T<C) $, which can be solved to find $C$, as follows: 
\begin{equation}
C2=F^{-1}\left (1-\varepsilon \right).
\label{cdf-lognormal}
\end{equation}

\subsection{Comparison of bandwidth provisioning approaches}
\label{comparison}

In this section, we compare the bandwidth provisioning approaches described above. The performance indicator is the empirical value of the performance criterion, which is denoted by $\hat{ \varepsilon}$ and defined as follows:
\begin{equation} 
\hat{ \varepsilon}= \frac{\# \left \{ A_{i}| A_{i} \geq CT\right \}}{n}  \textrm{ , } i\in 1\ldots n.  
\label{empepsilon}  
\end{equation}

In words, this empirical value is the percentage of all the data samples of the captured traffic which are measured larger than the estimated link capacity. Ideally, $\hat{ \varepsilon}$ would be equal to the target value of the performance criterion $\varepsilon$. The difference between $\hat{ \varepsilon}$ and $\varepsilon$ is due to the fact that the chosen traffic model is not accurately describing the real network traffic. A simple example of the described comparison approach is illustrated in Figure~\ref{time-domain}, in which we plot the captured data rate for a MAWI trace ($T=100$ msec)\footnote{Note that in all subsequent figures we have also included results for a Weibull model to get insights about bandwidth provisioning using a heavy-tailed distribution.}. The calculated capacity values from each approach when the target $\varepsilon$ is $0.01$ are $C1=344.8$ Mbps and $C2=444.3$ Mbps (represented by the horizontal lines in Figure~\ref{time-domain}). The empirical value can be calculated by using Equation~\ref{empepsilon}, which gives $\hat{ \varepsilon}_{1}=0.042$ and $\hat{ \varepsilon}_{2}=0.012$. Obviously, with the first approach the network operator would not be able to meet the target $\varepsilon=0.01$, while with the second approach the empirical value is close to the target.

\begin{figure}[h]
	\setlength{\belowcaptionskip}{-5pt}
	\centering
	\subcaptionbox{target  $\varepsilon = 0.5$}[.48\linewidth][c]{%
		\includegraphics[width=1\linewidth]{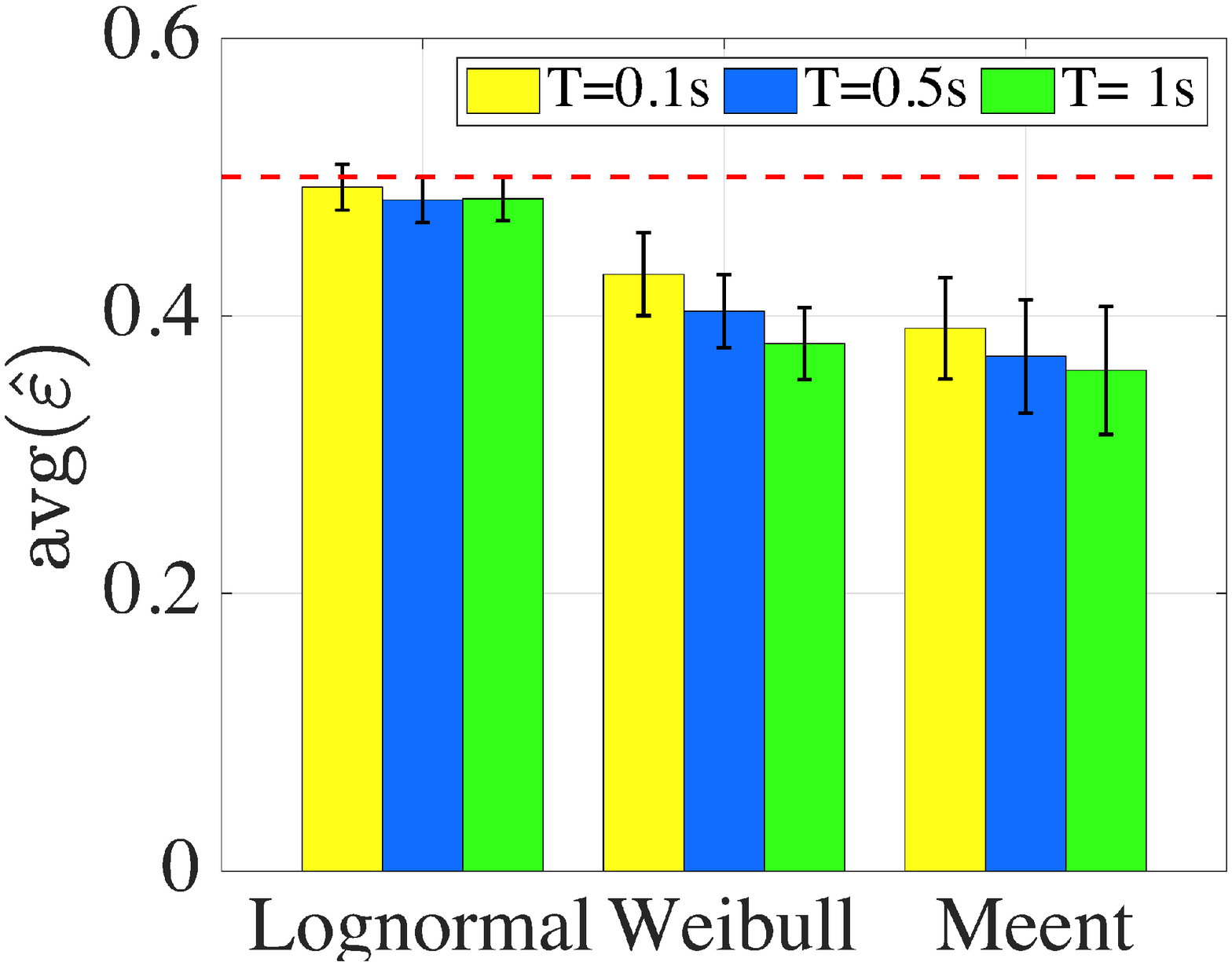}}\quad
	\subcaptionbox{target  $\varepsilon = 0.1$}[.48\linewidth][c]{%
		\includegraphics[width=1\linewidth]{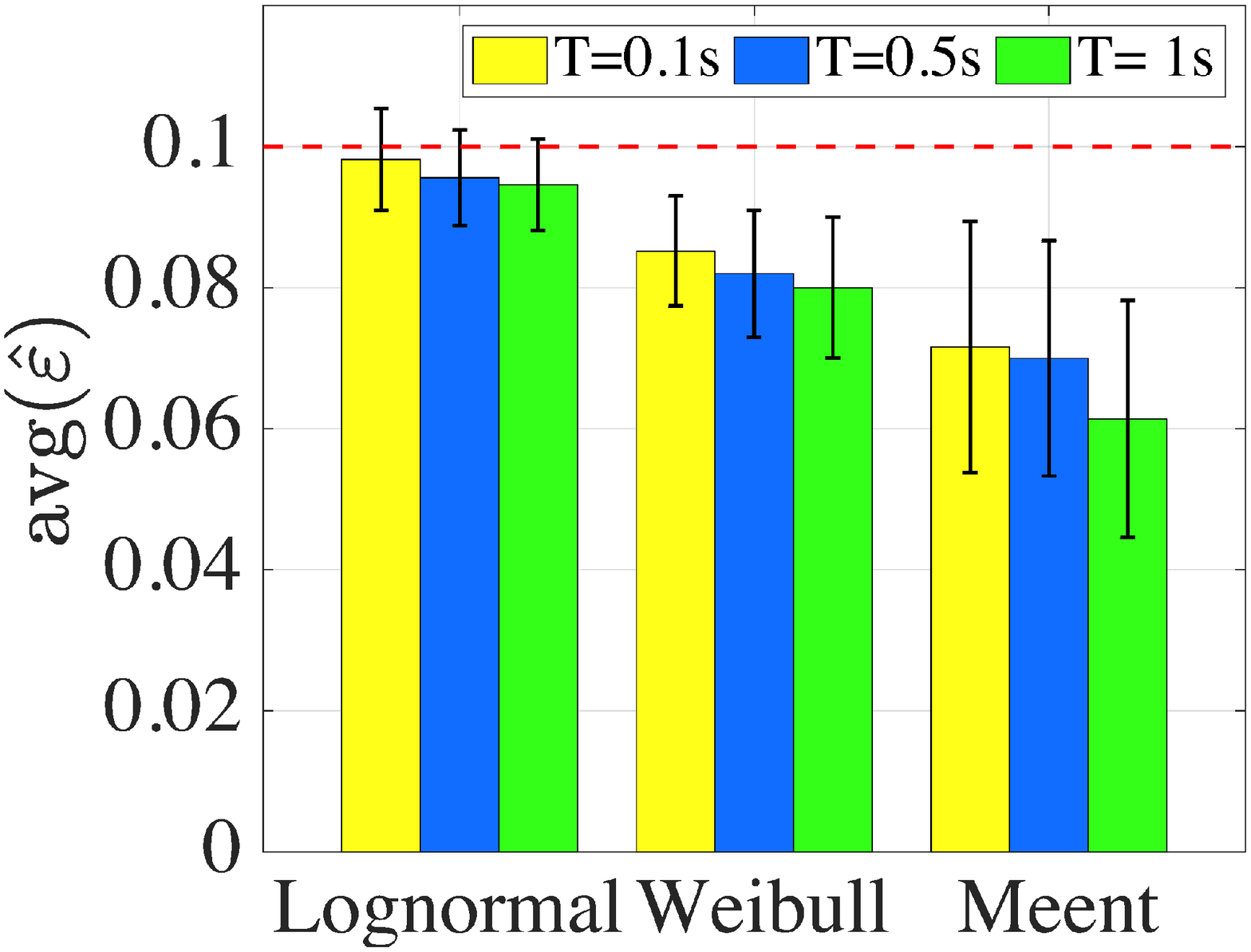}}\quad
	\caption{Link dimensioning based on log-normal, Weibull and Meent for 24 subtraces from 24-hour long MAWI trace.}
	\label{link-dim-24-hour-subtraces} 
\end{figure}

We next compare results of bandwidth provisioning calculations based on the (a) Meent's formula, (b) Weibull model and (c) proposed log-normal model. Figure~\ref{empEpsiRes}(a-d) shows the average of the empirical value (avg$(\hat{ \varepsilon})$) for all traces in each dataset at $T= 0.1$ sec, $T= 0.5$ sec and $T=1$ sec. The value of $T$ is chosen to be sufficiently small so that the fluctuations in the traffic can be modelled as well. Each model is tested for four different values of the performance criterion: $\varepsilon = 0.5$, $\varepsilon = 0.1$, $\varepsilon = 0.05$ and $\varepsilon = 0.01$. In Figure~\ref{empEpsiRes}(a-d) we clearly see that the log-normal model is able to satisfy the required performance criterion $\varepsilon$ at different aggregation time-scales for all datasets. In contrast, Meent's formula failed to allocate sufficient bandwidth, which results in missing the target performance criterion $\varepsilon$ for all datasets and target performance values, as depicted in Figure~\ref{empEpsiRes}(i-l) (see horizontal red line). The Weibull distribution performs better comparing to Meent's formula, but bandwidth provisioning using the log-normal model is far superior, as can be seen from Figures~\ref{empEpsiRes}(a-d) and~\ref{empEpsiRes}(e-h). 

We apply the same link dimensioning tests as discussed above on 24 subtraces (each one being 1-hour long) from the 24-hour long MAWI trace. Figure~\ref{link-dim-24-hour-subtraces} shows the avg$(\hat{ \varepsilon})$ for all subtraces at different timescale values. As shown in the figure, the log-normal model performs the best compared to the other two in estimating bandwidth allocation, with respect to the target performance criterion.

\textcolor{black}{Over long time periods (hours and days) the data is not stationary as it is subject to daily and weekly variations related to human activity. A single distribution cannot sensibly capture the behaviour of traffic over one day. In Figure~\ref{mawi-24-power-law-results}a we showed that for all 1-hour long traces the log-normal distribution is the best fit compared to all tested alternative distributions, therefore, a sensible procedure for the operator could be to fit individual log-normal distributions to smaller periods of the day. The traffic would then be modelled as a series of log-normal distributions where the mean and variance change between times of day.  A practical indicative procedure that an operator can follow to use the log-normal model for bandwidth provisioning and provide the respective analysis can be as in the following example. We divide the MAWI 24-hour long trace into 96 15-minute long subtraces starting from time 00:00 to 23:59. Then, we apply the bandwidth provisioning mechanism on these 96 samples by measuring the empirical value $\hat{ \varepsilon}$ for two target values: $\varepsilon = 0.5$ and $\varepsilon = 0.1$ using different models (log-normal, Weibull and Meent). Figure~\ref{bw-prov-long} shows that the log-normal model is able to achieve the target performance values much better than Weibull and Meent models (NRMSE $= 0.0396$ and $0.0052$ for targets $\varepsilon = 0.1$ and   $\varepsilon = 0.5$, respectively, in log-normal results). These results show that the log-normal model can accurately predict the proportion of time a link will exceed a given capacity. As an example, the operator could choose to provision bandwidth based on the fitted model for the peak time of the day.  It is worth pointing out that we do not intend our example to be fully worked systems for practical deployment (see future work in Section~\ref{sec:conclusion}).}

\begin{figure}[htp]
	\centering
	\includegraphics[width=1\linewidth, scale=0.35]{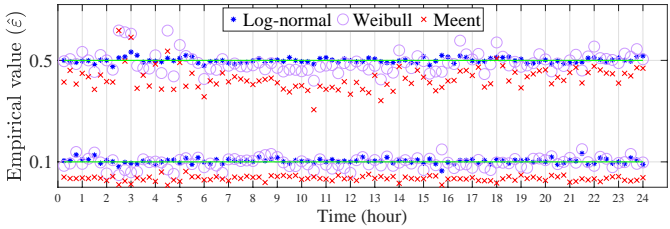}\quad
	\caption{\textcolor{black}{Link dimensioning for 96 15-minute subtraces from the 24-hour Mawi trace for two target  $\varepsilon$ values.}}
	\label{bw-prov-long} 	
\end{figure}

\begin{figure*}[t]
	\centering
	\subcaptionbox{CAIDA}[.29\linewidth][c]{%
		\includegraphics[scale=0.23]{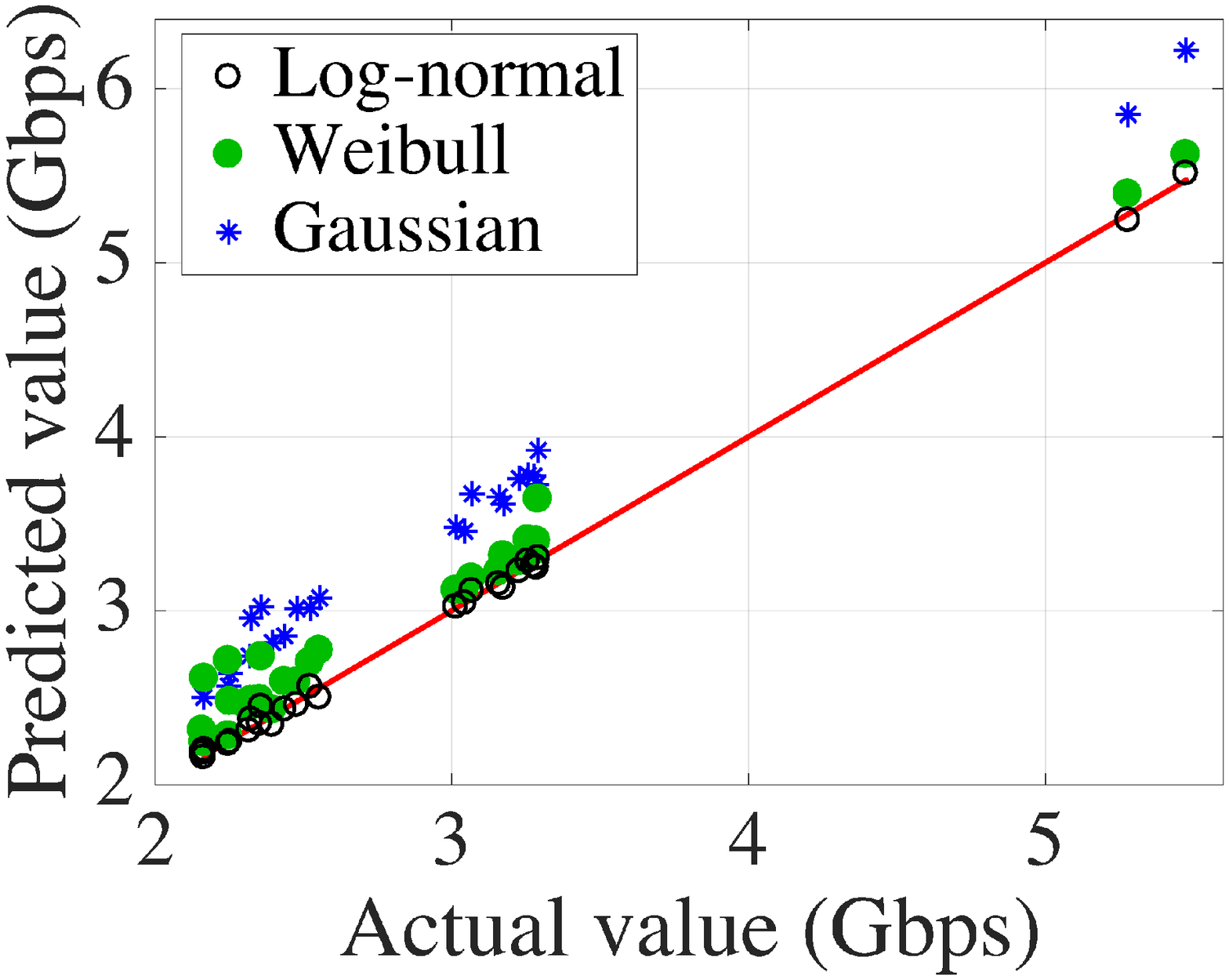}}\quad
	\subcaptionbox{Waikato}[.29\linewidth][c]{%
		\includegraphics[scale=0.23]{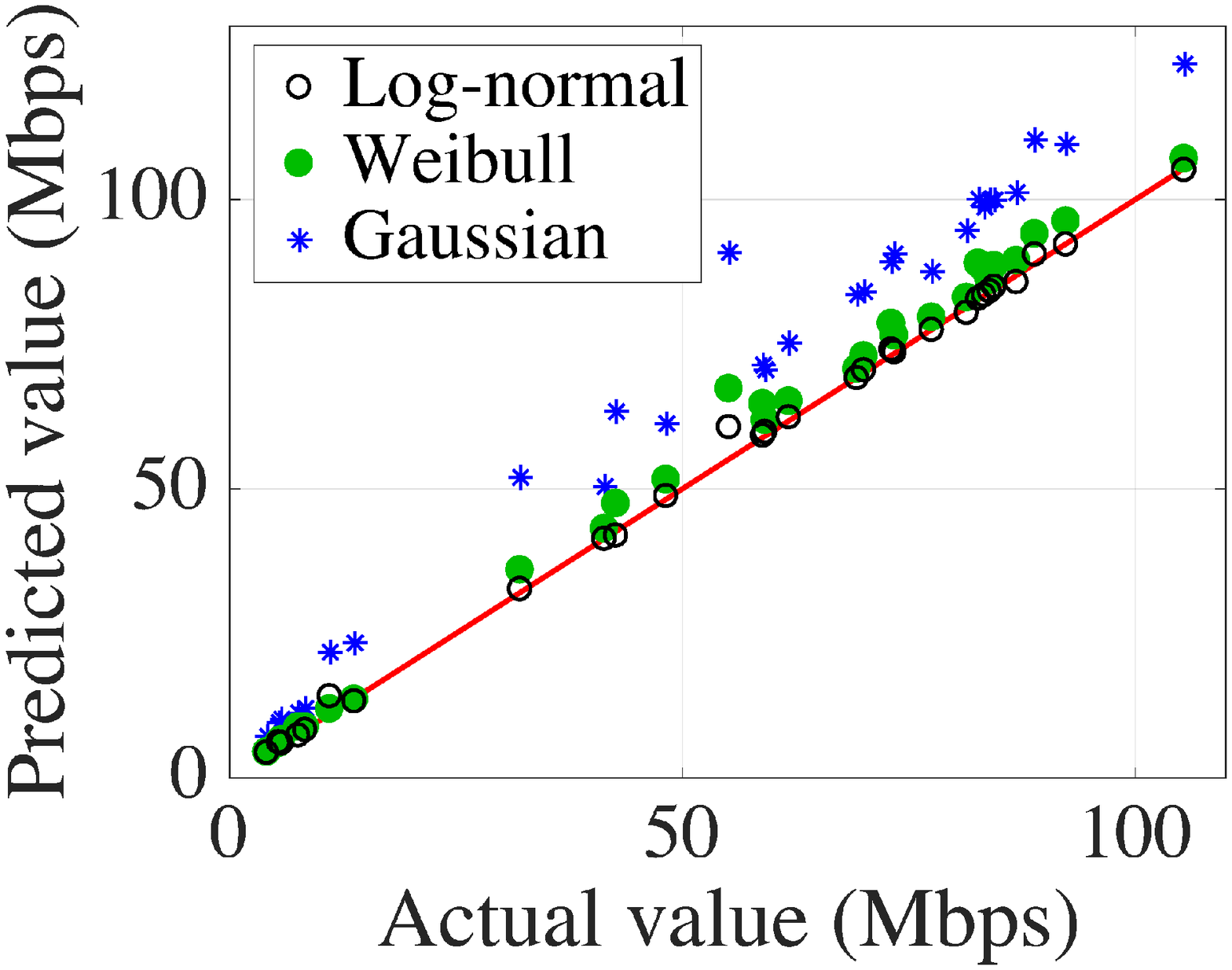}}\quad
	\subcaptionbox{Auckland}[.29\linewidth][c]{%
		\includegraphics[scale=0.23]{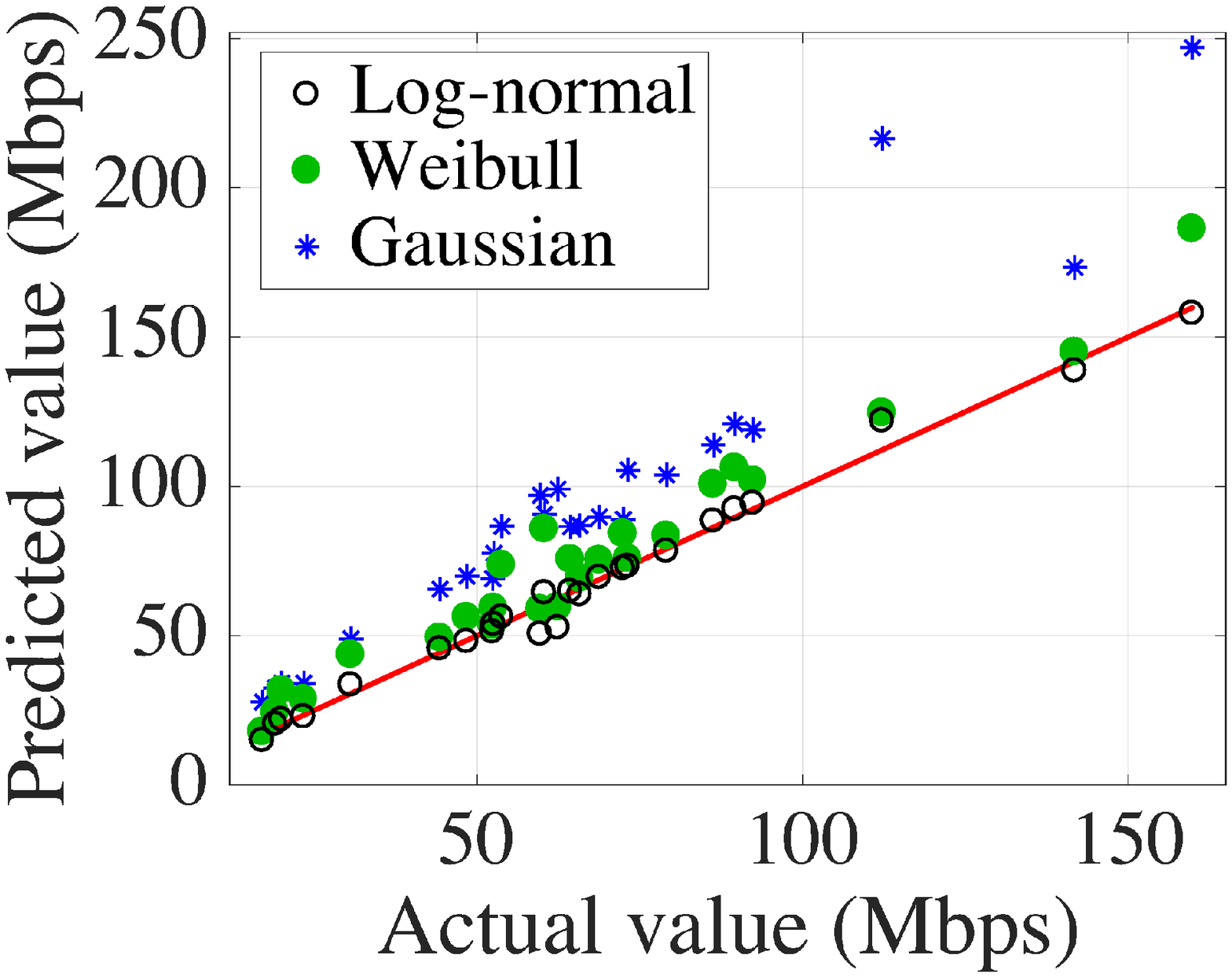}}
	\\
	\setlength{\belowcaptionskip}{-4pt}
	\subcaptionbox{Twente}[.29\linewidth][c]{%
		\includegraphics[scale=0.23]{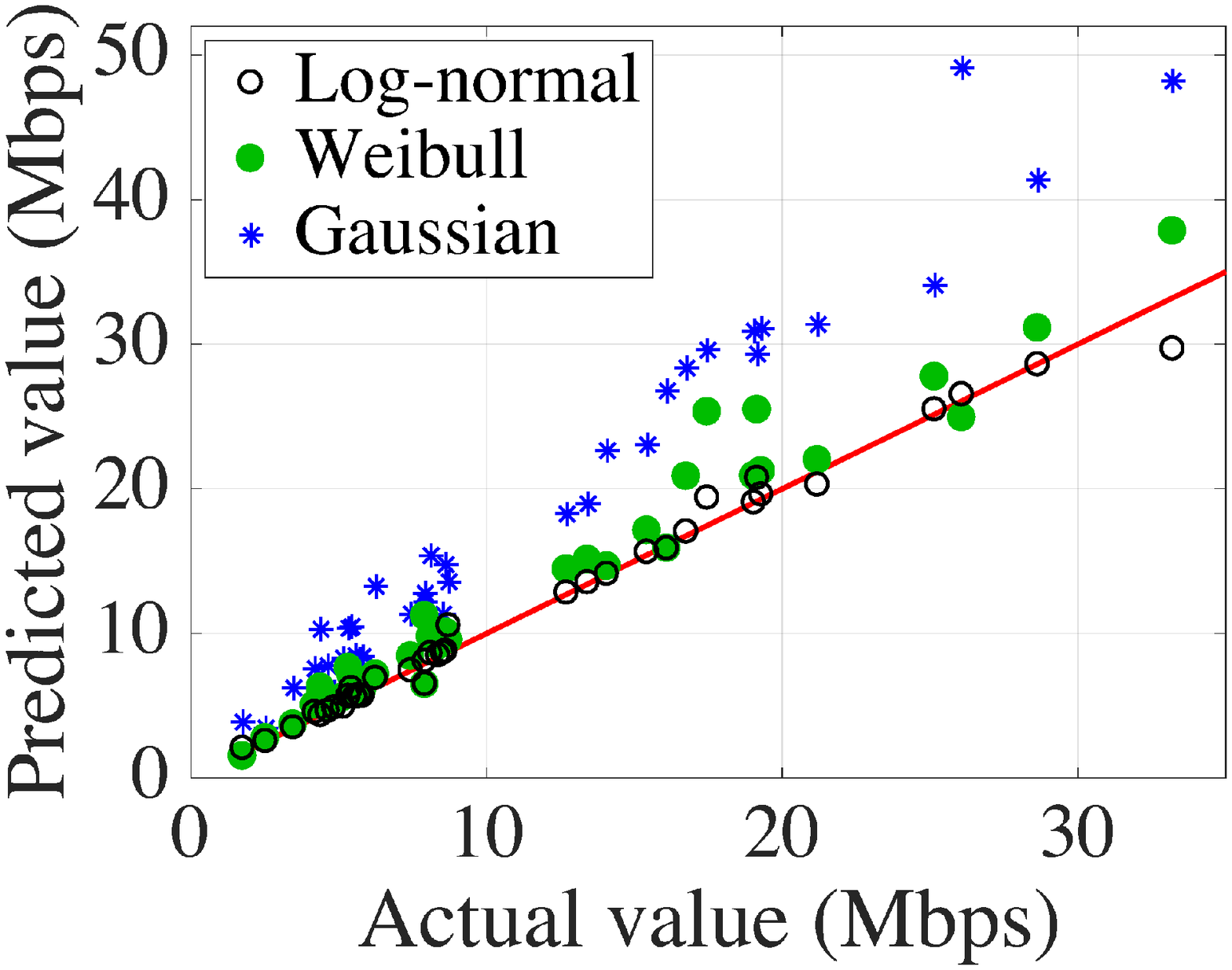}}\quad
	\subcaptionbox{MAWI}[.29\linewidth][c]{%
		\includegraphics[scale=0.23]{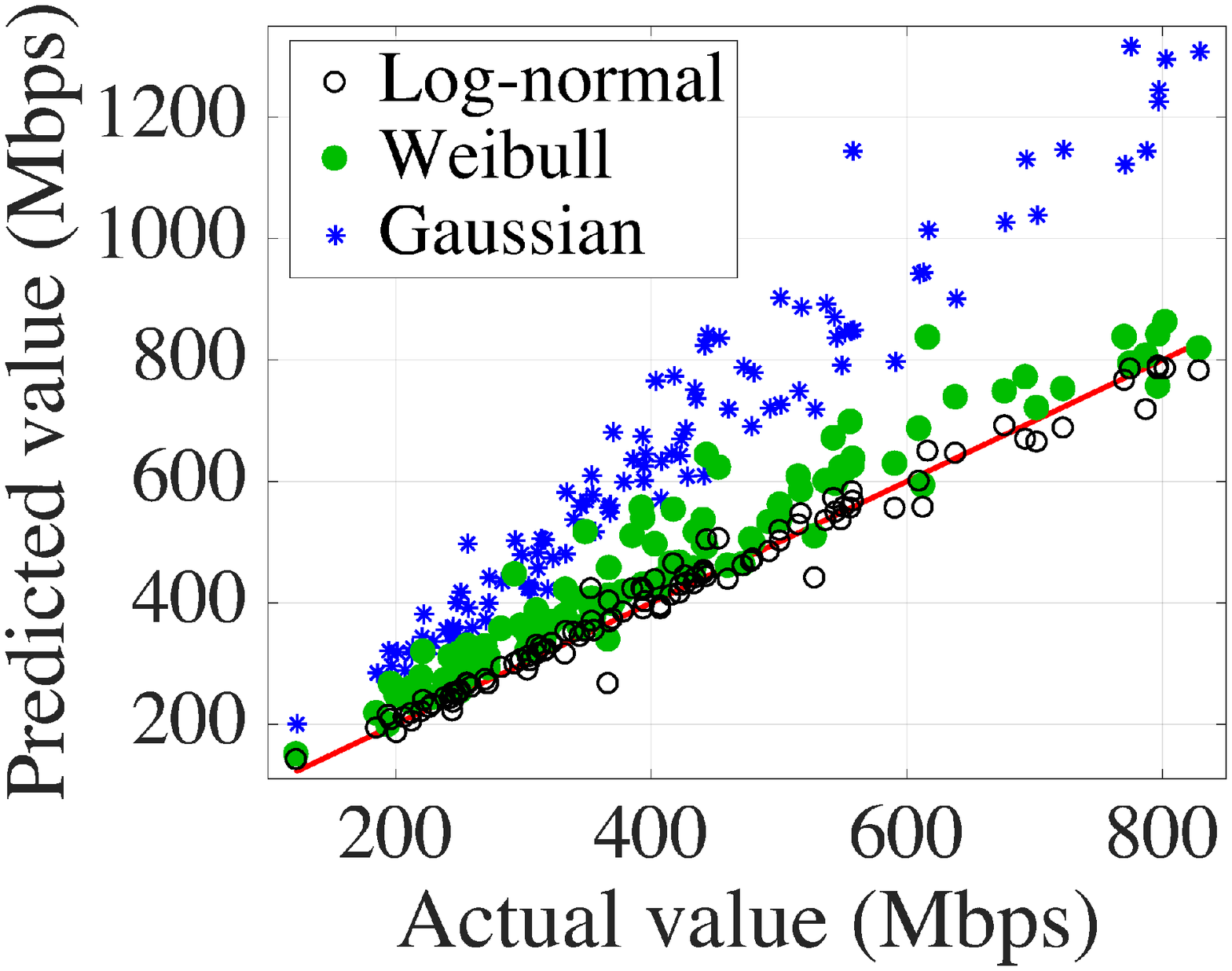}}\quad
	\subcaptionbox{24 subtraces in 24-hour MAWI trace}[.29\linewidth][c]{%
		\includegraphics[scale=0.23]{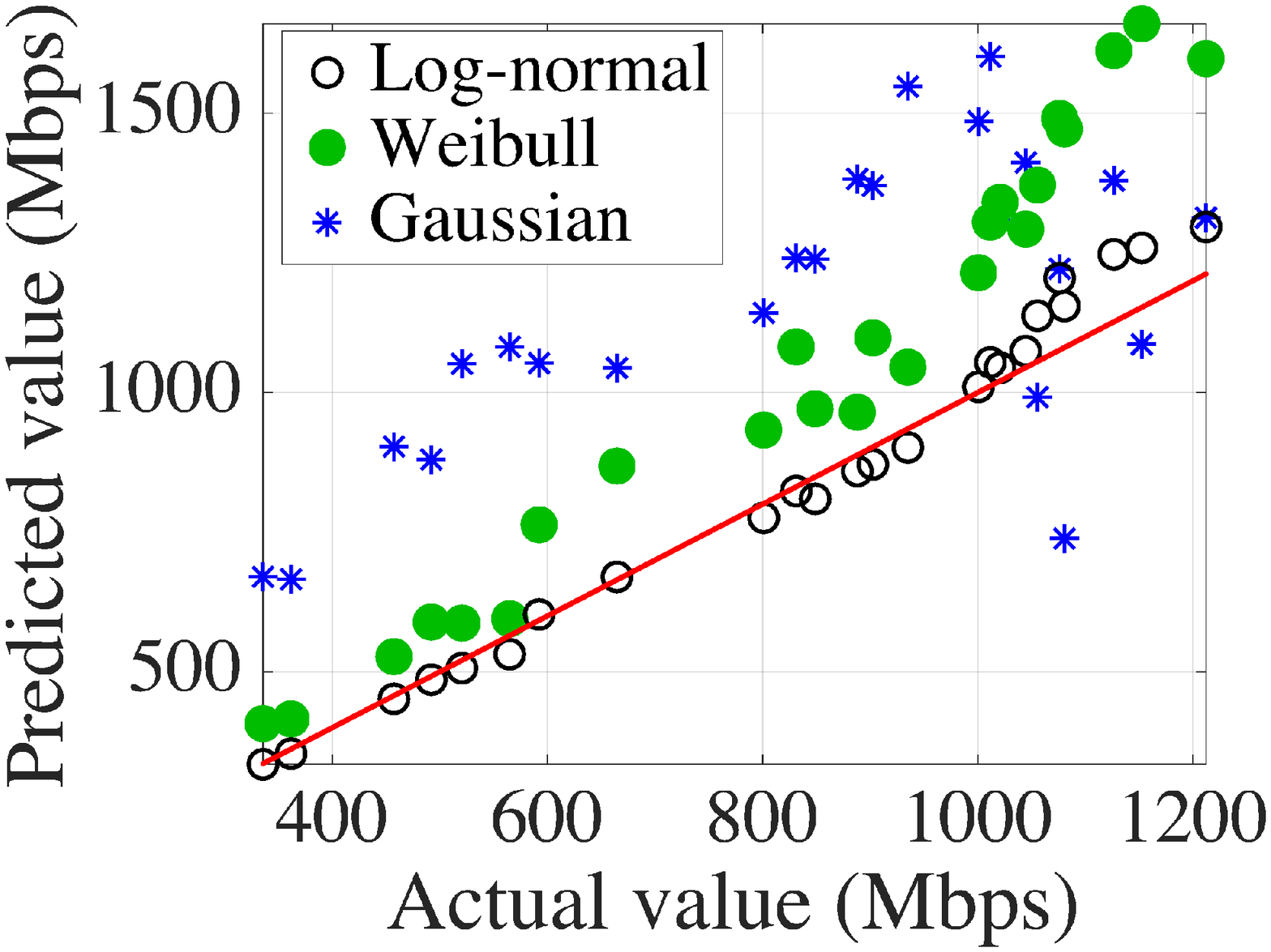}}\quad
	\caption{95th percentile values (actual vs predicted rates) based on log-normal, Weibull and Gaussian models. An ideal model would result in points in the plot area that fall exactly on the red line. \textcolor{black}{The results in (a-e) are for the 15-minute long traces in each dataset. The results in (f) are for the 24 subtraces in the 24-hour long MAWI trace.}}
	\label{percentile} 	 
\end{figure*}

\section{95th percentile pricing scheme based on log-normal model}
\label{sec:pricing}

Traffic billing is typically based on the 95th percentile method \cite{billPAM}. Traffic volume is measured at border network devices (typically aggregated at time intervals of 5 minutes) and bills are calculated according to the 95-percentile of the distribution of measured volumes; i.e. network operators calculate bills by disregarding occasional traffic spikes. Forecasting future bills, which is important for ISPs and clients, can be done using a model of the traffic calculated through previously sampled traffic. In this section, we apply our findings on Internet traffic modelling in predicting the cost of traffic according to the 95th percentile method.

For each network trace we calculate the actual 95th percentile of the traffic volume. The majority of the studied traffic traces were 15-minute long but operators typically use measurements traffic volumes for much longer periods, therefore we scale down the calculation of the 95th percentile by dividing each trace (900 seconds) into 90 groups (10 seconds length each). In reality, of course, the 95th percentile method would use traffic from different times of day with different means which would need to be modelled as separate log normal distributions with separate means and variances. This is not possible with the 15-minute long samples that form the focus of this paper. However, it remains a level playing field test for which of the distributions best captures the real underlying data.

We calculate the 95th percentile for the observed traffic. We then fit a Gaussian, Weibull and log-normal distribution to each trace (for $T=100$ msec) and calculate the 95th percentile of the fitted distribution. We plot the actual 95th percentile against the three predictions in Figure~\ref{percentile} with a red reference line to show where perfect predictions would be located. It is clear that the log-normal model provides much more accurate predictions of the 95th percentile than the Gaussian model. As with the bandwidth dimensioning case discussed in Section \ref{sec:provision}, the Weibull is better than the Gaussian model but worse than the proposed log-normal model. \\ \indent We employ the normalised root mean squared error (NRMSE) as a goodness of fit to the results in Figure~\ref{percentile}. NRMSE measures the differences between values predicted by a hypothetical model and the actual values. In other words, it measures the quality of the fit between the actual data and the predicted model. Table~\ref{percentileGoF} shows the NRMSE for all datasets and the three considered models. It is clear that the lowest NRMSE value is for the log-normal model, which is the best model compared to the Gaussian and Weibull ones.  \\ \indent \textcolor{black}{A simple procedure of how an operator might use the log-normal model in predicting the 95th percentile would be to fit individual log-normal distributions to smaller periods of the day.  The traffic is then being modelled as a series of log-normal distributions}. We apply the same 95th percentile test on each 1-hour long subtrace from the 24-hour long MAWI trace. For each 1-hour long subtrace, we calculated the 95th percentile by dividing each subtrace (1 hour) into 60 groups (1-minute long each). Figure~\ref{percentile}f shows the results for the 24 subtraces. The log-normal model is significantly more accurate in predicting the 95th percentile compared to the Weibull and Gaussian models. \textcolor{black}{This example is not intended to be a fully deployable system. It aims at highlighting the benefits of our findings. Several months of data would be necessary to investigate the applicability of this model (see future work in Section~\ref{sec:conclusion}). } 

\begin{table}[]
	\setlength{\belowcaptionskip}{-6pt}
	\scriptsize
	\caption{Goodness of fit (GOF) using Normalised Root Mean Squared Error (NRMSE)}
	\begin{tabular}{|c|c|c|c|l|l|}
		\hline
		Model/Dataset & CAIDA  & Waikato & Auckland & Twente & MAWI   \\ \hline
		Log-normal & 0.0399 & 0.0401  & 0.1058   & 0.0979 & 0.1528 \\ \hline
		Weibull    & 0.2410 & 0.1148  & 0.2984   & 0.2123 & 0.4145 \\ \hline
		Gaussian   & 0.5544 & 0.4193  & 0.6866   & 0.5741 & 0.9828 \\ \hline
	\end{tabular}
	\label{percentileGoF} 	
\end{table}

\section{Related work}
\label{sec:related}

Reliable traffic modelling is important for network planning, deployment and management; e.g. for traffic billing and network dimensioning. Historically, network traffic has been widely assumed to follow a Gaussian distribution. In~\cite{Gaussian-everywhere,Gaussian-revisited}, the authors studied network traces and verified that the Gaussianity assumption was valid (according to simple goodness-of-fit (GOF) tests they used) at two different timescales. In~\cite{BusyhourTraff}, the authors studied traffic traces during busy hours over a relatively long period of time and also found that the Gaussian distribution is a good fit for the captured traffic. Schmidt et al.~\cite{2014-ifip-conf} found that the degree of Gaussianity is affected by short and intensive activities of single network hosts that create sudden traffic bursts. All the above mentioned works agreed on the Gaussian or `fairly Gaussian' traffic at different levels of aggregations in terms of timescale and number of users. The authors in~\cite{TestingGaussianApproximation,heavy-sigcomm-2001} examined the levels of aggregation required to observe Gaussianity in the modelled traffic, and concluded that this can be disturbed by traffic bursts. The work in~\cite{iccTrafficcharact,12-GLOBECOM2002} reinforces the argument above, by showing existence of large traffic spikes at short timescales which result in high values in the tail. Compared to existing literature, our findings are based on a modern, principled statistical methodology, and traffic traces that are spatially and temporally diverse. We have tested several hypothesised distributions and not just Gaussianity.

An early work drawing attention to the presence of heavy tails in Internet file sizes (not traffic) is that of Crovella and Bestavros~\cite{self-sim97}. Deciding whether Internet flows could be heavy-tailed became important as this implies significant departures from Gaussianity. The authors in~\cite{heavytails2017} provided robust evidence for the presence of various kinds of scaling, and in particular, heavy-tailed sources and long-range dependence in a large dataset of traffic spanning a duration of 14 years. 

When modelling network traffic, many authors did not perform any tests for stationarity, assuming that their traces are realisations of a weakly stationary stochastic process~\cite{ieee-ton-94,self-similarity-Ethernet}. Other authors consider non-stationary behaviour of traffic observations~\cite{nonstationaryPoisson,NonstationarityIP}. The authors in~\cite{WideAreaInternetCharacteristics,failurePoisson} show that traffic patterns have almost deterministic daily variations resulting in clear non-stationary behaviour on a day timescale. The authors in~\cite{nonstationaryPoisson} demonstrated that multiplexed traffic on a high-speed link may have non-stationary behaviour and discussed possible causes of non-stationarity of traffic observations. They argue that this could be due to time-varying number of aggregated sources, routing changes or specific aggregation of a constant number of stationary sources. A common approach in traffic modelling is to choose sufficiently small blocks of observations such that observations in separate blocks are expected to be at least weakly stationary~\cite{stationaryModelingLS, ResDimension, StationarityPacketLossKurose}. For example, when testing for applicability of the Gaussian model to traffic modelling authors in~\cite{TestingGaussianApproximation} neglected a part of their trace claiming that it may introduce undesirable non-stationary behaviour. Authors in~\cite{NonstationarityIP} assumed that $5$-minute blocks of their traffic observations are sufficient to ensure intra-block~stationarity.

Understanding the traffic characteristics and how these evolve is crucial for ISPs for network planning and link dimensioning. Operators typically over-provision their networks. A common approach to do so is to calculate the average bandwidth utilisation~\cite{proveGaussianIFIP} and add a safety margin. As a rule of thumb, this margin is defined as a percentage of the calculated bandwidth utilisation~\cite{cisco}. Meent et al.~\cite{ieee-network-2009} proposed  a new bandwidth provisioning formula, which calculates the minimum bandwidth that guarantees the required performance, according to an underlying SLA. This approach relies on the statistical parameters of the captured traffic and a performance parameter. The underlying fundamental assumption for this to work is that the traffic the network operator sees follows a Gaussian distribution. Same approach has been used in~\cite{transaction2015}.

The 95th percentile method is used widely for network traffic billing. Dimitropoulos et al.~\cite{billPAM} have found that the computed 95th percentile is significantly affected by traffic aggregation parameters. However, in their approach they do not assume any underlying model of the traffic; instead, they base their study on specific captured traces. Stanojevic et al.~\cite{95percentileIMC} proposed the use of Shapley value for computing the contribution of each flow to the 95th percentile price of interconnect links. Works ~\cite{Tosendornot,bulkTransfers,Tuangou,requestmapping} propose calculating the 95th percentile using experimental approaches. Xu et al.~\cite{gaussianPercentile} assume that network traffic follows a Gaussian distribution ``through reasonable aggregation'' and propose a cost-efficient data centre selection approach based on the 95th percentile.


\section{Conclusion}
\label{sec:conclusion}

The distribution of traffic on Internet links is an important fundamental problem that has received relatively little attention.  We use a well-known, state-of-the-art statistical framework to investigate the problem using a large corpus of traces.  The traces cover several network settings including home user access links, tier 1 backbone links and campus to Internet links.  The traces are from times from  2002 to 2020  and are from a number of different countries.  We investigated the distribution of the amount of traffic observed on a link in a given (small) aggregation timescale which we varied from $5$ms to $5$s.  The hypotheses compared were that the traffic volume was heavy-tailed, that the traffic was log-normal and that the traffic was normal (Gaussian).  The vast majority of traces fitted the log-normal assumption best and this remained true for all timescales tried.  Where no distribution tested was a good fit this could be attributed either to the link being saturated (at full capacity) for a large part of the observation or exhibiting signs of link-failure (no or very low traffic for part of the observation).

We tested the data for the hypothesis of stationarity. Over long periods (hours and days) the data is not stationary as it is subject to daily and weekly behaviour related to human activity. Over a fifteen minute or one hour period our tests show that the data is stationary when aggregated at timescales of $500$ms to $5$s and is first-difference stationary when aggregated at smaller time-scales from $5$ms to $100$ms.

We investigate the impact of the distribution on two sample traffic engineering problems.  Firstly, we looked at predicting the proportion of time a link will exceed a given capacity.  This could be useful for provisioning links or for predicting when SLA violation is likely to occur.  Secondly, we looked at predicting the 95th percentile transit bill that ISP might be given.  For both of these problems the log-normal distribution gave a more accurate result than a heavy-tailed distribution or a Gaussian distribution.  We conclude that the log-normal distribution is a good (best) fit for traffic volume on normally functioning internet links in a variety of settings and over a variety of timescales, and further argue  that this assumption can make a large difference to  statistically predicted outcomes for applied network engineering problems.

\textcolor{black}{\noindent\textbf{Limitations and future work.} It is important to point out that the presented procedures on bandwidth provisioning and 95th-percentile pricing are not meant to be fully worked systems for practical deployment. This would require testing our model with larger data sets that last for days, weeks or even months. Instead, we intended to motivate the need for deriving good data models for Internet traffic volumes which would be crucial in developing real-world systems. As part of our future work, we will investigate the possibility of developing a bimodal distribution that can fit the anomalous traces and explore more candidate distributions for fitting Internet traffic volume data. Also, studying several months of data would be necessary to investigate the variability between hours of the day, days of the week and so on.}

\bibliographystyle{IEEEtran}
\bibliography{ref}

\begin{thebibliography}{10}
\providecommand{\url}[1]{#1}
\csname url@samestyle\endcsname
\providecommand{\newblock}{\relax}
\providecommand{\bibinfo}[2]{#2}
\providecommand{\BIBentrySTDinterwordspacing}{\spaceskip=0pt\relax}
\providecommand{\BIBentryALTinterwordstretchfactor}{4}
\providecommand{\BIBentryALTinterwordspacing}{\spaceskip=\fontdimen2\font plus
\BIBentryALTinterwordstretchfactor\fontdimen3\font minus
  \fontdimen4\font\relax}
\providecommand{\BIBforeignlanguage}[2]{{%
\expandafter\ifx\csname l@#1\endcsname\relax
\typeout{** WARNING: IEEEtran.bst: No hyphenation pattern has been}%
\typeout{** loaded for the language `#1'. Using the pattern for}%
\typeout{** the default language instead.}%
\else
\language=\csname l@#1\endcsname
\fi
#2}}
\providecommand{\BIBdecl}{\relax}
\BIBdecl

\bibitem{selfSimilarity95}
P.~Pruthi \emph{et~al.}, ``{Heavy-tailed on/off source behavior and
  self-similar traffic},'' in \emph{Proc. of ICC}, 1995.

\bibitem{self-sim97}
M.~E. Crovella \emph{et~al.}, ``{Self-similarity in World Wide Web traffic:
  evidence and possible causes},'' in \emph{IEEE/ACM ToN}, 1997.

\bibitem{heavy-tailed-2010-trans}
P.~Loiseau \emph{et~al.}, ``{Investigating Heavy-Tailed Distributions on a
  Large-Scale Experimental Facility},'' in \emph{IEEE/ACM ToN}, 2010.

\bibitem{95percentileIMC}
R.~Stanojevic \emph{et~al.}, ``{On Economic Heavy Hitters: Shapley Value
  Analysis of 95Th-percentile Pricing},'' in \emph{Proc. of IMC}, 2010.

\bibitem{Gaussian-everywhere}
R.~Meent \emph{et~al.}, ``{Gaussian traffic everywhere?}'' in \emph{Proc. of
  ICC}, 2006.

\bibitem{proveGaussianIFIP}
R.~d.~O.~Schmidt \emph{et~al.}, ``{Measurement-based network link
  dimensioning},'' in \emph{Proc. of IFIP Networking}, 2015.

\bibitem{Gaussian-revisited}
R.~d.~O.~Schmidt, R.~Sadre \emph{et~al.}, ``{Gaussian traffic revisited},'' in
  \emph{Proc. of IFIP Networking}, 2013.

\bibitem{2014-ifip-conf}
R.~d.~O.~Schmidt, R.~Sadre, N.~Melnikov \emph{et~al.}, ``{Linking network usage
  patterns to traffic Gaussianity fit},'' in \emph{Proc. of IFIP Networking},
  2014.

\bibitem{12-GLOBECOM2002}
X.~Yang, ``{Designing traffic profiles for bursty Internet traffic},'' in
  \emph{Proc. of GLOBECOM}, 2002.

\bibitem{clauset}
A.~Clauset \emph{et~al.}, ``{Power-law distributions in empirical data},''
  \emph{arXiv:0706.1062v2}, 2009.

\bibitem{our-infocom-paper}
M.~{Alasmar} \emph{et~al.}, ``{On the Distribution of Traffic Volumes in the
  Internet and its Implications},'' in \emph{Proc. of INFOCOM}, 2019.

\bibitem{Lauks2011}
G.~Lauks \emph{et~al.}, ``{Testing the Null Hypothesis of Stationarity of
  Internet Traffic},'' in \emph{Elektronika Ir Elektrotechnika}, 2011.

\bibitem{stationaryModelingLS}
D.~Moltchanov, ``{Modeling local stationary behavior of Internet traffic},'' in
  \emph{Journal of Communications Software and Systems}, 2008.

\bibitem{NonstationarityIP}
J.~Cao \emph{et~al.}, ``{On the Nonstationarity of Internet Traffic},'' in
  \emph{Proc. of SIGMETRICS}, 2001.

\bibitem{caidaRef}
\BIBentryALTinterwordspacing
``{The CAIDA UCSD Anonymized Internet Traces},'' 2016. [Online]. Available:
  \url{http://www.caida.org/data/passive/passive_dataset.xml}
\BIBentrySTDinterwordspacing

\bibitem{mawiRef}
\BIBentryALTinterwordspacing
``{Mawi Archive},'' 2018. [Online]. Available: \url{http://mawi.wide.ad.jp/}
\BIBentrySTDinterwordspacing

\bibitem{Meent2010Traces}
R.~Barbosa \emph{et~al.}, ``{Simpleweb/University of Twente Traffic Traces Data
  Repository},'' http://eprints.eemcs.utwente.nl/17829/, Tech. Rep., 2010.

\bibitem{waikatoRef}
\BIBentryALTinterwordspacing
``{WITS: Waikato Internet Traffic Storage},'' 2013. [Online]. Available:
  \url{https://wand.net.nz/wits/waikato/8/}
\BIBentrySTDinterwordspacing

\bibitem{aucklandRef}
\BIBentryALTinterwordspacing
``{WITS: Auckland X},'' 2009. [Online]. Available:
  \url{https://wand.net.nz/wits/auck/10/}
\BIBentrySTDinterwordspacing

\bibitem{Alstott}
J.~Alstott \emph{et~al.}, ``{powerlaw: a Python package for analysis of
  heavy-tailed distributions},'' in \emph{arXiv:1305.0215}, 2014.

\bibitem{moThesis}
M.~Alasmar, ``{Understanding the characteristics of Internet traffic and
  designing an efficient RaptorQ-based data transport protocol for modern data
  centres},'' in \emph{Ph.D. thesis, University of Sussex,
  https://sro.sussex.ac.uk/id/eprint/89371/}, 2019.

\bibitem{ResDimension}
M.~Mandjes and R.~van~de Meent, ``{Resource Dimensioning Through Buffer
  Sampling},'' in \emph{IEEE/ACM Transactions on Networking}, 2009.

\bibitem{transaction2015}
R.~d.~O.~Schmidt \emph{et~al.}, ``{Impact of Packet Sampling on Link
  Dimensioning},'' in \emph{Transactions on Network and Service Management},
  2015.

\bibitem{TestingGaussianApproximation}
J.~Kilpi \emph{et~al.}, ``{Testing the Gaussian Approximation of Aggregate
  Traffic},'' in \emph{Proc. of SIGCOMM}, 2002.

\bibitem{ieee-network-2009}
A.~Pras \emph{et~al.}, ``{Dimensioning network links: a new look at equivalent
  bandwidth},'' in \emph{IEEE Network}, 2009.

\bibitem{adf-test}
D.~Dickey and W.~Fuller, ``{Distribution of the Estimators for Autoregressive
  Time Series With a Unit Root},'' in \emph{JSTOR}, 1979.

\bibitem{pp-test}
P.~C. B.~Phillips and P.~Perron, ``{Testing for a Unit Root in Time Series
  Regression},'' in \emph{Journal of the American Statistical Association},
  1988.

\bibitem{kpss-test}
D.~Kwiatkowski \emph{et~al.}, ``{Testing the null hypothesis of stationarity
  against the alternative of a unit root},'' in \emph{Journal of Econometrics},
  1992.

\bibitem{statsmodels}
S.~Seabold \emph{et~al.}, ``{statsmodels: Econometric and statistical modeling
  with python},'' in \emph{9th Python in Science Conference}, 2010.

\bibitem{firstOrderDiff}
R.~Hyndman \emph{et~al.}, ``{Forecasting: Principles and Practice},'' in
  \emph{OTexts: [Online] https://otexts.com/fpp2/}, 2018.

\bibitem{cisco}
\BIBentryALTinterwordspacing
``{Best Practices in Core Network Capacity Planning},'' online, accessed
  December 2020. [Online]. Available:
  \url{https://www.cisco.com/c/en/us/products/collateral/routers/wan-automation-engine/white_paper_c11-728551.html}
\BIBentrySTDinterwordspacing

\bibitem{NetworkLinkDimensioning}
M.~Alasmar \emph{et~al.}, ``Network link dimensioning based on statistical
  analysis and modeling of real internet traffic,'' \emph{arXiv:1710.00420},
  2017.

\bibitem{billPAM}
X.~Dimitropoulos \emph{et~al.}, ``{On the 95-Percentile Billing Method},'' in
  \emph{Proc. of PAM}, 2009.

\bibitem{BusyhourTraff}
Garc\'{\i}a-Dorado \emph{et~al.}, ``{Characterization of the busy-hour traffic
  of IP networks based on their intrinsic features},'' in \emph{Computer
  Networks}, 2011.

\bibitem{heavy-sigcomm-2001}
A.~B. Downey, ``{Evidence for Long-tailed Distributions in the Internet},'' in
  \emph{Proc. of SIGCOMM Workshop on Internet Measurement}, 2001.

\bibitem{iccTrafficcharact}
H.~Abrahamsson \emph{et~al.}, ``{Traffic characteristics on 1Gbit/s access
  aggregation links},'' in \emph{Proc. of ICC}, 2017.

\bibitem{heavytails2017}
R.~Fontugne and et. al., ``Scaling in internet traffic: A 14 year and 3 day
  longitudinal study, with multiscale analyses and random projections,''
  \emph{IEEE/ACM Transactions on Networking}, 2017.

\bibitem{ieee-ton-94}
W.~E. Leland \emph{et~al.}, ``{On the self-similar nature of Ethernet
  traffic},'' in \emph{IEEE/ACM ToN}, 1994.

\bibitem{self-similarity-Ethernet}
W.~Willinger \emph{et~al.}, ``{Self-similarity through high-variability:
  statistical analysis of Ethernet LAN traffic},'' in \emph{IEEE/ACM ToN},
  1997.

\bibitem{nonstationaryPoisson}
T.~{Karagiannis} \emph{et~al.}, ``{A nonstationary Poisson view of Internet
  traffic},'' in \emph{Proc. of INFOCOM}, 2004.

\bibitem{WideAreaInternetCharacteristics}
K.~{Thompson} \emph{et~al.}, ``{Wide-area Internet traffic patterns and
  characteristics},'' in \emph{IEEE Network}, 1997.

\bibitem{failurePoisson}
V.~{Paxson} \emph{et~al.}, ``{Wide area traffic: the failure of Poisson
  modeling},'' in \emph{IEEE/ACM Transactions on Networking}, 1995.

\bibitem{StationarityPacketLossKurose}
M.~{Yajnik} \emph{et~al.}, ``{Measurement and modelling of the temporal
  dependence in packet loss},'' in \emph{Proc. of INFOCOM}, 1999.

\bibitem{Tosendornot}
L.~Golubchik and et. al., ``{To send or not to send: Reducing the cost of data
  transmission},'' in \emph{Proc. of INFOCOM}, 2013.

\bibitem{bulkTransfers}
N.~Laoutaris \emph{et~al.}, ``{Inter-datacenter Bulk Transfers with
  Netstitcher},'' in \emph{Proc. of SIGCOMM}, 2011.

\bibitem{Tuangou}
I.~Castro \emph{et~al.}, ``{Using Tuangou to Reduce IP Transit Costs},'' in
  \emph{IEEE/ACM Transactions on Networking}, 2014.

\bibitem{requestmapping}
H.~Xu \emph{et~al.}, ``{Joint request mapping and response routing for
  geo-distributed cloud services},'' in \emph{Proc. of INFOCOM}, 2013.

\bibitem{gaussianPercentile}
------, ``{Cost efficient datacenter selection for cloud services},'' in
  \emph{Proc. of ICCC}, 2012.

\end{thebibliography}

\end{document}